\pdfoutput=1
\documentclass[longauth]{aa} 

\usepackage{natbib}
\bibpunct{(}{)}{;}{a}{}{,} 
\usepackage{graphicx}
\usepackage{txfonts}
\usepackage{lscape}
\begin{document}

   \title{Molecular line emission in NGC1068 imaged with ALMA}

   \subtitle{II. The chemistry of the dense molecular gas}

   \author{S. Viti,
          \inst{1}
          S. Garc{\'{\i}}a-Burillo,\inst{2}
          A. Fuente,\inst{2}
          L. K. Hunt,\inst{3}
          A. Usero,\inst{2}
          C. Henkel,\inst{4,5}
          A. Eckart,\inst{6,4}
          S. Martin,\inst{7}
          M. Spaans,\inst{8}
          S. Muller,\inst{9}
          F. Combes,\inst{10}
          M. Krips,\inst{7}
          E. Schinnerer,\inst{11}
          V. Casasola,\inst{12}
          F. Costagliola,\inst{13}
          I. Marquez,\inst{13}
          P. Planesas,\inst{2}
          P. P. van der Werf,\inst{14}
          S. Aalto,\inst{9}
          A. J. Baker,\inst{15}
          F. Boone,\inst{16}
          L. J. Tacconi\inst{17}}

   \institute{Department of Physics and Astronomy, UCL, Gower St., London, WC1E 6BT,UK \\
              \email{sv@star.ucl.ac.uk}
         \and
              Observatorio Astronomico Nacional (OAN)-Observatorio de Madrid, Alfonso XII, 3, 2814-Madrid, Spain\\
\and
INAF - Osservatorio Astrofisico di Arcetri, Largo E. Fermi 5, 50125-Firenze, Italy \\
\and
 Max-Planck-Institut für Radioastronomie, Auf dem Hügel 69, 53121, Bonn, Germany \\
\and
Astronomy Department, King Abdulaziz University, P. O. Box 80203, Jeddah 21589, Saudi Arabia \\
\and
 I. Physikalisches Institut, Universität zu Köln, Zülpicher Str. 77, 50937, Köln, Germany \\
\and
 Institut de Radio Astronomie Millimétrique (IRAM), 300 rue de la Piscine, Domaine Universitaire de Grenoble, 38406-St.Martin d’Hères, France \\
\and
Kapteyn Astronomical Institute, University of Groningen, PO Box 800, 9700 AV, Groningen, The Netherlands \\
         \and
 Department of Radio and Space Science with Onsala Observatory, Chalmers University of Technology, 439 94-Onsala, Sweden \\
\and
 Observatoire de Paris, LERMA, 61 Av. de l’Observatoire, 75014-Paris, France \\
\and
 Max-Planck-Institut für Astronomie, Königstuhl, 17, 69117-Heidelberg, Germany \\
\and
 INAF - Istituto di Radioastronomia \& Italian
ALMA Regional Centre, via Gobetti 101, 40129, Bologna, Italy \\
\and
 Instituto de Astrofíisica de Andalucía (CSIC), Glorieta de la Astronomía s/n Granada, 18080, Spain \\
\and
 Leiden Observatory, Leiden University, PO Box 9513, 2300 RA Leiden, Netherland \\
\and
 Department of Physics and Astronomy, Rutgers, The State University of New Jersey, Piscataway, NJ 08854, USA \\
\and
 Université de Toulouse, UPS-OMP, IRAP, 31028, Toulouse, France \\
\and
 Max-Planck-Institut für extraterrestrische Physik, Postfach 1312, 85741-Garching, Germany}

   \date{Received ; accepted }

 
  \abstract
   {}
   {We present a detailed analysis of ALMA Bands 7 and 9 data of CO, HCO$^+$, HCN and CS, augmented with Plateau de Bure Interferometer (PdBI) data of the $\sim$ 200 pc circumnuclear disk (CND) and the $\sim$ 1.3 kpc starburst ring (SB ring) of NGC~1068, a nearby ($D$ = 14 Mpc) Seyfert 2 barred galaxy. We aim at determining the physical characteristics of the dense gas present in the CND 
and whether the different line intensity ratios we find within the CND as well as between the CND and the SB ring are due to excitation effects (gas density and temperature differences) or to a different chemistry.
   }
   {We estimate the column densities of each species in Local Thermodynamic Equilibrium (LTE). We then compute large one-dimensional non-LTE radiative transfer grids (using RADEX) 
by using first only the CO transitions, and then all the available molecules in order to constrain the densities, temperatures and column densities within the CND. We finally present a preliminary set of chemical models to determine the origin of the gas.  
   }
   {We find that in general the gas in the CND is very dense ($>$ 10$^5$ cm$^{-3}$) and hot (T$>$ 150K), with differences especially in the temperature across the CND. The AGN position 
has the lowest CO/HCO$^+$, CO/HCN and CO/CS column density ratios. RADEX analyses seem to indicate that there is 
chemical differentiation across the CND. We also find differences between the chemistry of the SB ring and some regions of the CND; the SB ring is also much colder and less dense than the CND. Chemical modelling does not succeed in reproducing all the molecular ratios with one model per region, suggesting the presence of 
multi-gas phase components. }
   {The LTE, RADEX and chemical analyses all indicate that more than one gas-phase component  is necessary to uniquely fit all the available molecular ratios within the CND. A higher number of molecular transitions at the ALMA resolution is necessary to quantitatively determine the physical and chemical characteristics of these components. }

   \keywords{Galaxies: active - Galaxies: individual: NGC1068 – Galaxies: ISM – Galaxies: nuclei – Molecular processes - Radio lines: galaxies}
\titlerunning{The chemistry of dense gas in NGC1068}
\authorrunning{Viti et al.} 
   \maketitle
%

\section{Introduction}
\label{sec:intro}
Galaxy evolution is now thought to depend on feedback mechanisms whereby star
formation is inhibited at some point in the galaxy's lifetime either by an
active galactic nucleus (AGN) or by stellar feedback through winds and
supernovae. Up to now it has been very difficult to capture directly the
observational signatures of {\em mechanical} feedback, but recent molecular emission-line
observations  are opening a new window on this aspect of galaxy evolution by the detection of molecular 
outflows \citep[e.g.,][hereafter Paper I]{Aalt12,Feru10,Veil13,Comb13,Cico14,Garc14}.
Signs of  {\em radiative} feedback in galactic nuclei are complicated by the interplay of a
central starburst with an AGN.  Most of the gas in galaxy centers
resides in molecules, and because most galaxy centers tend to be obscured
by dust, it is crucial to identify the chemical nature of the gas involved
and to determine the physical conditions that are most prone to lead to a starburst; 
hence
sub-millimeter molecular tracers have become one of the most important
probes of physical conditions in galaxy nuclei.  Through comparison with
models, molecular abundances and line ratios can be used to infer physical
conditions but also to constrain the chemical reactions that drive molecule
formation and destruction in irradiated regions.  Molecular abundances in starburst and AGN-dominated galaxies 
are usually 
interpreted through models for Photon-Dominated Regions (PDRs) and X-ray
Dominated Regions (XDRs), and are among the best diagnostics for disentangling
massive star formation and the influence of an AGN \citep{Meij07,Baye09}.
More recently, in an attempt to interpret the observations of molecules that may arise from very dense 
and possibly shocked gas,  chemical gas-grain models as well as shocks models have also been used to interpret 
molecular lines in these galaxies \citep[e.g.,][]{Alad13}.  

Massive stars are usually forming in huge
concentrations of dense (n(H$_2$)$\sim$ 10$^5$ cm$^{-3}$) and warm ($T_k\ge$ 50 K) molecular gas, which are able to survive disruptive forces (winds or radiation)
from nearby newly formed stars, longer than the gas in the local interstellar medium \citep[e.g.,][]{Zinn07}. 
Molecules constitute a significant fraction of the interstellar gas and, due to their high critical densities, trace the dense 
regions where
star formation takes place \citep[e.g.,][]{Omont07}. Moreover, in general and at large scales,
the overall abundance of molecules is determined by the density of the gas, which favours their formation, and the energetic processes, such as UV or X-ray radiation,  which in general tend to destroy molecules. 
Subject to a proper interpretation, observations of molecules can be used to (i) trace the matter that is
the reservoir or leftover of the star formation process; (ii) trace the process of star formation
itself; and (iii) determine the influence of newly-formed stars or an AGN on their
environments and hence the galaxy energetics.

Although $^{12}$CO is the most common tracer of the global molecular reservoir, for the nearest sources we now have an inventory of species 
that allow us to study in detail the different gas components within a galaxy as well as its energetics 
\citep[e.g][]{Garc02,Wang04,User04,Fuen08,Aalt11,Mart11,Alad13,Costa11,Kame11,Watanabe14}.
A common trend observed in recent surveys is a chemical
diversity and complexity that is hard to explain by a
one-component gas phase; as with our own Galaxy, it is becoming clear that indeed, 
different molecules will trace different regions of a galaxy \citep[e.g.,][]{Meie05,Meie12,User06,Garc10}, from the cold, relatively low density molecular gas traced by CO, to highly shocked regions,  traced by e.g., SiO, 
and finally to very high density star forming clouds (traced by e.g., CS and possibly CH$_3$OH). 
Often high abundances of particular species such as HCO$^+$ are automatically attributed to the presence of strong PDRs, since this molecule is indeed abundant in the presence of a strong UV field, regardless of the metallicity, 
cosmic ray ionisation rate or density of the galaxy \citep{Baye09}. Nevertheless this species is also highly enhanced when the cosmic ray or X-ray ionisation rates are high, regardless of the UV field \citep{Baye11,Meij05}. 
This apparent degeneracy is common to most molecules and again reflects the fact that chemistry is a non linear process influenced by a combination of energetics and physical conditions.   

Galaxies characterized by both AGN and vigorous circumnuclear star formation present a complex combination of energetic processes
(UV--rays, X--rays plus shocks associated with AGN feedback) that can take
over the processing of the interstellar medium (ISM)  in the circumnuclear regions \citep[e.g.,][]{Comb06}.  
For these objects the connection between the starburst and the central AGN is still unknown, although large amounts of dense circumnuclear gas imply that the two regions must `interact'. 
There is no unique molecular tracer for such dense gas in these complex systems; theoretical models \citep[e.g.,][]{Baye08,Baye09} show that all the molecules that ought to be detectable in AGN plus starburst systems are also 
characteristics of `pure' starburst galaxies. High-resolution imaging of molecular line emission is a key to disentangling the contribution of the  starburst from that of the AGN component in composite systems. Observations seem to indicate that abundance ratios of HCN/HCO$^+$ may help trace the dominating energetic processes \citep[e.g.,][]{Krip08,Krip11,Aalt11,Loen08}; 
however the chemistry of both these species is highly dependent on the physical conditions of the gas as well as on its energetics.
Galaxies where starbursts and AGN are both present therefore require both high spatial resolution and targeted multi-species, multi-line observations in 
order to disentangle and characterize the energetics and chemical effects 
of the molecular gas.
\par
 One of the most studied examples of a composite starburst/AGN galaxy is NGC~1068, a prototypical nearby (D $\sim$ 14 Mpc) Seyfert 2 galaxy, subject of numerous observational campaigns focused on the study of the fuelling of its central region and related feedback activity using molecular line observations \citep[e.g.,][]{User04,Isra09,Kame11,Hai12,Alad13}.  Interferometric observations of CO, as well as HCN, HCO$^+$ CS and SiO \citep[e.g.,][]{Tacc94,Schi00,Garc10,Krip11,Tsai12} clearly show that molecular tracers of dense gas are essential in order to spatially resolve the distribution, kinematics and excitation of the circumnuclear gas of NGC~1068 and to study  the relationship between the $r\sim1-1.5$~kpc starburst (SB) ring and the $r\sim200$~pc circumnuclear disk (CND) located around the AGN. A summary of interferometric
observations of NGC~1068 can be found in the accompanying Paper I, and references therein. We summarize below the main results of Paper I and the aims of the present work.

\subsection{Summary of Paper I and aims of this paper}\label{summary}

In order to probe the bulk of the dense molecular gas in the $r\sim2$~kpc disk of NGC~1068 we carried out ALMA Cycle 0 observations in Bands 7 and 9 of several molecular transitions, namely CO (3-2) and (6-5), HCO$^+$ (4-3), HCN (4-3) and CS (7-6)
within the $r\sim200$ pc CND as well as in the $\sim$ 1--1.5 kpc SB ring. The data and their reduction are detailed in Paper I.
We list below the main observational results of Paper I that are relevant to this work:
\begin{itemize}
\item
We found evidence from several tracers that the kinematics of the dense molecular gas ($n(H_{\rm 2})>$10$^5$cm$^{-3}$) from $r\sim50$~pc out to $r\sim$400~pc are shaped by a massive ($M_{\rm gas}\sim2.7\times10^{7}~M_{\odot}$) outflow, which reveals the signature of {\em mechanical} feedback produced by the AGN. The outflow involves a sizeable fraction ($\geq50\%$) of the total gas reservoir in this region.

\item
Molecular line ratios derived from the ALMA data show significant differences between the SB ring and the CND. Furthermore, line ratio maps at the
20--35~pc resolution of ALMA show up to order-of-magnitude changes inside the CND. Changes are correlated  with the UV/X-ray illumination of the molecular gas at the CND by the AGN. Overall, these results hint at a strong AGN {\em radiative} feedback on the excitation/chemistry of the molecular gas.
\end{itemize}

In the present paper we concentrate on the chemical analysis of the gas within the CND and towards a particularly prominent star-forming knot of the SB ring,  with the aim of quantifying the chemical differentiation and of determining the chemical origin of such differentiation. This is the highest resolution multi-line/multi-species chemical study of this object to date, and allows us, for the first time, to spatially resolve and study sub-regions within the CND.

This paper is subdivided as follows: in Section 2 we report the molecular line ratios and briefly describe the sub-regions of the CND, as determined by our ALMA observations in Paper I; in Section 3 we analyse the data by performing Local Thermodynamical Equilibrium  (LTE) calculations, as well as radiative transfer modelling by the use of RADEX, developed by Van der Tak et al. (2007). In Section 4 we attempt to interpret the data via chemical modelling; in Section 5 we list our conclusions.

\section{Molecular line ratios}

\subsection{Methodology}\label{global}

\begin{figure*}[tbh!]
\centering
\includegraphics[width=18cm]{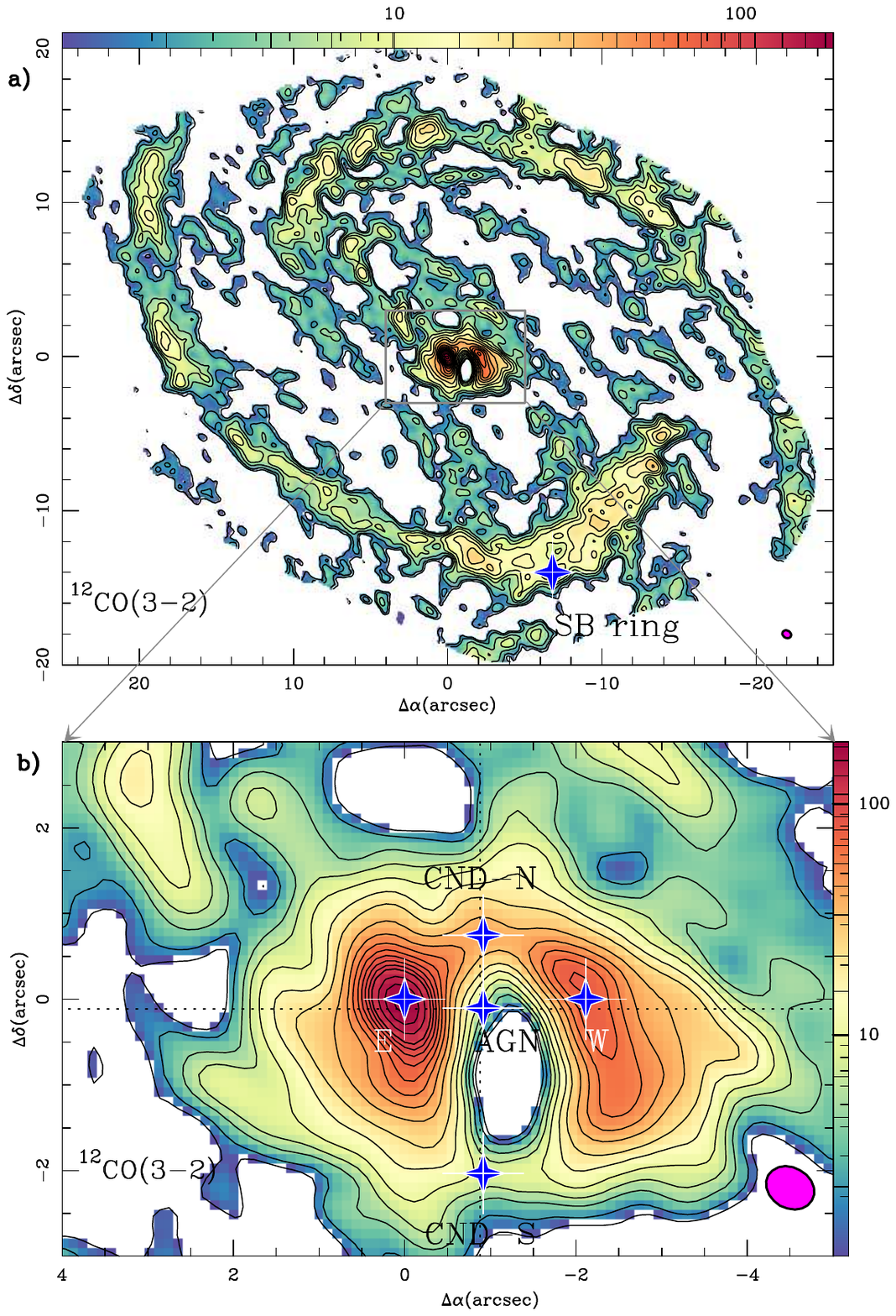}
\caption{CO (3-2) velocity-integrated intensity map, measured in Jy/beam km/s over a 460 km s$^{-1}$ window,  obtained with ALMA (see Paper I Figure 4 for more details). The central position is the phase tracking center (J2000( RA$_0$, DEC$_0$) = (02$^h$42$^m$40$^s$.771), (-00$^0$00$^{'}$47$^{"}$.84)). The Top panel: Map encompassing the circumnuclear disk (CND) and the starburst ring. Bottom panel: same as above but zooming in the CND region. The filled ellipses in the lower right corners represent the spatial resolution of the observations. }
\label{intco}
\end{figure*}

\begin{figure*}[tbh!]
\centering
\includegraphics[width=18cm]{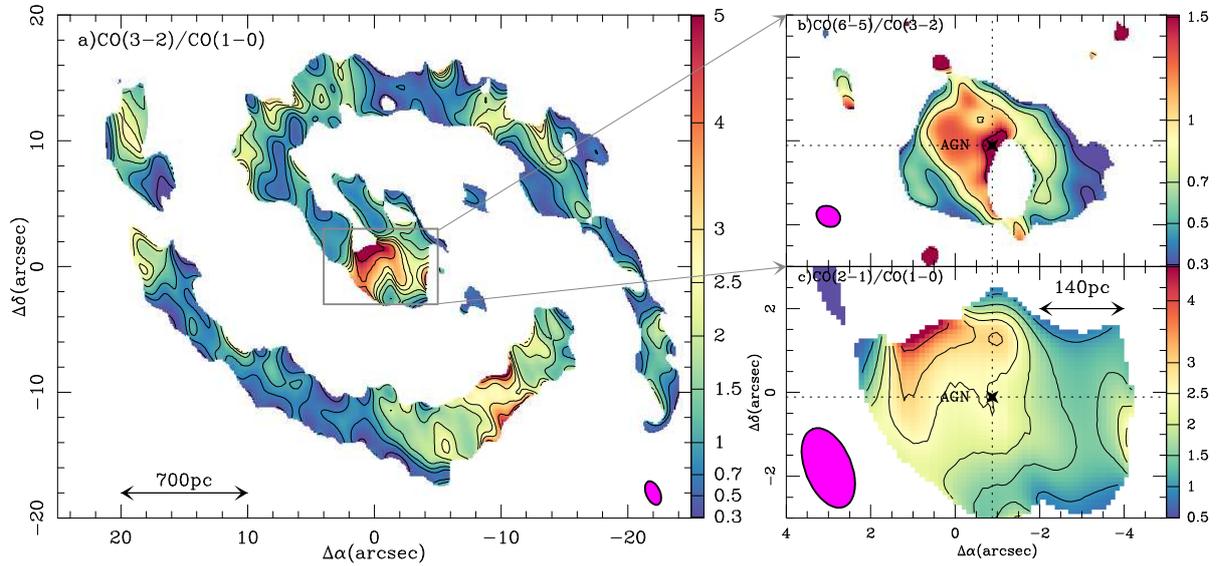}
\caption{Left: Velocity-integrated intensity ratios CO(3-2)/CO(1-0) in the CND as well as in the SB ring. Right: CO(6-5)/CO(3-2) and CO(2-1)/CO(1-0) in the CND. The filled ellipses represent the spatial resolutions used to derive the line
ratio maps: 2$^"\times$1.1$"$ PA = 22$^0$ for CO(3-2)/CO(1-0) and CO(2-1)/CO(1-0);  0.6$^"\times$0.5$^"$ PA = 60$^0$   for CO(6-5)/CO(3-2)}
\label{ratio1}
\end{figure*}

\begin{figure*}[tbh!]
\centerline{%
\includegraphics[width=0.7\textwidth]{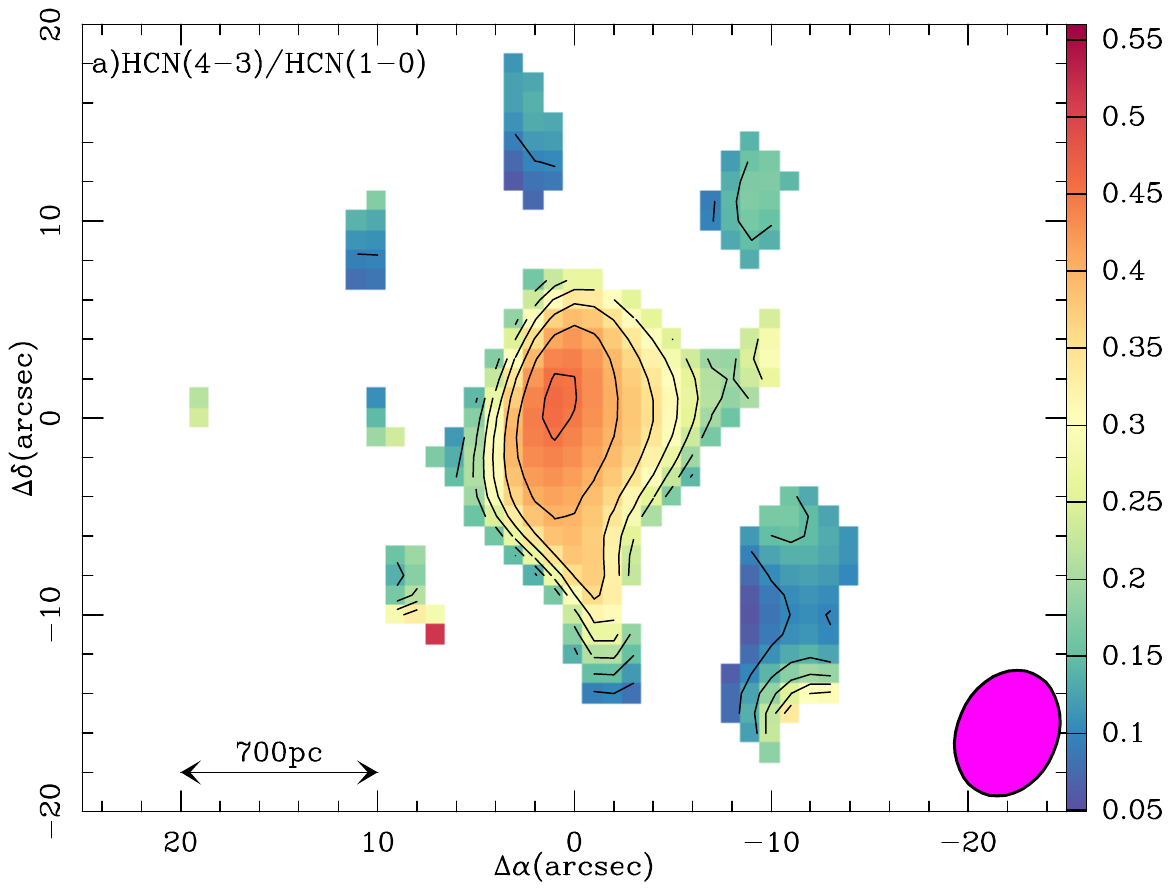}
\includegraphics[width=0.7\textwidth]{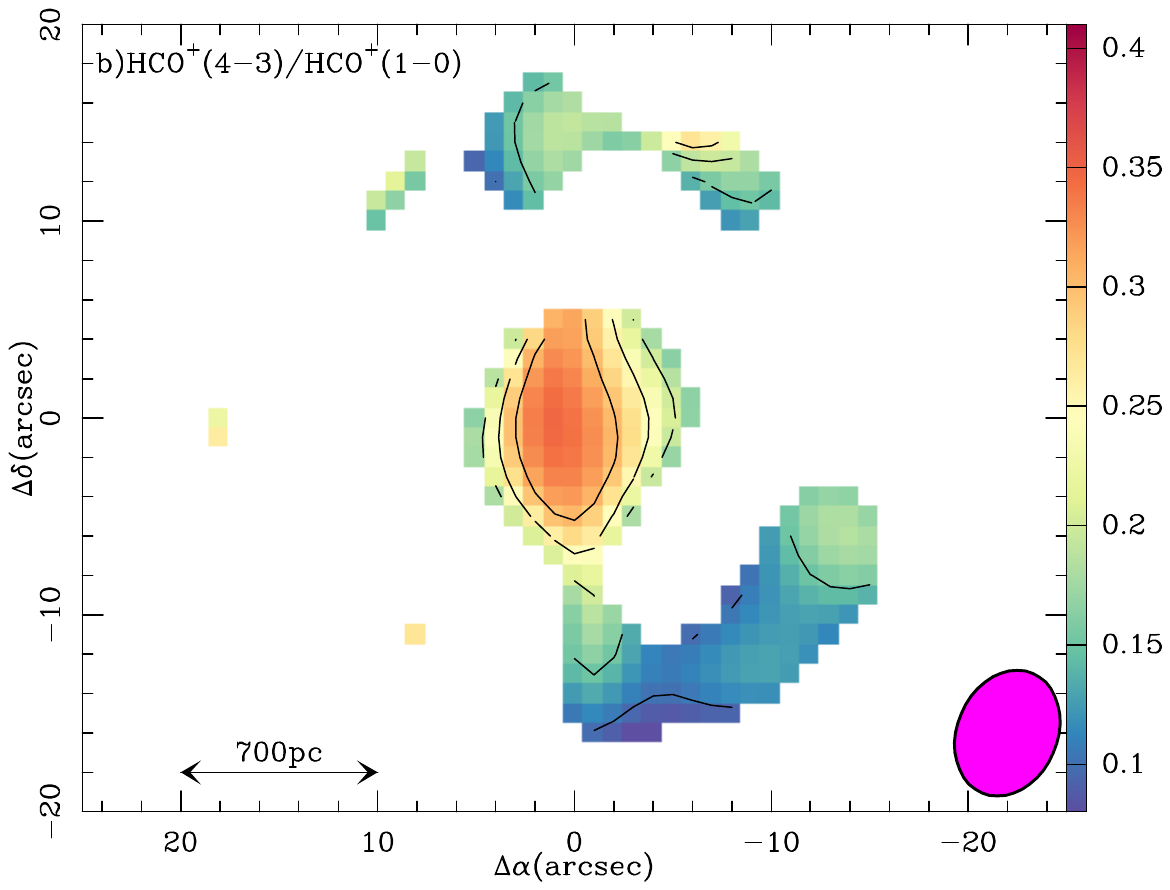}
}%
\caption{Velocity-integrated intensity ratios HCN(4-3)/HCN(1-0) (Left) and HCO$^+$(4-3)/HCO$^+$(1-0) (Right) across the CND as well as in the SB ring. The filled ellipses in the lower right corners represent the spatial resolutions used to derive the line
ratio maps: 6.6$"\times$5.1$^"$ PA = 157$^0$, for both panels.}
\label{ratio2}
\end{figure*}

\begin{figure*}[tbh!]
\centering
\includegraphics[width=20cm]{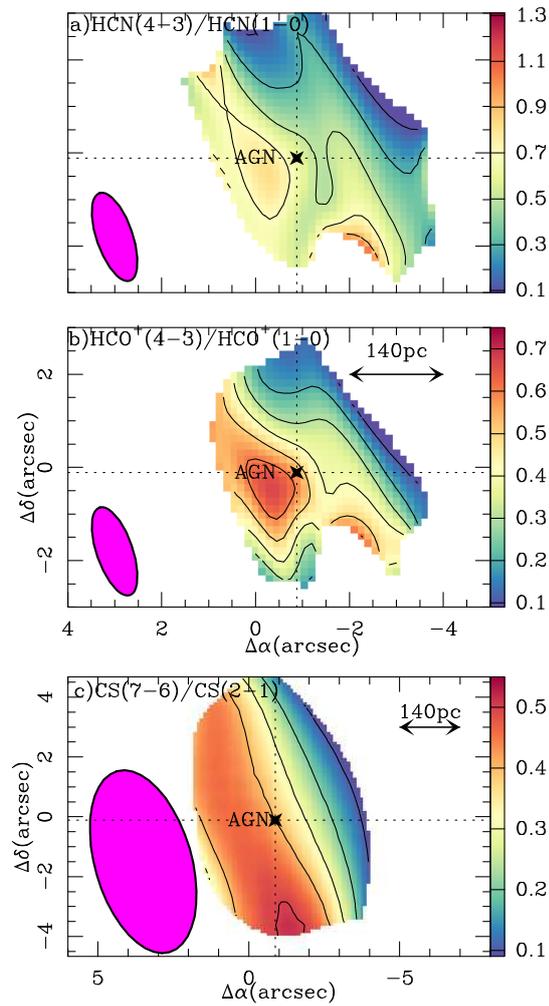}
\caption{Zoomed in velocity-integrated intensity ratios HCN(4-3)/HCN(1-0), HCO$^+$(4-3)/HCO$^+$(1-0) and CS(7-6)/CS(2-1) in the CND. The filled ellipses represent the spatial resolutions used to derive the line ratio maps: 2$^"\times$0.8$^"$, PA = 19$^0$ for HCN(4-3)/HCN(1-0) and HCO$^+$(4-3)/HCO$^+$(1-0); 6.6$^"\times$3.2$^"$ PA = 16$^0$ for CS(7-6)/CS(2-1).}
\label{ratio3}
\end{figure*}

\begin{figure*}[tbh!]
\centering
\includegraphics[width=20cm]{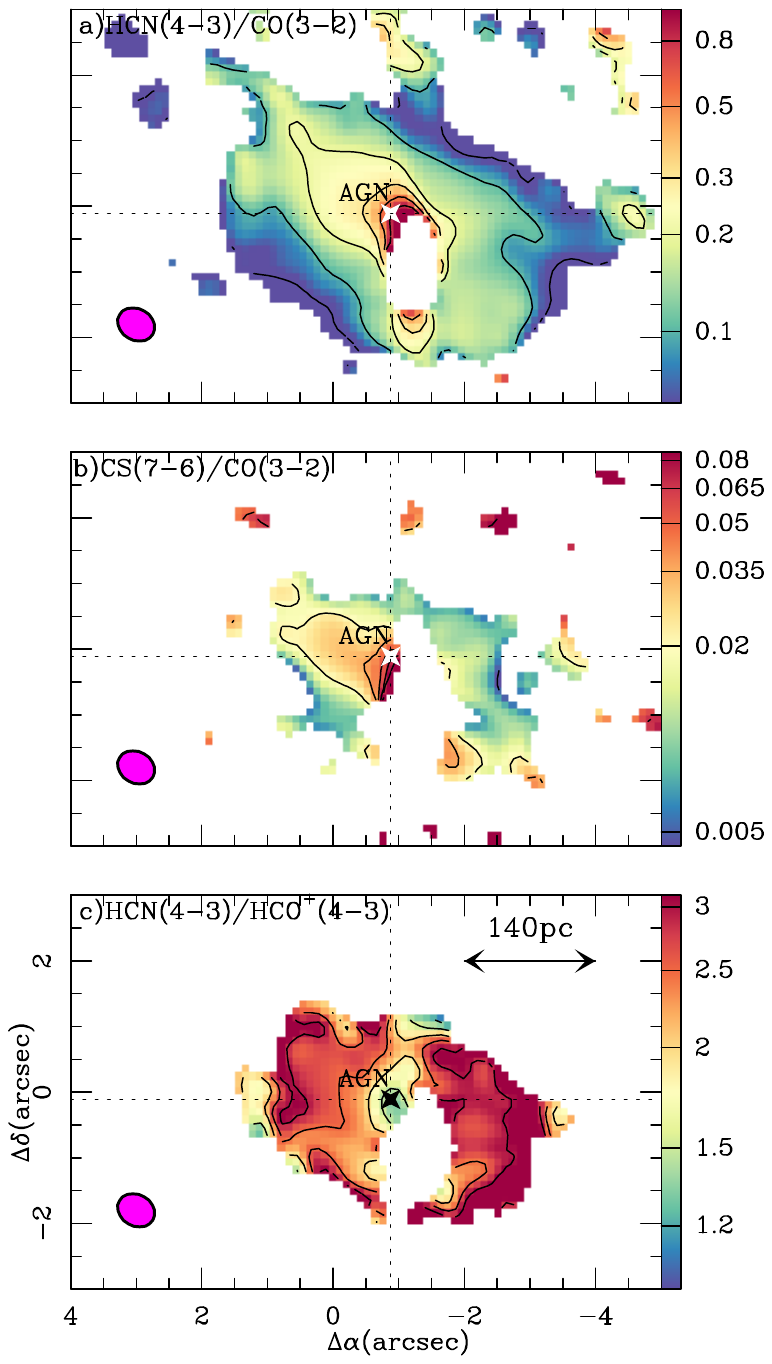}
\caption{HCN(4-3)/CO(3-2), CS(7-6)/CO(3-2) and HCN(4-3)/HCO$^+$(4-3) velocity-integrated intensity ratios in the CND. The filled ellipses represent the spatial resolutions used to derive the line ratio maps:  0.6$^"\times$0.5$^"$ PA = 60$^0$, for all panels.}
\label{ratio4}
\end{figure*}

\begin{figure*}[tbh!]
\centerline{%
\includegraphics[width=0.7\textwidth]{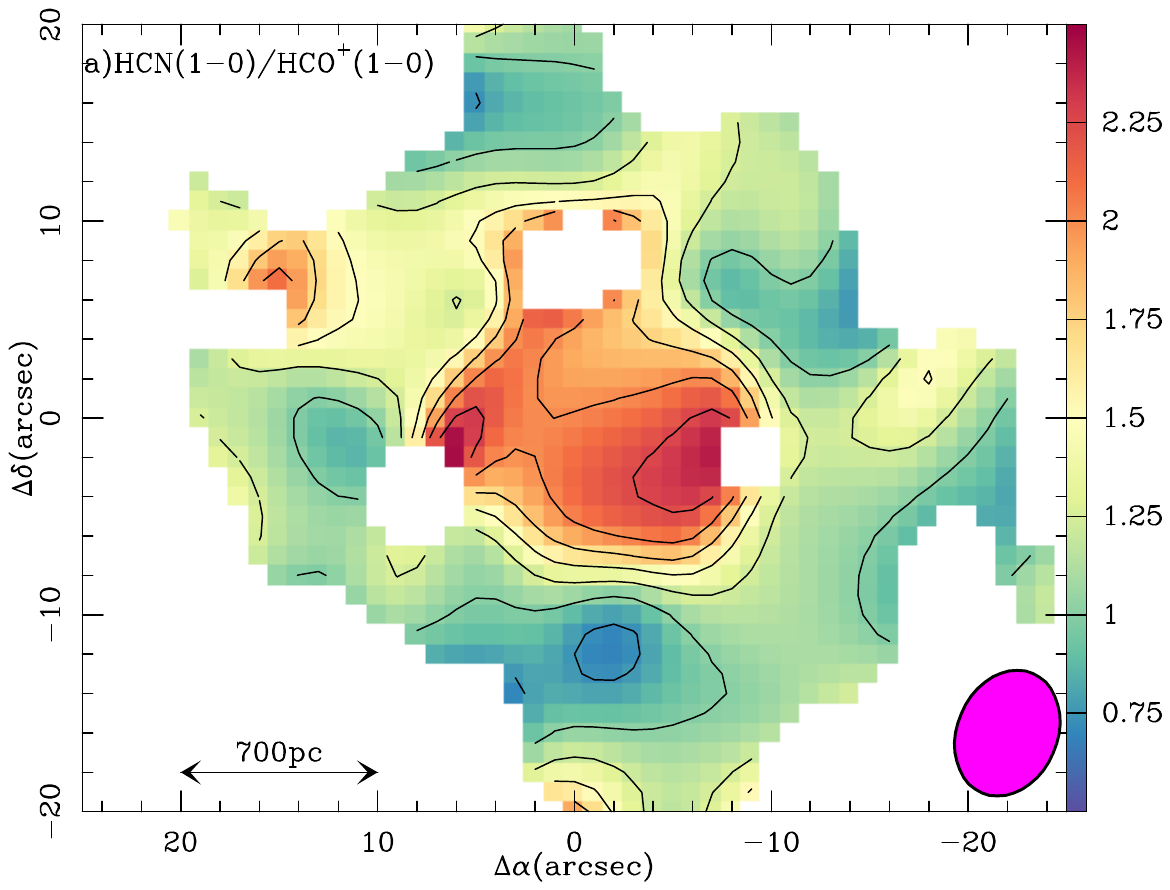}
\includegraphics[width=0.7\textwidth]{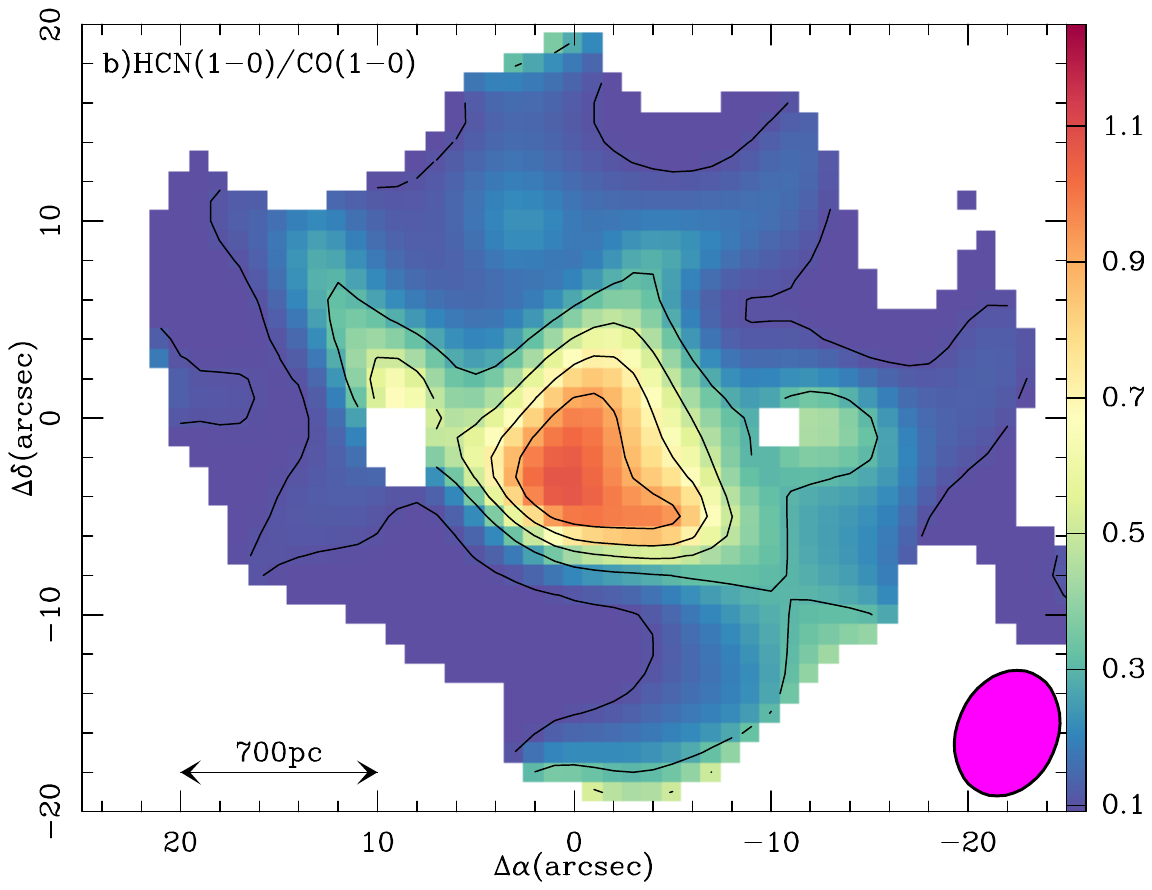}
}%
\caption{The HCN(1-0)/HCO$^+$(1-0) velocity-integrated intensity ratio (Left) and the HCN(1-0)/CO(1-0) velocity-integrated intensity ratio (Right) in the SB ring. The filled ellipses represent the spatial resolutions used to derive the line
ratio maps: 6.6$^"\times$5.1$^"$ PA = 157$^0$.}
\label{ratio5}
\end{figure*}

Figure~\ref{intco}
shows the velocity-integrated intensity map of CO(3-2) in the disk of NGC~1068 obtained with ALMA. This is a reproduction of Figure 4 from Paper I and can be used as a reference `chart' for the present paper.
In Figures~\ref{ratio1} to ~\ref{ratio5} we present the molecular line ratios of the velocity-integrated intensities, in T$_{\rm mb}\times\Delta {\rm v}$ units (K km/s), used for our study of the CND and the SB ring. 
Prior to deriving the line ratios, all the maps
have been first referred to the same tracking center and then degraded to the lower/common 
spatial resolution of the data used to derive each line ratio. The latter is performed in the plane of the sky by an adequate gaussian convolution kernel specifically adapted for each line ratio.

We show and subsequently use in our analysis line ratios which can be considered as {\em independent}, given the limited number of {\em different} intensities available for each region. To that purpose we combined the ALMA maps presented in Paper I with a set of interferometric images of NGC~1068 obtained by the IRAM Plateau de Bure Interferometer (PdBI) for the lower-$J$ transitions of the same molecular species observed by ALMA, but with spatial resolutions which are comparatively lower in the PdBI maps. In particular, some of the ratios make use of the PdBI data for CO(1-0) and CO(2-1) \citep[taken from][]{Schi00}, HCO$^+$(1-0) and HCN(1-0) \citep[taken from][Usero et al. 2014, in prep]{Garc08} and CS(2-1) transitions \citep[taken from][]{Tacc97}.

The adopted methodology for all the analyses, except for that in Section 3.2.3, required that all interferometric images, obtained with original spatial resolutions that range from 20 to 350~pc for the CND (see Table 1 for details), and from 100 pc (for the CO(3-2)/CO(1-0) ratio) to 400 pc (for the rest) for the SB ring, had first to be degraded down to the lower/common spatial resolution of the two transitions involved in each case. For the SB ring, the downgrade to 400~pc in spatial resolution is required in order to recover the emission from the SB ring of the HCN (1-0) and HCO$^+$ (1-0) PdBI data. Intensities were calculated by integrating inside a common 460~km~s$^{-1}$ velocity window (-230, 230~km~s$^{-1}$ relative to the systemic velocity). We also used a common 3$\sigma$ clipping threshold on the intensities when we derived the ratios in order to enhance the reliability of the images. 
For the sake of simplicity, the figures are all centered around the same nominal position of the phase tracking center adopted in Paper I 
(J2000( RA$_0$, DEC$_0$) = (02$^h$42$^m$40$^s$.771), (-00$^0$00$^{'}$47$^{"}$.84)).

None of the maps used to derive the line ratios include short-spacing correction. However we estimate that the amount of flux filtered in the interferometric images used in this work starts to be significant only on scales $>6\arcsec$. This is well above the lowest spatial resolution of any of the line ratio maps used in this work, where we purposely use fluxes extracted from {\em single} apertures that range from  $\sim0.5\arcsec$ to $\sim5\arcsec$ in order to minimize the bias due to the lack of short spacings.

In our analysis of line ratios we also use single-dish (IRAM 30m) estimates of the $^{12}$C/$^{13}$C isotopic line ratios for CO, HCN and HCO$^+$ to constrain the opacity range of the solutions \citep[][Usero, priv. communication]{Papa99,User04}.

\subsection{The ALMA `topology' of NGC1068}\label{CND}

From Figures 2 to 6 and also summarizing the findings from Paper I,
we find that, overall  molecular line ratios show significant differences between the SB ring and the CND. In particular, ratios that measure the excitation of CO, HCN and HCO$^+$ (like the 3--2/1--0 ratio of CO or the 4--3/1--0 ratio of HCN or HCO$^+$) are a factor of 2--5 higher in the CND on average. Furthermore, {\em mixed} ratios that involve transitions of different species having distinctly different critical densities (like HCN(1-0)/CO(1-0) or HCN(1-0)/HCO$^+$(1-0)) are also a factor  2--5 higher in the CND relative to the SB ring. Taken at face value this result hints at a comparatively  higher excitation and/or enhanced abundances of dense gas tracers in the CND. Disentangling excitation from chemistry effects is crucial if we are to interpret the observed differences between the SB and the CND, and it is one of the aims of this paper.

As discussed in Paper I, molecular line ratios at the full resolution of our ALMA data ($\sim$ 0.3"--0.5", or 20--35 pc at the assumed distance of 14 Mpc) allow us to measure significant differences within the CND. In the following we define five spatially resolved sub-regions inside the CND, denoted as E Knot, W Knot, AGN, CND-N, and CND-S (see Figure~\ref{intco}, lower panel). The E and W Knots coincide with the two emission peaks in the CO ALMA maps of the CND. The AGN knot corresponds to the position of the central engine. The CND-N and CND-S knots are located north and south of the AGN locus in the CND ring, close to the contact points of the jet-ISM working surface; they are therefore prime candidates to explore shock-driven chemistry related to the molecular outflow discussed in Paper I. We list the  coordinates of all sub-regions  in Table~\ref{subregions}.  
The remarkable differences in all line ratios across the CND sub-regions may be associated with an uneven degree of illumination of molecular gas by UV/X-ray photons from the AGN (see discussion in Section~7 of Paper I).

As already mentioned, line ratios show a notable differentiation between the CND and the SB ring taken as a whole. In an attempt to
optimize the information derived from the high-resolution data used in this work, we have identified a representative star-forming knot along the SB ring, where a maximum number of molecular lines are detected. We have taken this position as a reference in our comparison with the conditions derived individually for the CND knots. This helps putting the results on an equal footing thanks to the use of similar apertures.
The chosen SB ring knot is located in the southern region of the pseudo-ring (see Figure~\ref{intco}) and coincides with a strong Pa$\alpha$ emission peak. We take this knot as representative, based on similar line ratios measured in an ensemble of 12 molecular star-forming knots detected in the dense gas tracers of the SB ring (see Section~5 of Paper I). Furthermore, the use of fluxes extracted from a {\em single} $\leq5\arcsec$ aperture is preferred to adopting line ratio values averaged over the whole SB ring on scales $>20\arcsec$, as we thus minimize the amount of flux filtered out in the interferometric images used in this work, which starts to be relevant only on scales $>6\arcsec$, as discussed in Section.~\ref{global}.

As a frame of reference, we list in Table~\ref{res} the original aperture sizes in parsec for each transition prior to the spatial resolution downgrade. We report in Table~\ref{line-int} the individual line intensities and in Table~\ref{ratios-trans}, in the common (lowest) spatial resolution,
all the intensity ratios with their errors for each subregion within the CND, as well as in the SB ring knot.
The uncertainties in Table~\ref{ratios-trans} have been estimated from the errors in the individual intensities which we take to be 15\% for Band 7 data, 30\% for Band 9 data, and 10\% for the PdBI data.

\begin{table*}
\caption{Original spatial resolution in parsecs for all (ALMA as well as PdBI) molecular transitions used in this paper. We assume a distance of 14 Mpc.}
\label{res}
\begin{tabular}{cccc}
\hline
Transition & Spatial resolution  & Average Spatial Resolution (pc) & Instrument \\ 
\hline
CO (1-0) &  $2\arcsec\times1.1\arcsec$ (140~pc$\times$77~pc) &109 & PdB$^1$\\
CO (2-1)  & $2\arcsec\times1.1\arcsec$ (140~pc$\times$77~pc) &109 & PdB$^1$ \\
CO (3-2) &  $0.6\arcsec\times0.5\arcsec$ (42~pc$\times$35~pc)  & 39 & ALMA Band 7 \\
CO (6-5) &  $0.4\arcsec\times0.2\arcsec$ (28~pc$\times$14~pc)   & 21 & ALMA Band 9 \\
HCO$^+$ (1-0) high-res. & $2\arcsec\times0.8\arcsec$ (140~pc$\times$56~pc)	 & 98 & PdBI$^2$ \\
HCO$^+$ (1-0) low-res.  & $6.6\arcsec\times5.1\arcsec$ (460~pc$\times$350~pc)	 & 405 & PdBI$^2$ \\
HCO$^+$ (4-3) & $0.6\arcsec\times0.5\arcsec$ (42~pc$\times$35~pc) & 39 & ALMA Band 7 \\
HCN(1-0) high-res. & $2\arcsec\times0.8\arcsec$ (140~pc$\times$56~pc)	 & 98 & PdBI$^2$ \\
HCN(1-0) low-res.  & $6.6\arcsec\times5.1\arcsec$ (460~pc$\times$350~pc)	 & 405 & PdBI$^2$ \\
HCN (4-3) & $0.6\arcsec\times0.5\arcsec$ (42~pc$\times$35~pc) & 39 &ALMA Band 7 \\
CS (2-1)   & $6.6\arcsec\times3.2\arcsec$ (462~pc$\times$224~pc)    & 343 & PdBI$^3$ \\
CS (7-6) & $0.6\arcsec\times0.5\arcsec$ (42~pc$\times$35~pc) & 39 &  ALMA Band 7 \\
\hline
\multicolumn{3}{c}{$^1$ \citet{Schi00}; $^2$ \citet{Garc08};  $^3$ \citet{Tacc97}}
\end{tabular}
\end{table*}

\begin{table*}[tbh!]
\caption{Coordinates, and RA and DEC offsets, relative to the AGN,
of the five regions within the CND and the star-forming knot of the SB ring}
\label{subregions}
\begin{tabular}{|c|cc|c|}
\hline
Name & RA & DEC & ($\Delta$RA, $\Delta$DEC) (") \\
\hline
E Knot & 02:42:40.771 & -00:00:47.84 & (0.9,0.1)  \\
W Knot & 02:42:40.630 &  -00:00:47.84 & (-1.3,0.1) \\
AGN & 02:42:40.710 &  -00:00:47.94 & (0,0) \\
CND-N &  02:42:40.710 & -00:00:47.09 & (0,0.85) \\
CND-S & 02:42:40.710 & -00:00:49.87 & (0,0.07) \\
SB ring & 02:42:40.317 & -00:01:01.84 & (-5.9,-14) \\
\hline
\end{tabular}
\end{table*}

\begin{table*}
\caption{Molecular velocity-integrated intensities, measured in K km/s, for each region within the CND as well as in the starburst ring.  The errors on the intensities are not listed but for our analysis are taken to be 15\% for the ALMA Band 7 data, 30\% for Band 9 data, and 10\% for the PdBI data (see Table~\ref{ratios-trans}). 
For the CND, all the high resolution (35 pc) data have been reduced to a spatial resolution of 100 pc (1.4"). The conversion to Jy/beam is 1 K = 0.215 Jy/beam for the CO(3-2) transition. }
\label{line-int}
\begin{tabular}{c|ccccc|c}
\hline
Molecule & E Knot & W Knot & AGN & CND-N & CND-S & SB ring \\
\hline
CO(1-0) & 597.33 & 500.47 & 310.87 & 203.57 & 298.23 & 115.29 \\
CO(2-1) & 1480.28 & 823.21 & 773.35 & 554.65 & 482.57 & -- \\
CO(3-2) & 2346.72 & 1587.95 & 911.03 & 833.28 & 429.84 & 160.51 \\
CO(6-5) & 2712.7 & 937.20 & 1024.85 & 783.62 & 366.60 & -- \\
HCO$^+$(1-0) & 404.37 & 236.20 & 152.15& 178.47 & 91.69 & 20.22 \\ 
HCO$^+$(4-3) & 251.18 & 82.18 & 82.52 & 58.09 & 27.73 & 2.22 \\
HCN(1-0) & 821.32 & 466.06 & 291.49 & 312.71 & 143.12 & 19.84 \\
HCN(4-3) & 639.47 & 233.47 & 174.29 & 130.02 & 75.58 & 2.22 \\
CS(2-1) & 18.46 & 20.28 & 20.97 & 21.07 & 14.54 & 0.56 \\
CS(7-6) & 8.26 & 5.23 & 8.33 & 7.61 & 6.89 & -- \\
\hline
\end{tabular}
\end{table*}

\begin{table*}
\caption{Molecular ratios of velocity-integrated intensities (and their error in brackets - see Section 2.2), measured in K km/s, rounded to the first decimal point in most cases, for each region within the CND as well as in the starburst ring. 
For the CND, all the high resolution (0.5" or 35 pc) data have been reduced to a spatial resolution of 100 pc (1.4"). Note that for the CS(7-6)/(2-1) ratio the common spatial resolution is 350 pc. For the SB ring the CO(3-2)/(1-0) ratio was measured at the common resolution of 100 pc, while the rest of the ratios at 400 pc. }
\label{ratios-trans}
\begin{tabular}{c|ccccc|c}
\hline
Ratios & E Knot & W Knot & AGN & CND-N & CND-S & SB ring \\
\hline
CO(2-1)/(1-0) & 2.5 (0.4) & 1.6 (0.2) & 2.5 (0.4)& 2.7 (0.4) & 1.6 (0.2) &  --- \\
CO(3-2)/(1-0)& 3.9 (0.7) & 3.2 (0.6) &2.9 (0.5) &4.1 (0.7) &  1.4 (0.3) & 1.4(0.3) \\
CO(6-5)/(3-2) & 1.2 (0.4)&  0.6 (0.2) &1.1 (0.4)& 0.9 (0.3)&  0.9 (0.3) & --- \\
HCO$^+$(4-3)/HCO$^+$(1-0) & 0.6 (0.1) & 0.3 (0.1) &0.5 (0.1)& 0.3 (0.1)& 0.3 (0.1)& 0.1 (0.01) \\
 HCN(4-3)/CO(3-2) & 0.3 (0.1)& 0.1 (0.03) & 0.2 (0.04)& 0.16 (0.03)& 0.18 (0.03)&  0.01 (0.003) \\
HCN(4-3)/HCN(1-0) & 0.8 (0.1)& 0.5 (0.1) & 0.6 (0.1)& 0.4 (0.1)&  0.5 (0.1) & 0.1 (0.02) \\
HCN(4-3)/HCO$^+$(4-3) & 2.5 (0.5) & 2.8 (0.6) & 2.1 (0.4) & 2.2 (0.5) & 2.7 (0.6) & 1 (0.2) \\  
CS(7-6)/CO(3-2) & 0.004 (0.001) & 0.0033 (0.001) &  0.0091 (0.002)  & 0.0091 (0.0i02) & 0.0160 (0.003) & --- \\
CS(7-6)/(2-1) & 0.4 (0.1) & 0.25 (0.04) &0.4 (0.1)  &0.4 (0.1) & 0.5 (0.1) & --- \\
\hline
\end{tabular}
\end{table*}

In the next two Sections we will perform different analyses with the common aim of (i) estimating the column densities of each species across the CND and SB ring knot (by combining LTE analyses and RADEX computations), (ii) determining the gas density and temperature of each region within the CND, as well as the SB ring knot (via RADEX modelling), and by combining the results from (i) and (ii), (iii) attempt to determine the origin of the chemistry by the use of the chemical and shock model UCL\_CHEM \citep{Viti04,Viti11}.  

\section{Data analysis: the physical characteristics of the gas within the CND and SB ring}
\subsection{LTE analysis}
\label{subsec:lte}

\noindent

In an environment that is in LTE the total
column densities of the observed species can be obtained from observations of a single transition, provided  the kinetic temperature ($T_k$) of the gas is known and that the emission in that transition is optically thin.
In LTE at temperature $T_k$ the column density of the upper level $u$ of a particular transition is related to the total column density via the Boltzmann equation:

\begin{equation}
N=\frac{N_u Z}{g_ue^{-E_u/kT_k}},
\label{eq:N}
\end{equation}
where $N$ is the total column density of the species, $Z$
is the partition function, $g_u$ is the statistical weight of the level $u$ and $E_u$ is its energy above the ground state. If optically thin, and assuming a filling factor of unity, the column density $N_u$ is related to the observed antenna temperature:
\begin{equation}
N_u = \frac{8\pi k \nu^2 I}{hc^3A_{ul}}
\label{eq:Nu}
\end{equation}
where $I$ is the integrated line intensity (in K km s$^{-1}$) over frequency.

\par For all our analyses, where we are working with ratios of
transitions with similar beams, line ratios under these approximations can be considered meaningful, and Equation~\ref{eq:N} is sufficient; in Table~\ref{cd1and2} we present the
LTE total column densities for each species calculated for several kinetic temperatures (from 50 to 250 K, although the LTE calculations were performed from 10 to 300 K) 
and from each transition, for each region of the CND, as well as for the SB ring. Note that we have made the implicit assumption that all our transitions are optically thin. In reality this is likely not the case and hence the derived LTE column densities are a lower limit: if (i) 
the emitting gas is in true LTE, (ii) each transition were optically thin,  and (iii) the beam and velocity filling factors were the same for each transition, then the same total column density at a specific temperature should be obtainable from each transition. This is clearly not the case (and in fact differences in molecular ratios of different transitions can be quite instructive); for CO, in
most cases, the differences are within an order of magnitude, with the exception of the column density derived from the CO (6-5) line;  at times these
differ by slightly more than an order of magnitude when compared  with the column density derived from the $J$=1-0 line, especially in the case of CND-S. However for HCO$^+$ and HCN the $J$=4--3 transition gives a much lower column density than the $J$=1--0 one, suggesting that the high $J$ lines may be optically thick (see Section 3.2.1), or sub-thermally excited. 
From Table~\ref{cd1and2} we note that for most temperatures the E Knot is characterized by larger column densities than the other regions.
Within the CND, the least chemically rich region seems to be CND-S. Nevertheless, the differences in column densities across the CND rarely exceeds
one order of magnitude.  In general when comparing the CND and the SB ring in Table~\ref{cd1and2}, we find that
the SB ring contains gas with lower column densities by at least one order of magnitude compared to the CND, in agreement with the analysis of the molecular ratios presented in Paper I.

We may be able to better evaluate
the chemical richness across the CND and potential differences between the latter and the SB ring by looking at column density ratios derived exclusively from the transitions
observed with ALMA: \\
{\it Within the CND}: Krips et al. (2008, 2011) RADEX simulations give an upper limit for the {\em average} kinetic temperature within the CND of 200 K; as we shall see later (see Section~\ref{subsec:radex}) our RADEX analysis of the individual components within the CND give temperatures in the range of $\sim$ 60--250 K.  Therefore taking an intermediate kinetic temperature of 150 K, the ratio of the CO to HCN 
column densities derived, respectively,  from CO(3-2) and HCN(4-3) lines (whose levels have similar lower and upper energies) varies from $\sim$ 2700 in the E Knot to $>$ 5000 in the W Knot,
hinting to the highest and lowest abundance of the high density tracer HCN in the E Knot and W Knot respectively.
The ratio of the CO to HCO$^+$ column densities derived, respectively,  from CO(3-2) and HCO$^+$(4-3) is lowest in the E Knot ($\sim$ 11,700) and AGN ($\sim$ 13,700) and highest in the W Knot ($\sim$ 23,300), with the ratios in the CND-S and CND-N both quite high as well (19,500 and 17,800). Thus the AGN and the E Knot could be the regions where most energetic activity is occurring since HCO$^+$ can be enhanced by cosmic ray, X-ray or UV activity. 

We only have a single high $J$ transition of CS observed with ALMA. If we make the very simplistic assumption that high $J$ CS preferentially trace dense, shielded 
and hot (due to shocks) gas, then we can use the ratio of the CO to CS column densities derived respectively from the CO(6-5) and CS(7-6) across the CND to 
obtain an idea of the diversity in the hotter gas across this region.
We find that this ratio is lowest for the AGN and the CND-S ($\sim$ 5500--6000) and highest for the E Knot (33,000), which could be an indication that the AGN and the CND-S are characterized by a large fraction of hot (shocked) gas, while the E Knot is the coolest region in the CND. 
Of course the above analysis is hampered by the fact that the ALMA transitions may suffer more of optical depth effects and hence the LTE approximation may be less valid. \\
{\it CND vs SB ring}: Both the column density ratios, CO/HCN and CO/HCO$^+$, estimated from the ALMA data are much higher in the SB ring ($\sim$ 50,000 and $\sim$ 90,000 respectively) than in the CND (the highest being $\sim$ 5000 and $\sim$ 23,000 respectively), indicating that both HCN and HCO$^+$ are more abundant in the CND (unless vibrational excitation by infrared photons is in fact responsible for the different ratios, a possibility we shall explore in a future paper). The fact that CO (and hence H$_2$) is abundant but other common species remain underabundant may be an indication that the gas has
a relatively low density (traced by CO but not by higher density tracers, such as HCN and CS); this explanation would also 
be consistent with the relatively low HCN/HCO$^+$ ratio found in the SB ring, HCN being a high density tracer.

In conclusion, a simple LTE analysis provide us with the first clear indication that chemical differentiatin exists across the CND and between the CND and the SB ring. In the next subsection we present the results obtained by producing a
rotation diagram of CO (for which we have four transitions) for all the regions within the CND to assess whether each subregion of the CND can in fact
be modelled by a one gas phase component.

\begin{table*}
\small
\caption{Total beam averaged column densities (in cm$^{-2}$) for CO, HCO$^+$, HCN and CS derived using single transitions for a range of temperature assuming LTE and optically thin emission for the CND components, as well as for the SB ring. The transitions observed with ALMA were degraded to a resolution of $\sim$ 100 pc. a(b) stands for a$\times$10$^{b}$.}
\label{cd1and2}
\begin{tabular}{|c|ccccc|c|}
\hline
 T(K)  & E Knot & W Knot & AGN & CND-N & CND-S & SB ring\\
\hline 
\multicolumn{7}{|c|}{CO J=1-0} \\
\hline
  50.0    &1.4(18)  & 1.2(18)  &7.5(17) & 4.9(17) & 7.2(17) & 2.8(17) \\
 100.0    &2.7(18)  & 2.3(18)  & 1.4(18) & 9.3(17) & 1.4(18) & 5.3(17) \\
 150.0    &4.0(18)  & 3.4(18)  & 2.1(18) & 1.4(18) & 2.0(18)  &7.7(17) \\
 200.0   & 5.3(18)  & 4.4(18)  & 2.8(18) & 1.8(18) & 2.6(18)  &1.0(18) \\
 250.0   & 6.6(18)  & 5.5(18)  & 3.4(18) & 2.2(18) & 3.3(18)  &1.3(18) \\
\hline
\multicolumn{7}{|c|}{CO J=2-1} \\
\hline
  50.0    &1.1(18)  & 6.2(17)  &5.8(17) & 4.2(17) & 3.6(17) & \\
 100.0    &2.7(18)  & 1.0(18)  & 9.9(17) & 7.1(17) & 6.2(17)&  \\
 150.0    &2.7(18)  & 1.5(18)  & 1.4(18) & 1.0(18) & 8.7(17)&  \\
 200.0   & 3.5(18)  & 1.9(18)  & 1.8(18) & 1.3(18) & 1.1(18)&  \\
 250.0   & 4.3(18)  & 2.4(18)  & 2.2(18) & 1.6(18) & 1.4(18)&  \\
\hline
\multicolumn{7}{|c|}{CO J=3-2} \\
\hline
  50.0    &1.1(18)  & 7.4(17)  &4.2(17) & 3.9(17) & 2.0(17) & 7.5(16)\\
 100.0    &1.6(18)  & 1.1(18)  & 6.1(17) & 5.6(17) & 2.9(17)& 1.1(17) \\
 150.0    &2.1(18)  & 1.4(18)  & 8.2(17) & 7.5(17) & 3.9(17)& 1.4(17)  \\
 200.0   & 2.7(18)  & 1.8(18)  & 1.0(18) & 9.4(17) & 4.9(17)& 1.8(17) \\
 250.0   & 3.2(18)  & 2.2(18)  & 1.2(18) & 1.1(18) & 5.9(17)& 2.2(17) \\
\hline
\multicolumn{7}{|c|}{CO J=6-5} \\
\hline
  50.0    &1.7(18)  & 5.8(17)  &6.3(17) & 4.8(17) & 2.3(17) & \\
 100.0    &1.0(18)  & 3.6(17)  & 4.0(17) & 3.0(17) & 1.4(17)&  \\
 150.0    &1.1(18)  & 3.7(17)  & 4.0(17) & 3.1(17) & 1.4(17)&  \\
 200.0   & 1.2(18)  & 4.0(17)  & 4.4(17) & 3.4(17) & 1.6(17)&  \\
 250.0   & 1.3(18)  & 4.5(17)  & 5.0(17) & 3.8(17) & 1.8(17)&  \\
\hline
\multicolumn{7}{|c|}{HCO$^+$ J=1-0} \\
\hline
  50.0    &1.3(15)  & 7.4(14)  &4.8(14) & 5.6(14) & 2.9(14)& 6.4(13) \\
 100.0    &2.4(15)  & 1.4(15)  & 9.2(14) & 1.1(15) & 5.5(14)&1.2(14)  \\
 150.0    &3.6(15)  & 2.1(15)  & 1.4(15) & 1.6(15) & 8.2(14)&1.8(14)  \\
 200.0   & 4.8(15)  & 2.8(15)  & 1.8(15) & 2.1(15) & 1.1(15)&2.4(14)  \\
 250.0   & 5.9(15)  & 3.5(15)  & 2.2(15) & 2.6(15) & 1.3(15)&3.0(14)  \\
\hline
\multicolumn{7}{|c|}{HCO$^+$ J=4-3} \\
\hline
  50.0    &1.1(14)  & 3.5(13)  &3.5(13) & 2.5(13) & 1.2(13)& 9.5(11) \\
 100.0    &1.4(14)  & 4.6(13)  & 4.6(13) & 3.2(13) & 1.5(13)&1.2(12) \\
 150.0    &1.8(14)  & 6.0(13)  & 6.0(13) & 4.2(13) & 2.0(13)&1.6(12)   \\
 200.0   & 2.2(14)  & 7.4(13)  & 7.4(13) & 5.2(13) & 2.5(13)&2.0(12)  \\
 250.0   & 2.7(14)  & 8.8(13)  & 8.8(13) & 6.2(13) & 3.0(13)&2.4(12)  \\
\hline
\multicolumn{7}{|c|}{HCN J=1-0} \\
\hline
  50.0    &1.3(16)  & 7.5(15)  &4.7(15) & 5.0(15) & 2.3(15)& 3.2(14) \\
 100.0    &2.5(16)  & 1.4(16)  & 9.0(15) & 9.6(15) & 4.4(15)&6.1(14)  \\
 150.0    &3.7(16)  & 2.1(16)  & 1.3(16) & 1.4(16) & 6.5(15)&9.0(14) \\
 200.0   & 5.0(16)  & 2.8(16)  & 1.8(16) & 1.9(16) & 8.6(15)&1.2(15)  \\
 250.0   & 6.2(16)  & 3.5(16)  & 2.3(16) & 2.4(16) & 1.1(16)&1.5(15)  \\
\hline
\multicolumn{7}{|c|}{HCN J=4-3} \\
\hline
  50.0    &4.6(14)  & 1.7(14)  &1.3(14) & 9.4(13) & 5.5(13)&1.6(12)  \\
 100.0    &6.0(14)  & 2.2(14)  & 1.6(14) & 1.2(14) & 7.1(13)&2.1(12)  \\
 150.0    &7.8(14)  & 2.9(14)  & 2.1(14) & 1.6(14) & 9.3(13) &2.7(12) \\
 200.0   & 9.7(14)  & 3.6(14)  & 2.7(14) & 2.0(14) & 1.2(14)& 3.4(12) \\
 250.0   & 1.2(15)  & 4.3(14)  & 3.2(14) & 2.4(14) & 1.4(14)&4.1(12)  \\
\hline
\multicolumn{7}{|c|}{CS J=2-1} \\
\hline
  50.0    &3.3(14)  & 3.7(14)  &3.8(14) & 3.8(14) & 2.6(14)& 1.0(12) \\
 100.0    &6.2(14)  & 6.8(14)  & 7.0(14) & 7.0(14) & 4.9(14)& 1.9(13) \\
 150.0    &9.1(14)  & 5.0(14)  & 1.0(15) & 1.0(15) & 7.2(14)&2.8(13)  \\
 200.0   & 1.2(15)  & 1.3(15)  & 1.4(15) & 1.4(15) & 9.4(14)&3.7(13)  \\
 250.0   & 1.5(15)  & 1.6(15)  & 1.7(15) & 1.7(15) & 1.2(15)&4.5(13)  \\
 300.0   & 1.8(15)  & 1.9(15)  & 2.0(15) & 2.0(15) & 1.4(15)&  \\
\hline
\multicolumn{7}{|c|}{CS J=7-6} \\
\hline
  50.0    &2.4(13)  & 1.5(13)  &5.2(13) & 2.2(13) & 2.0(13) &         \\
 100.0    &2.4(13)  & 1.5(13)  & 5.4(13) & 2.3(13) & 2.0(13)&         \\
 150.0    &3.0(13)  & 1.9(13)  & 6.5(13) & 2.7(13) & 2.5(13)&   \\
 200.0   & 3.5(13)  & 2.2(13)  & 7.7(13) & 3.2(13) & 2.9(13)&   \\
 250.0   & 4.1(13)  & 2.6(13)  & 9.0(13) & 3.8(13) & 3.4(13)&   \\
\hline
\end{tabular}
\end{table*}

\subsubsection{CO rotation diagram}
\label{subsec:rot}
Since we have more than two CO transitions for each region within the CND, and making the assumptions that LTE conditions hold, that the lines
are optically thin, and that we are mostly in the Rayleigh-Jeans regime, we can use Equation~\ref{eq:N} to construct a rotation diagram which relates the column density per statistical weight
of a number of molecular energy levels to their
energy above the ground state. The assumption that the lines are optically thin may of course not be valid, especially as, in dense gas,  
the $^{12}$CO is well known to be often optically thick.
The rotation diagram is a plot of the natural
logarithm of $N_u/g_u$ versus $E_u/k$ and, if the above mentioned assumptions hold, 
is a straight line with a negative slope of $1/T_{rot}$. The temperature inferred from such a diagram is often referred to as the rotational temperature and it is expected to be equal to the kinetic temperature if all the levels are thermalized \citep{Gold99}.

\begin{figure*}[tbh!]
\centerline{%
\includegraphics[width=0.4\textwidth]{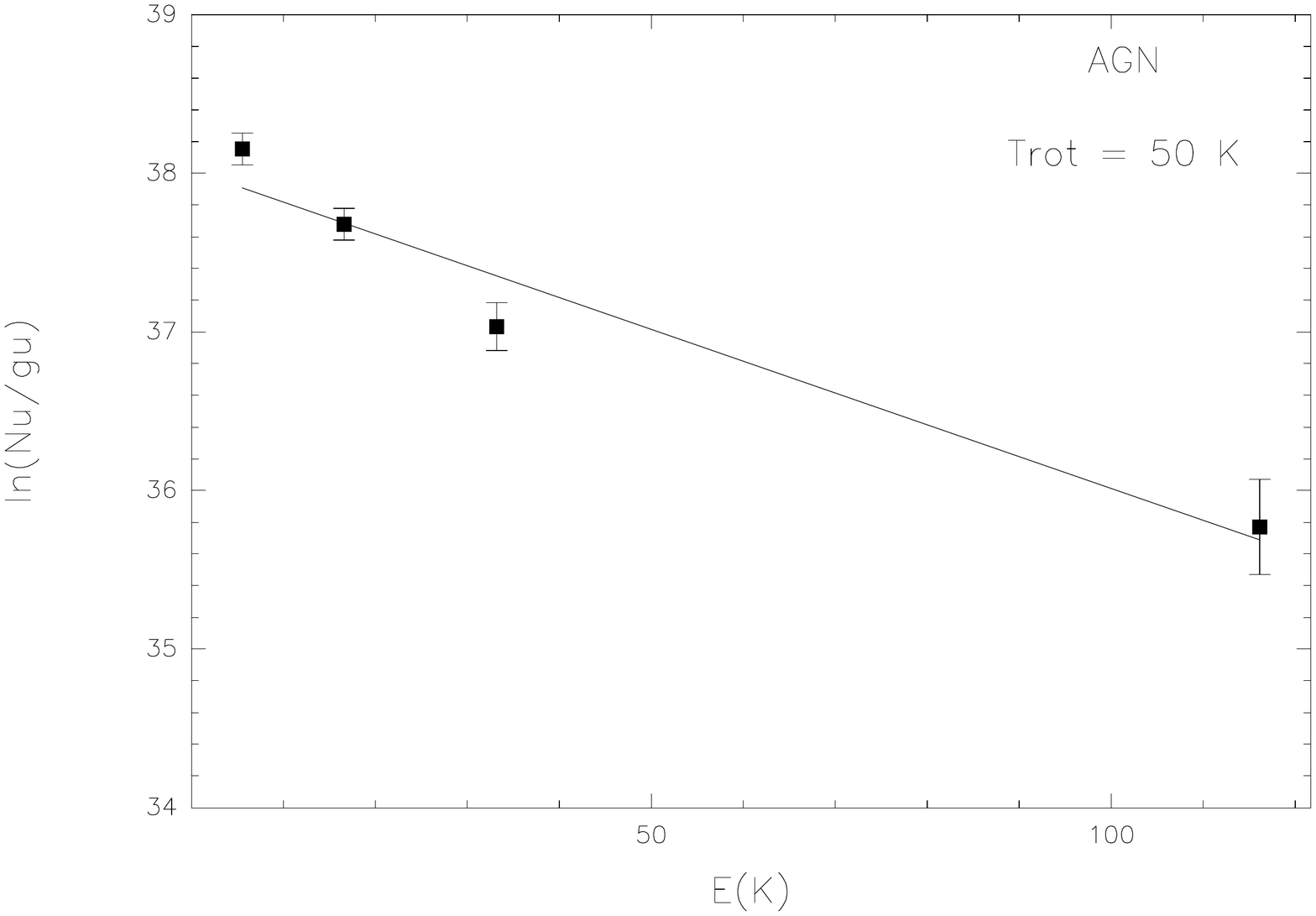}
\includegraphics[width=0.4\textwidth]{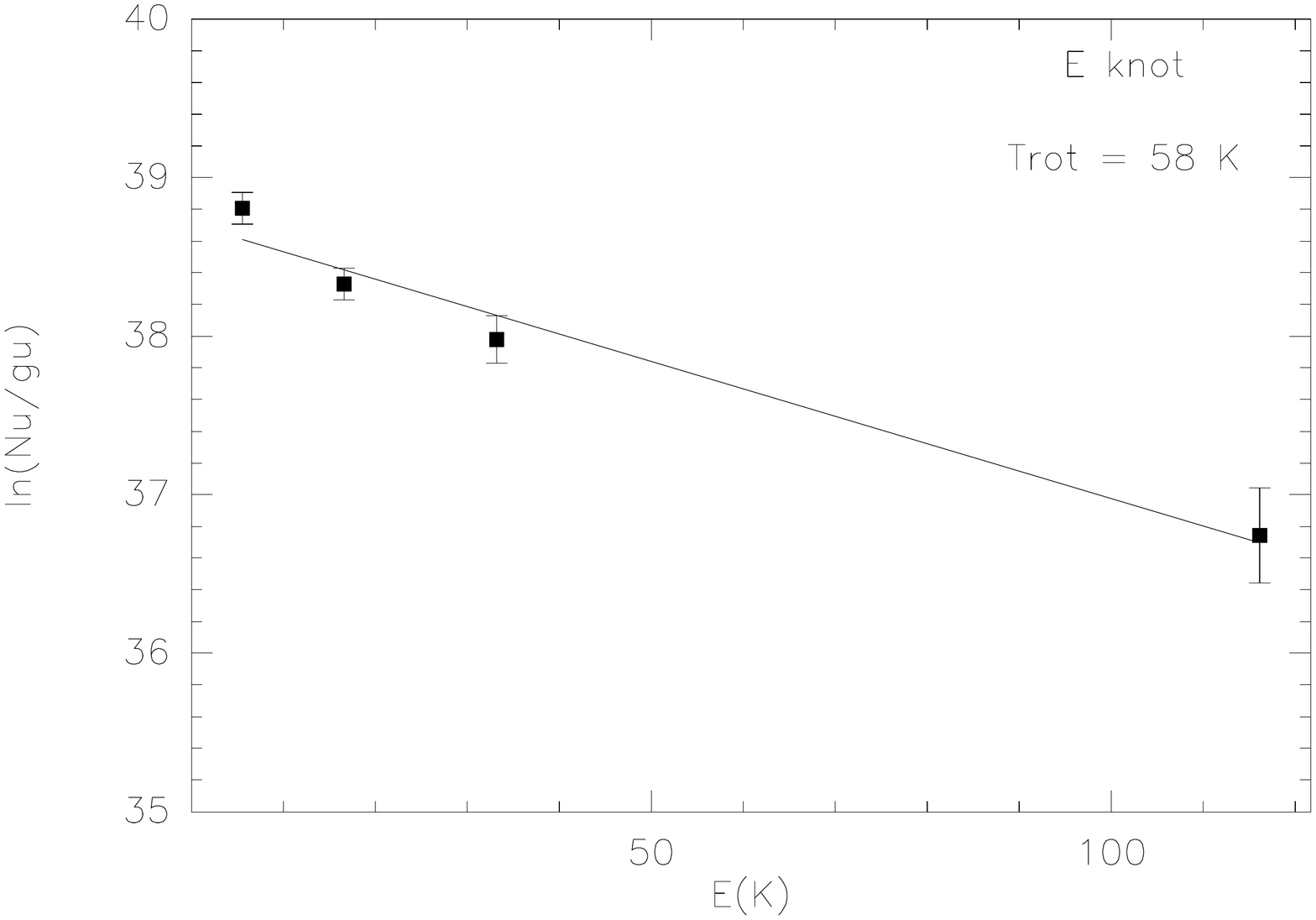}
\includegraphics[width=0.4\textwidth]{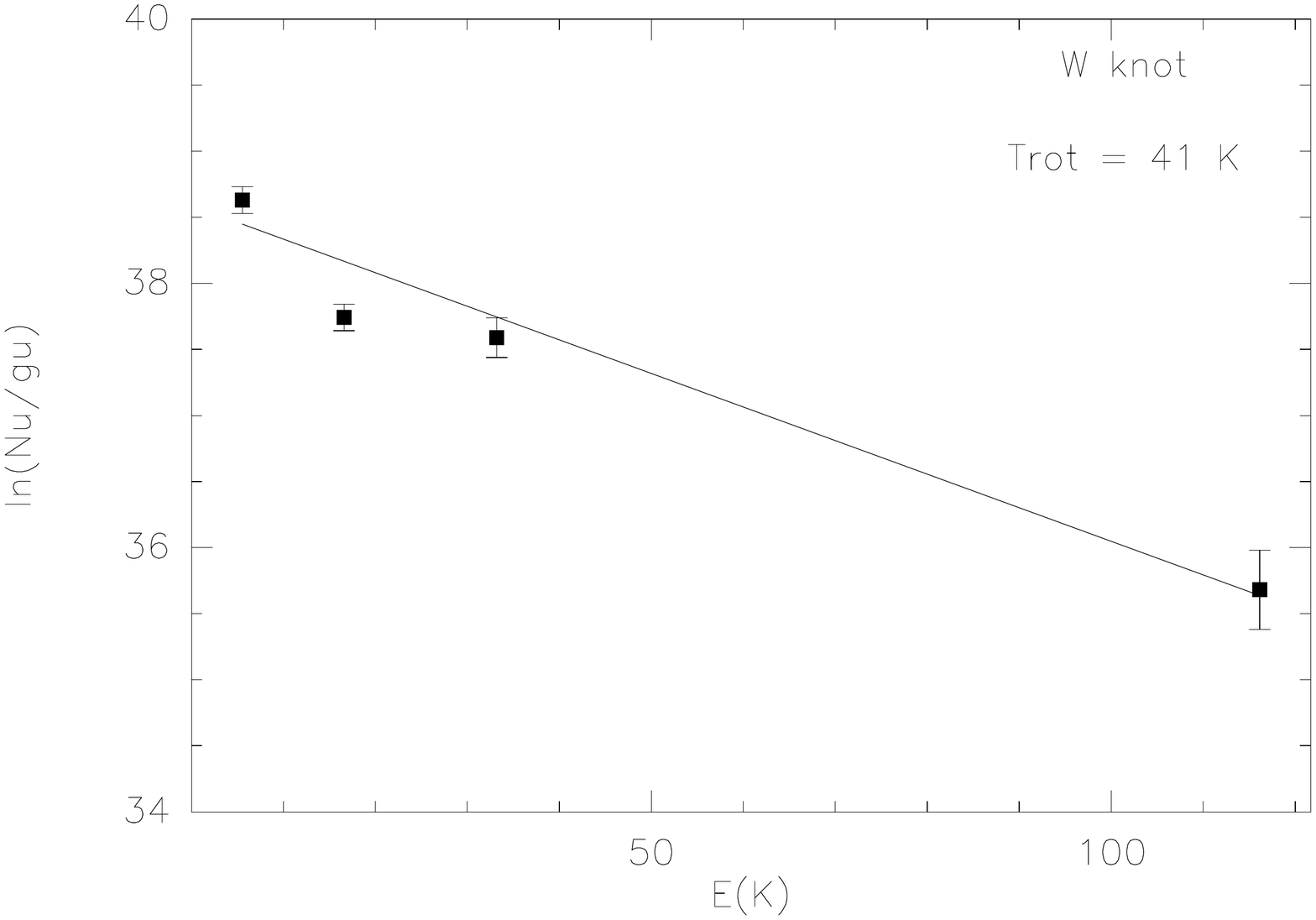}
}%
\centerline{%
\includegraphics[width=0.4\textwidth]{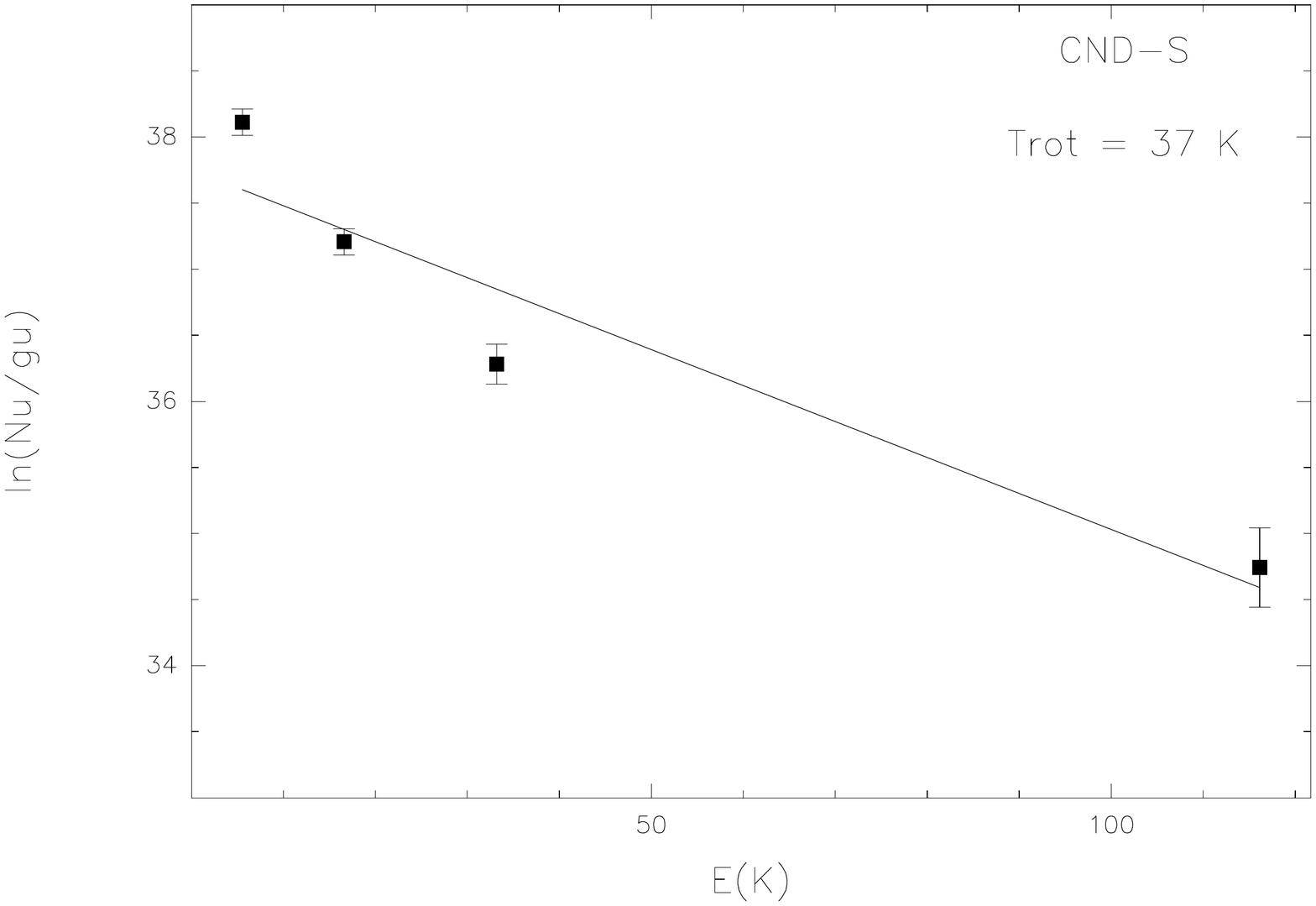}
\includegraphics[width=0.4\textwidth]{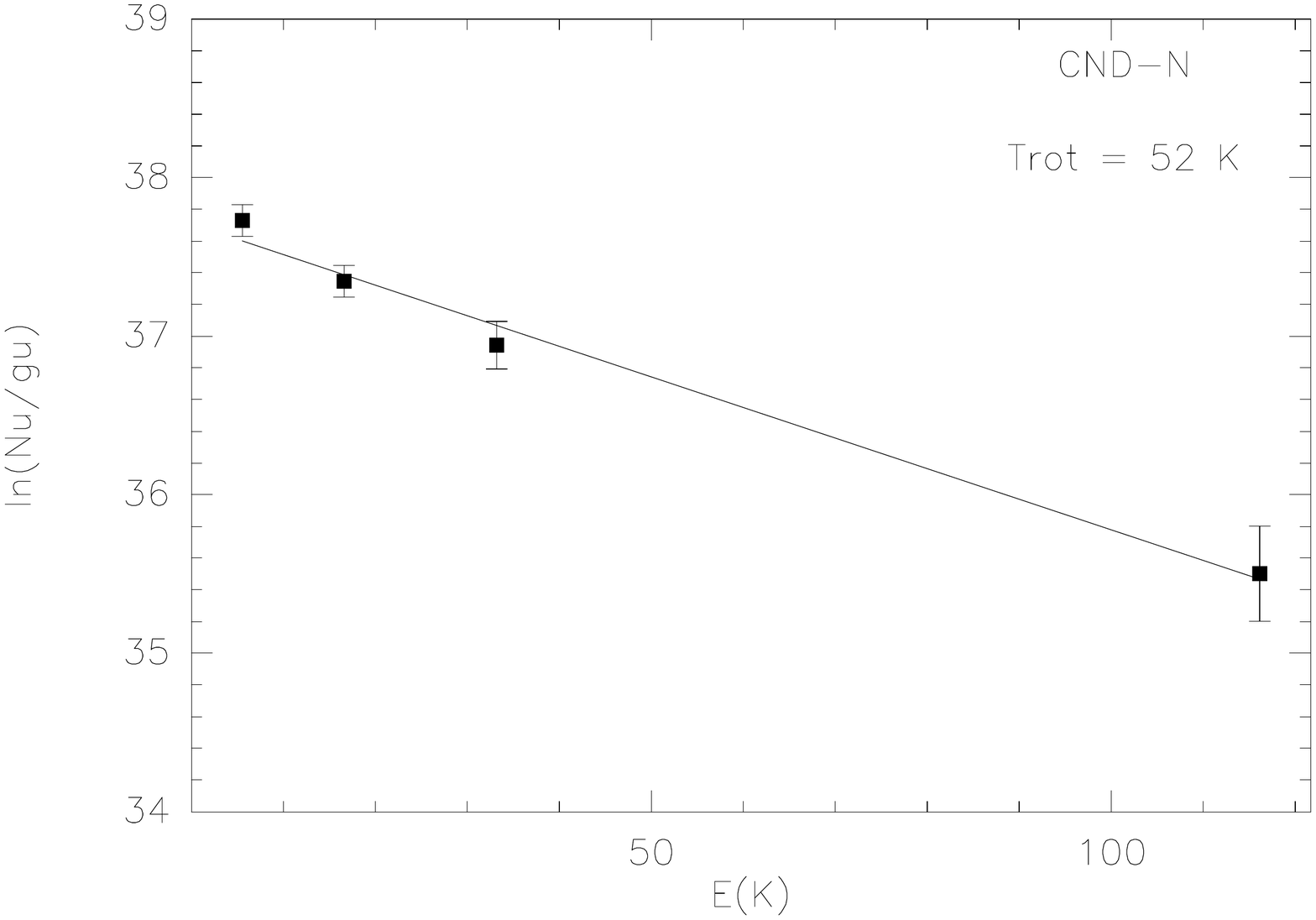}
}%
\caption{Rotation diagrams for CO in the AGN (Top Left); E Knot (Top Middle); W Knot (Top Right); CND-S (Bottom Left); CND-N (Bottom Right). The black solid squares show the data with error bars, which correspond to those of the integrated intensities (errors are of the order of 10\%--30\%).}
\label{rotdia}
\end{figure*}

Figure~\ref{rotdia} shows the rotation diagrams (with error bars - note that the errors are the absolute uncertainty in a natural log of $N_u/g_u$) for each of our regions, namely, E Knot, W Knot, AGN, CND-N and CND-S. The uncertainties were calculated (via standard propagation of errors) $assuming$ the error on the level column densities are 10\% for the first two transitions, 15\% for the $J$=3--2 and 30\% for the highest $J$. Deviations from these errors would not be large enough to dramatically change the results. 
The derived rotational temperatures are, respectively, 58 K, 41 K, 50 K, 52 K, and 37 K. The errors on the rotational temperatures are around 5-7 K. However, 
the AGN and CND-S regions may be better represented by a two component fit (not shown) which gives 25 K and 67 K for the AGN region, and 15 K and 54 K for the CND-S region. The rotational temperatures derived from the rotation diagrams ought to be considered as lower limits of the          
excitation and kinetic temperatures \citep{Gold99}.
The CO column density can also be quantified using the rotation diagram: we find that the CND-N and CND-S have the lowest CO column density at $\sim$2-3$\times$10$^{16}$ cm$^{-2}$, while the AGN, the W knot and the E Knot CO column densities are respectively 3$\times$10$^{16}$, 5$\times$10$^{16}$ and 7$\times$10$^{16}$ cm$^{-2}$. Compared to the values from Table~\ref{cd1and2}, they are at least one order of magnitude lower. 
This can be due to a combination of effects: firstly,  referring to Equation 24 in \cite{Gold99}, and assuming optically thin emission, this would lead to the ordinate of the rotation diagram to be considerably
below its correct value \citep{Gold99}. Secondly, as already mentioned, a single component does not always fit the data; taking into consideration a two component fit would boost the column density in all cases. The main conclusion to take away from this rotation diagram analysis is that, even within each region of the CND, the gas can not be characterized by a single phase.

\subsection{RADEX analysis of the CND}
\label{subsec:radex}
\citet{Krip11} performed an extensive and detailed analysis of the northern region of the CND with the radiative transfer code RADEX developed by \citet{VanderTak07}. They used 
ratios of several transitions of CO, HCN and HCO$^+$ and found the
molecular gas to be quite warm ($T_{kin} \geq$ 200 K) and dense ($n$(H$_2$) $\sim$ 10$^4$ cm$^{-3}$).  

In this section, in a similar manner to \citet{Krip11}, we perform a RADEX analysis  with all the available molecular ratios for each region of the CND. Of course such modelling is limited by the assumption that all the molecular transitions sample the same gas. Although, by definition we adopt such assumption, we note that the different critical densities of the available transitions, together with different physical and energetic conditions required for each chemical species to form (see Section 5) make this assumption a crude one. Nevertheless, if variations in temperature and density within each subregion of the CND are not too steep, a RADEX analysis should still be able to give us an indication of the 
average gas density and temperature and the likely range of column densities present. 

In order not to duplicate the \citet{Krip08,Krip11} analyses, we shall not discuss in detail the results of each simulation, nor the `goodness' and limitations of the best $\chi^2$ fit(s),  but, rather, 
concentrate on the {\it differences} in the physical characteristics among the subregions of the CND obtained from the best RADEX fits.
RADEX offers three different possibilities for the escape probability method: a uniform sphere, an expanding sphere (LVG), and a plane parallel slab. We used all three methods but, as \citet{Krip11}, we  did not find significant differences for the best fits; we therefore present the results from uniform sphere models.
The collisional rates used for the RADEX calculations were taken from the LAMBDA database \citep{Scho05} and are for the H$_2$ as collisional partner. We have used a background temperature of 2.7 K; this may not be completely appropriate for the AGN region.

We have performed three different grids of models, which we discuss in the next three subsections. 

\subsubsection{CO fitting}
We first used all our CO data as well as the $^{12}$CO/$^{13}$CO line intensity ratios of 14, 10 and 14 for the CO(1-0), CO(2-1), and CO(3-2) transitions taken from \citet{Papa99}. The isotopic line intensity ratios were measured with a very large beam of 14--21" ($>$950 pc), and are hence averages for the entire CND; by adopting the same $^{12}$CO/$^{13}$CO line intensity ratio for each region within the CND we are favouring, a priori, a common opacity which may not be a valid assumption (see below). As an intrinsic $^{12}$C/$^{13}$C ratio we used 40; this ratio is well known to vary across clouds within our Galaxy \citep[e.g.,][]{Mila05}, as well among galaxies \citep[e.g.,][]{Henk14}. For NGC~1068, \citet{Alad13} finds an upper limit to the isotopic ratios $^{12}$C/$^{13}$C of 49, while \citet{Krip11} find a ratio of 29. We did vary this ratio for a subset of our RADEX grid and did not find large enough differences to substantially change our conclusions. 

We construct a large grid of spherical models with the following parameters:
\begin{enumerate}
\item Gas density: n(H$_2$) = 10$^3$--10$^7$ cm$^{-3}$
\item Kinetic temperature: T$_{kin}$ = 10--500 K
\item Column density/Line-width: N($^{12}$CO)/$\Delta V$ = 10$^{14}$–10$^{20}$ cm$^{-2}$/km~s$^{-1}$
\end{enumerate}

\noindent
for a total of over 14,000 models. 

We use all possible combinations of
independent CO ratios:
2-1/1-0; 3-2/1-0; 6-5/3-2, $^{12}$CO(1-0)/$^{13}$CO(1-0), $^{12}$CO(2-1)/$^{13}$CO(2-1), $^{12}$CO(3-2)/$^{13}$CO(3-2).  Using all the available 
interferometric data has the advantage that we have more ratios than parameters to fit  but has the disadvantage that the ALMA observations used for the ratios involving the 3-2 and the 6-5 lines are $degraded$ to a linear resolution of $\sim$100 pc: within such a region it is very unlikely that we have a single homogeneous gas component: what we are deriving here is an average of the gas density and temperature, $provided$ that all the transitions used are tracing similar gas components within the $\sim$100 pc beam. 
We shall always report results from a reduced $\chi^2$ fitting, where, $K$, the number of degrees of freedom is taken to be equal to $N - n$ where $N$ is the number of intensity ratios, and $n$ is the number of fitted parameters; to be accurate, our reduced 
$\chi^2$ is in fact strictly speaking not a standard $\chi^2$ but a minimization function which we define as: \\
\begin{equation}
\chi^{2}_{red} = \frac{1}{K}\sum\limits_{i=1}^6[log(R_o)-log(R_m)]^2/\sigma^2
\end{equation}
where $R_o$ and $R_m$ are the observed and modelled ratio, respectively and $\sigma$ represents the uncertainty of the observed ratio.

\begin{figure*}[tbh!]
\centerline{%
\includegraphics[width=0.4\textwidth]{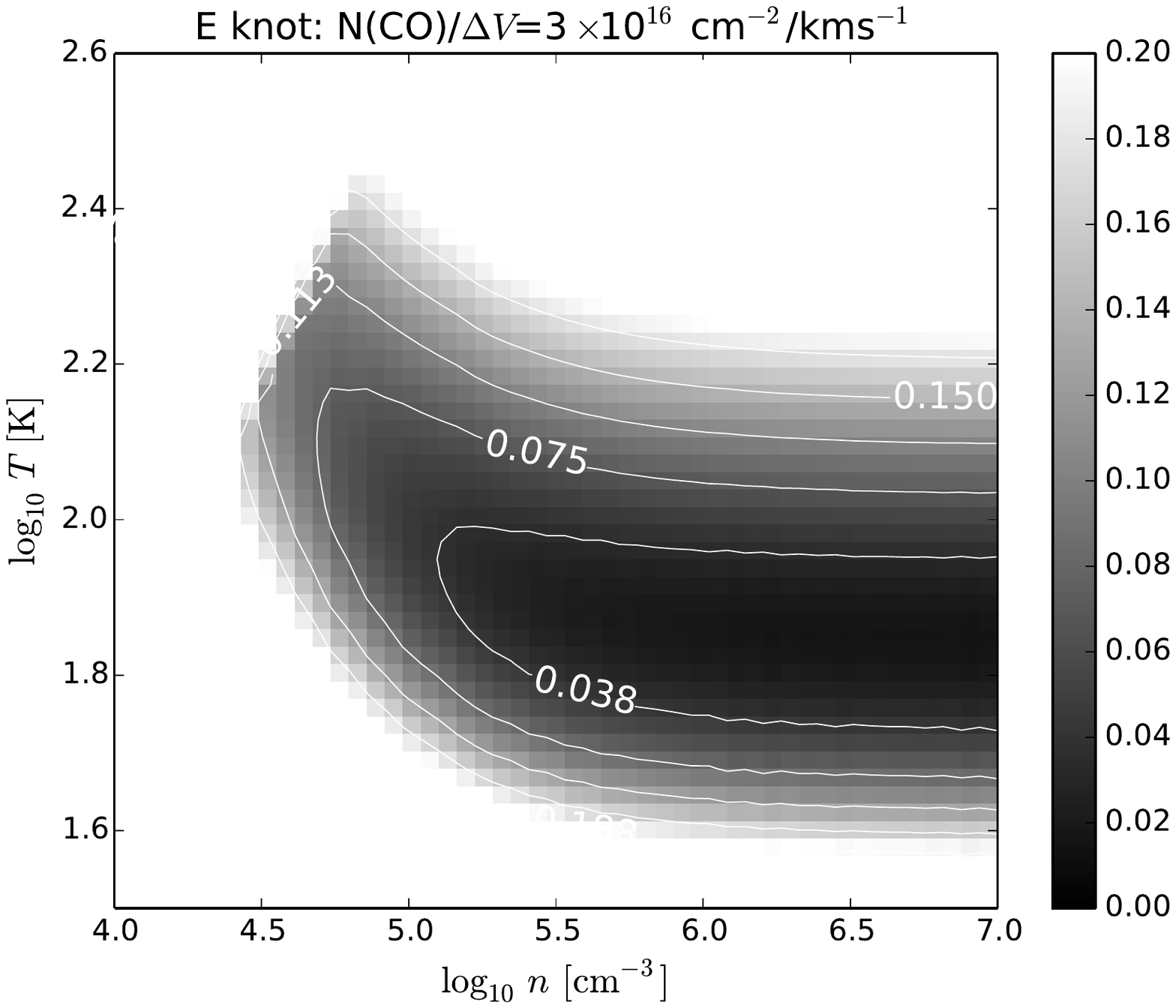}
\includegraphics[width=0.4\textwidth]{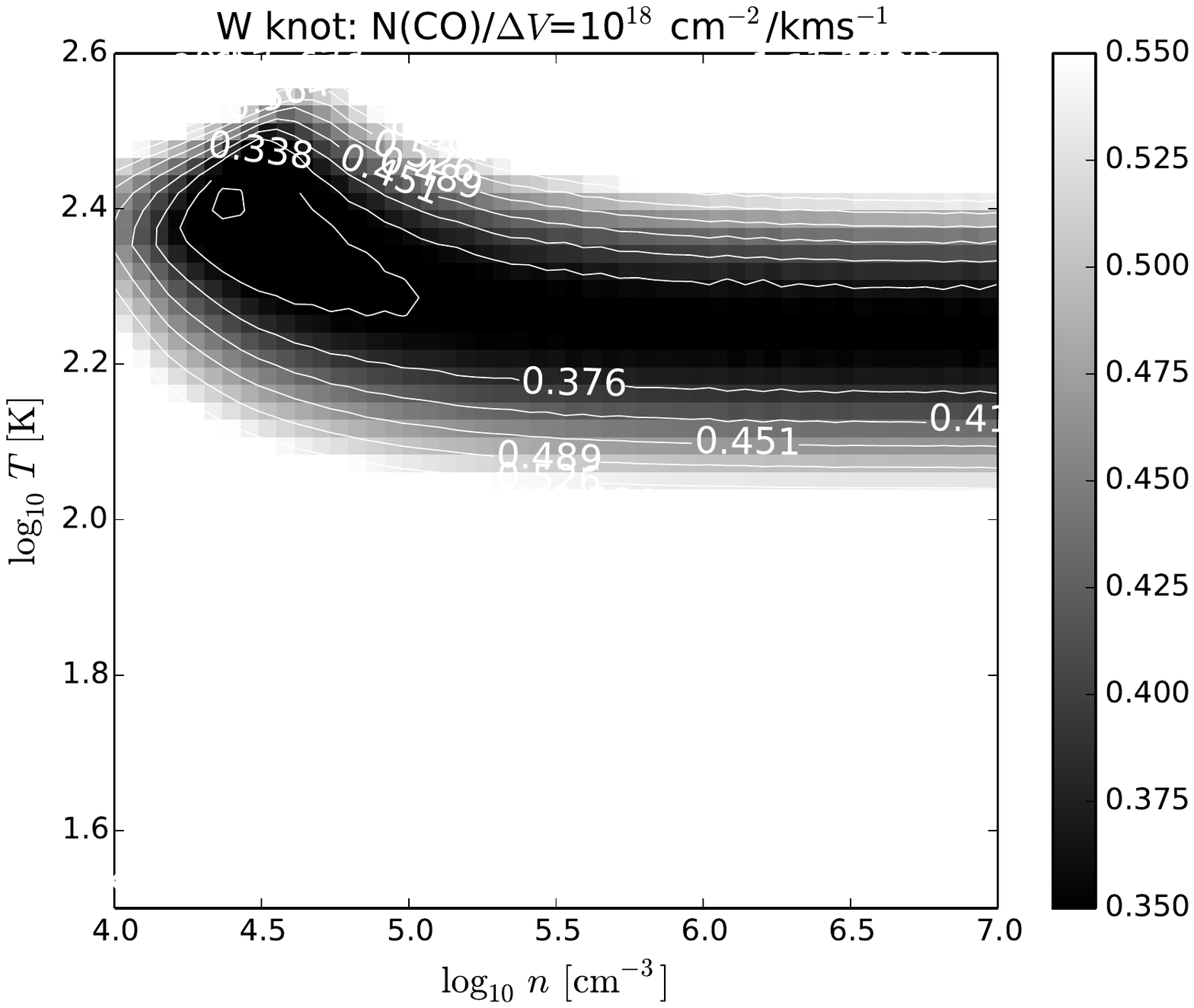}
\includegraphics[width=0.4\textwidth]{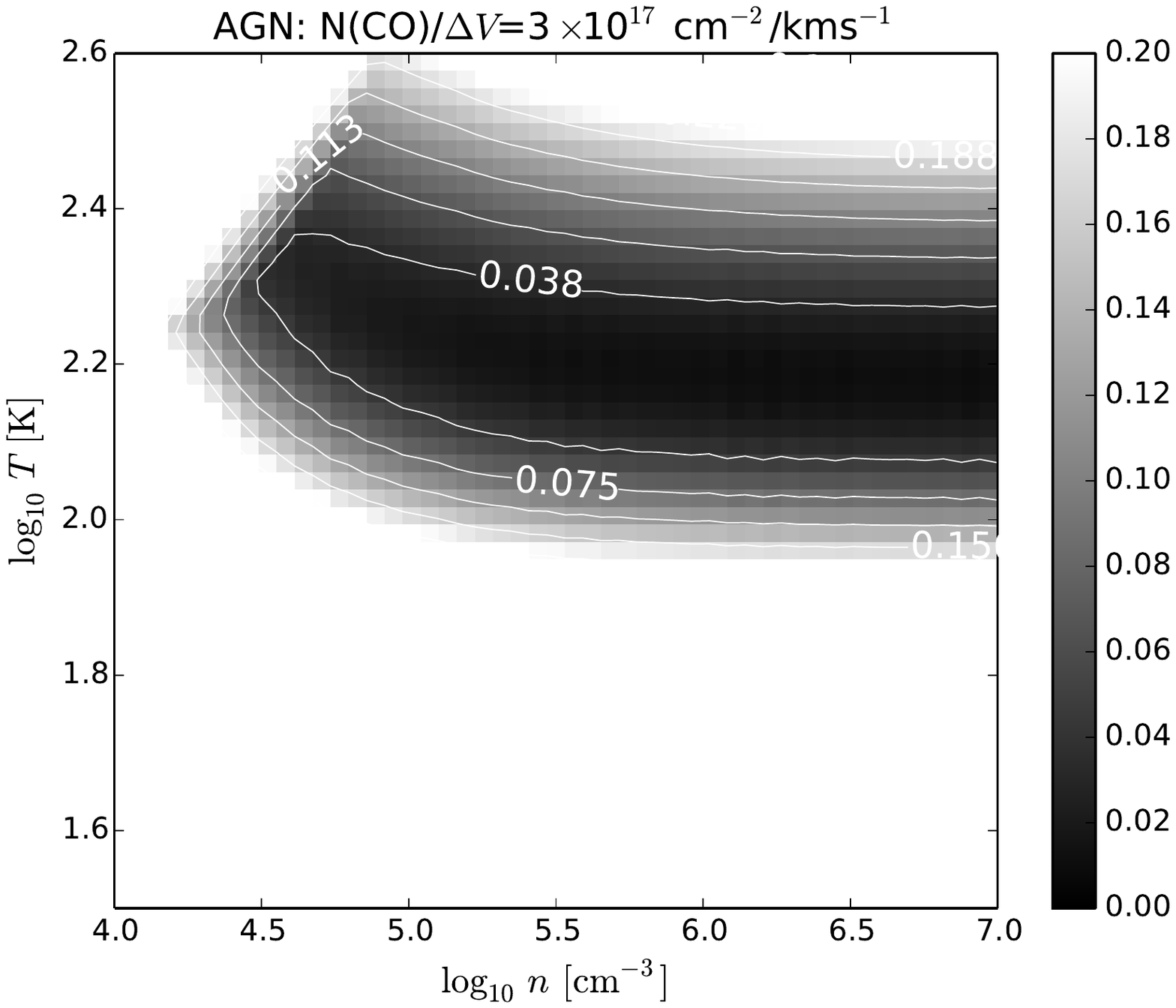}
}%
\centerline{%
\includegraphics[width=0.4\textwidth]{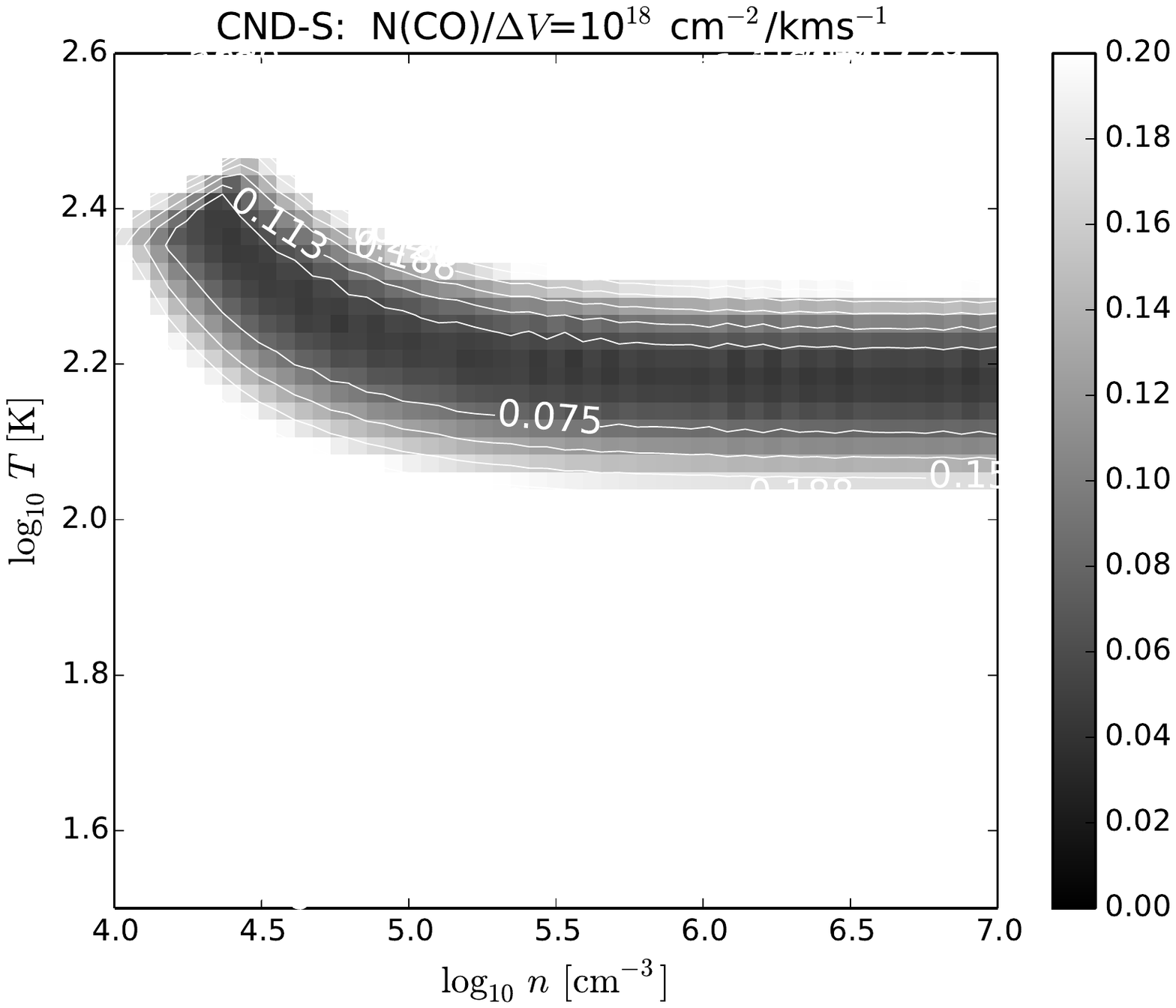}
\includegraphics[width=0.4\textwidth]{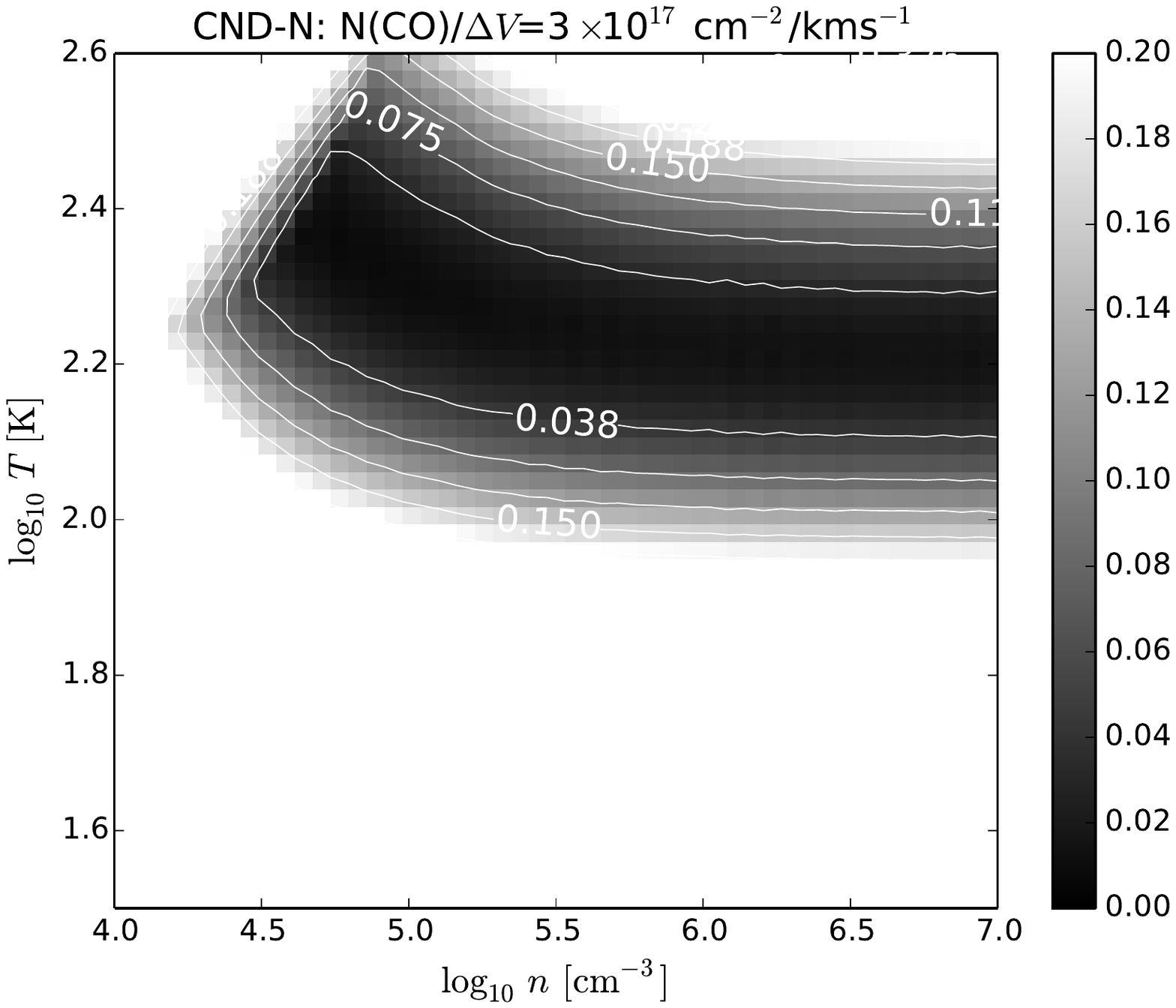}
\includegraphics[width=0.4\textwidth]{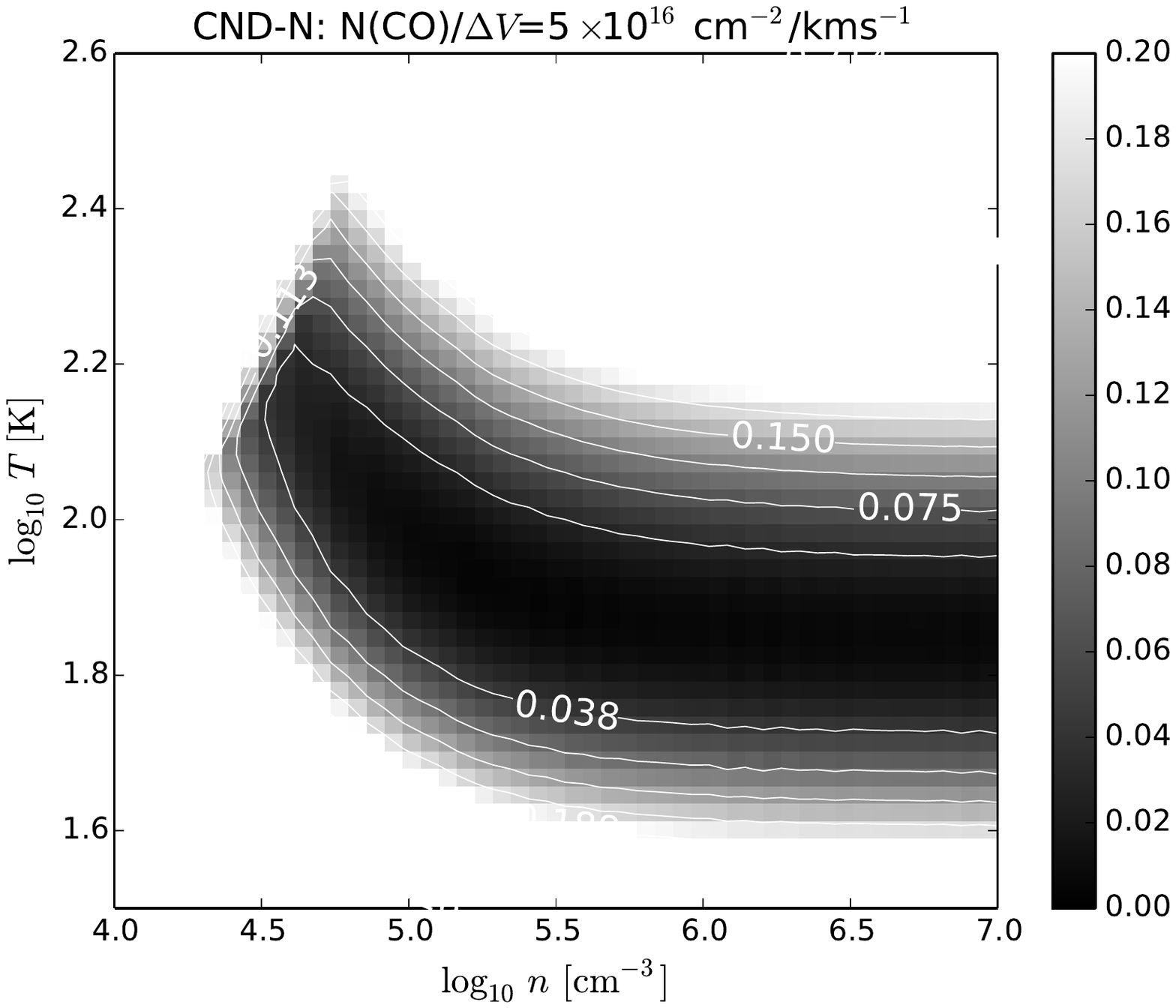}
}
\caption{$\chi^2$ fit results from the RADEX CO simulations of the excitation conditions within the E Knot (Top Left), W Knot (Top Middle), AGN (Top Right), CND-S (Bottom Left) and CND-N (Bottom Middle and Right).} 
\label{co-radex}
\end{figure*}

Examples of the best fits are shown in Figure~\ref{co-radex}. Table~\ref{radex_co} summarizes the best fit CO column densities (divided by the line-width), gas density and gas kinetic temperatures for each region within the CND. 

\begin{table}
\caption{Physical and chemical characteristics of the five selected subregions within the CND (see Figure 6) as derived by a RADEX analysis of the CO emission only.}
\label{radex_co}
\begin{tabular}{c|ccc}
\hline
Region & $N$(CO)/$\Delta V$ & $n$(H$_2$) & $T_k$  \\
 &   (cm$^{-2}$)/(km s$^{-1}$) & (cm$^{-3}$) & (K) \\
\hline
E Knot & 3$\times$10$^{16}$ & $>$ 10$^{5}$ & 60--80 \\
W Knot & 10$^{18}$ &  10$^{4}$--10$^{5}$ & 160--250 \\ 
AGN & 3$\times$10$^{17}$ & $>$ 3$\times$10$^{4}$ & 120--200 \\
CND-S & 10$^{18}$&  $>$ 10$^4$ & 160 \\
CND-N & 3$\times$10$^{17}$  & $>$ 3$\times$10$^{4}$ & 120--200\\
       & 5$\times$10$^{16}$ & $>$ 3$\times$10$^{4}$ & 60--100 \\  
\hline 
\end{tabular}
\end{table}

The opacity regimes probed by our RADEX solutions vary among the regions but also within the best density and temperature solutions: for example in the E Knot, the best fit model plotted in Figure~\ref{co-radex} (Top Left), for a density of 10$^5$ cm$^{-3}$ and a temperature of 60~K, gives an opacity,
$\tau$, of $\sim$ 0.2 for the $J$ = 1--0 line (which may be surprising given the low $^{12/13}$CO (1-0) observed ratio) to a $\tau$ of just below 2 for $J >$ 4.
Toward the AGN, on the other hand, the best fit model plotted in Figure~\ref{co-radex} (Top Right), gives an opacity which is still low for the $J$ = 1--0 line but up to almost 6 for the 6--5 line, for temperatures of the order of 200 K and densities of the order of 10$^5$ cm$^{-3}$. The RADEX solutions confirms what we found via the rotation diagram, namely that not all our transitions for each position are optically thin and that a single gas-phase component is not sufficient to match all the CO transitions within each sub region of the CND.

Although the conclusions we can draw from these fits are qualitative in nature, we find that: i) for all the regions, but the E Knot, 
within the CND, the gas is hot with an average $T_k$ $>$ 150 K. Such high temperatures are probably a consequence of X-rays or cosmic rays, and/or mechanical heating as it is unlikely that the UV would penetrate sufficiently within the gas to heat it to such high average temperatures. In fact, for cosmic rays alone to heat the gas to such temperatures, ionization rates ought be as high as 10$^{-14}$ s$^{-1}$ \citep{Baye11} or even 10$^{-13}$ s$^{-1}$ \citep{Meij11}.   The E Knot seems to have the lowest temperature ($<$ 80 K).
The hottest components are the W Knot, the AGN and, according to one solution (see Figure~\ref{co-radex}, Bottom Middle Panel), the CND-N. 
ii) Although the gas densities for each component
are less well constrained than the temperatures, the gas in the CND is on average dense ($>$ 10$^4$ cm$^{-3}$), in agreement with the findings by Krips et al. (2011); 
iii) The column densities for CO are around 10$^{17}$--10$^{19}$ cm$^{-2}$. 

\subsubsection{A RADEX grid computed  with multi-species ratios}

We now, independently,  attempt to find the best density
and temperature of the gas by using ratios of all the available molecules and their isotopes. For our first grid we use {\it all} the independent ratios: 
CO(2-1/1-0); CO(3-2/1-0); CO(6-5/3-2); $^{12}$CO(1-0)/$^{13}$CO(1-0), $^{12}$CO(2-1)/$^{13}$CO(2-1), $^{12}$CO(3-2)/$^{13}$CO(3-2), HCN(4-3/1-0), HCO$^+$(4-3/1-0), CS(7-6/2-1), HCN(4-3)/HCO$^+$(4-3), HCN(4-3)/CO(3-2), 
CS(7-6)/CO(3-2), H$^{12}$CO$^+$(1-0)/H$^{13}$CO$^+$(1-0), and H$^{12}$CN(1-0)/H$^{13}$CN(1-0).
We have used the same line H$^{12}$CN(1-0)/H$^{13}$CN(1-0) and  H$^{12}$CO($^+$1-0)/H$^{13}$CO$^+$(1-0) ratios of, respectively, 18 and 24  \citep[][Usero, priv. communication]{Papa99,User04}  for each region within the CND.
In order to use four different species, we need to constrain a set of parameters, namely the CO column density and the ratios 
among molecular abundances. We have varied the gas density and temperature as in our first RADEX grid in Section 3.2.1 
for several molecular column density ratios (see Table~\ref{sets}) spanning from $N$(CO)/$\Delta V$ = 5$\times$10$^{15}$ to 10$^{18}$ cm$^{-2}$/kms$^{-1}$, $N$(HCO$^+$)/$\Delta V$ = 10$^{10}$ to 10$^{13}$ cm$^{-2}$/kms$^{-1}$, $N$(HCN)/$\Delta V$ = 10$^{11}$ to 10$^{14}$ cm$^{-2}$/kms$^{-1}$, $N$(CS)/$\Delta V$ = 10$^{11}$ to 10$^{14}$ cm$^{-2}$/kms$^{-1}$ (resulting in a grid of over 10000 models).  We define a set of models to be a subgrid of the same individual column densities for each linewidth. Of course these Sets are not exhaustive and do not cover all the possible column densities ratios, but they were guided by the range of column densities derived in LTE. 
As for the previous RADEX grid, we shall not discuss the results in detail,  as similar RADEX analyses have been performed by \citet{Krip08,Krip11} and our {\it average} densities and temperatures confirm their general conclusions. Here we
mainly concentrate on variations within the CND.
However, we note that our reduced $\chi^2$ fits are much poorer than the ones
obtained for the CO fits (an indication of the presence of multiple gas components within each region); hence,  for this RADEX grid we shall report results from the log of the reduced $\chi^2$ fitting. We note also that for each region, we obtain always more than one solution in density and temperature for the same set of column densities (see later), as well as the same density and temperature for more than one set of column densities. 

\begin{figure*}[tbh!]
\centerline{%
\includegraphics[width=0.4\textwidth]{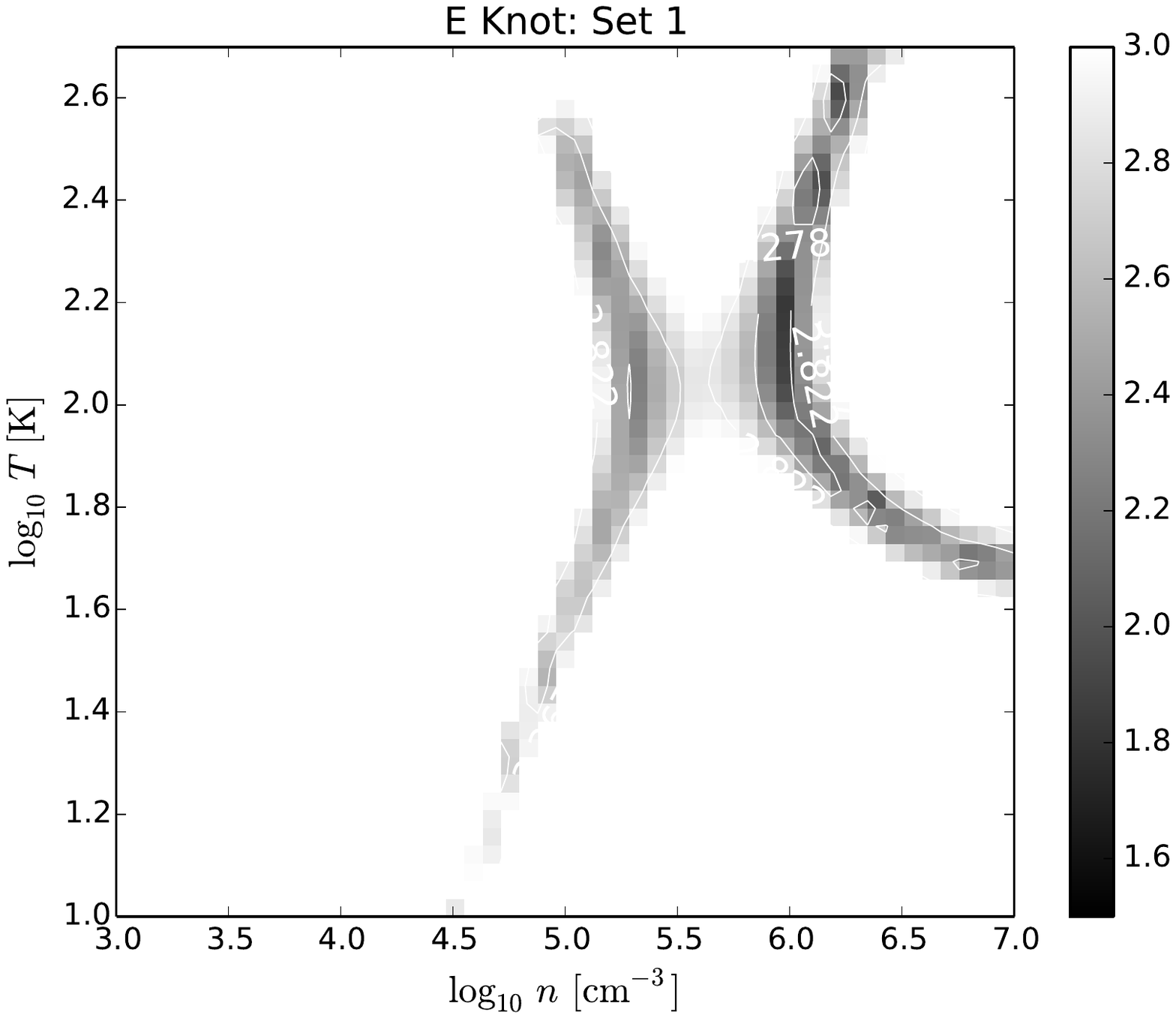}
\includegraphics[width=0.4\textwidth]{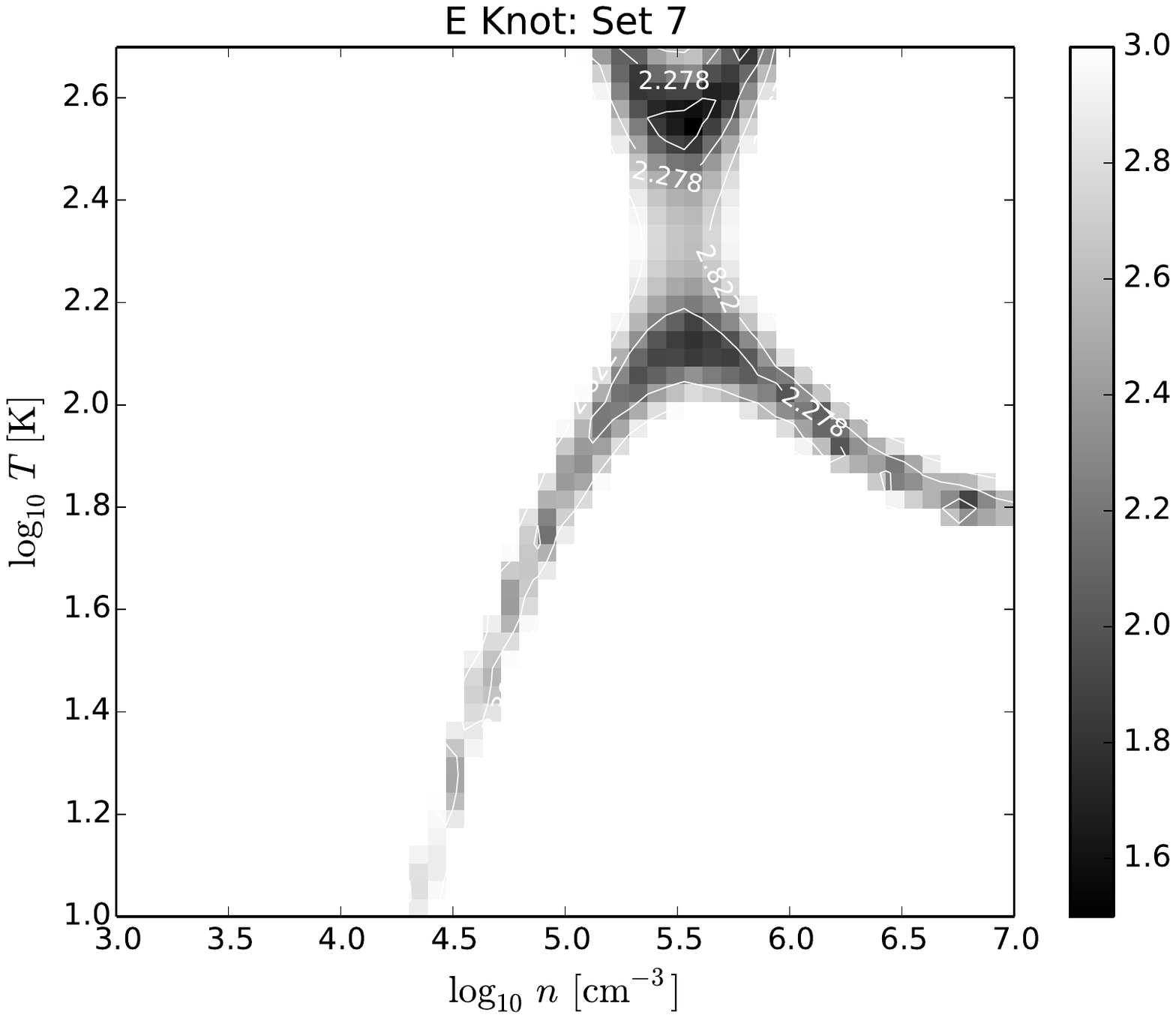}
\includegraphics[width=0.4\textwidth]{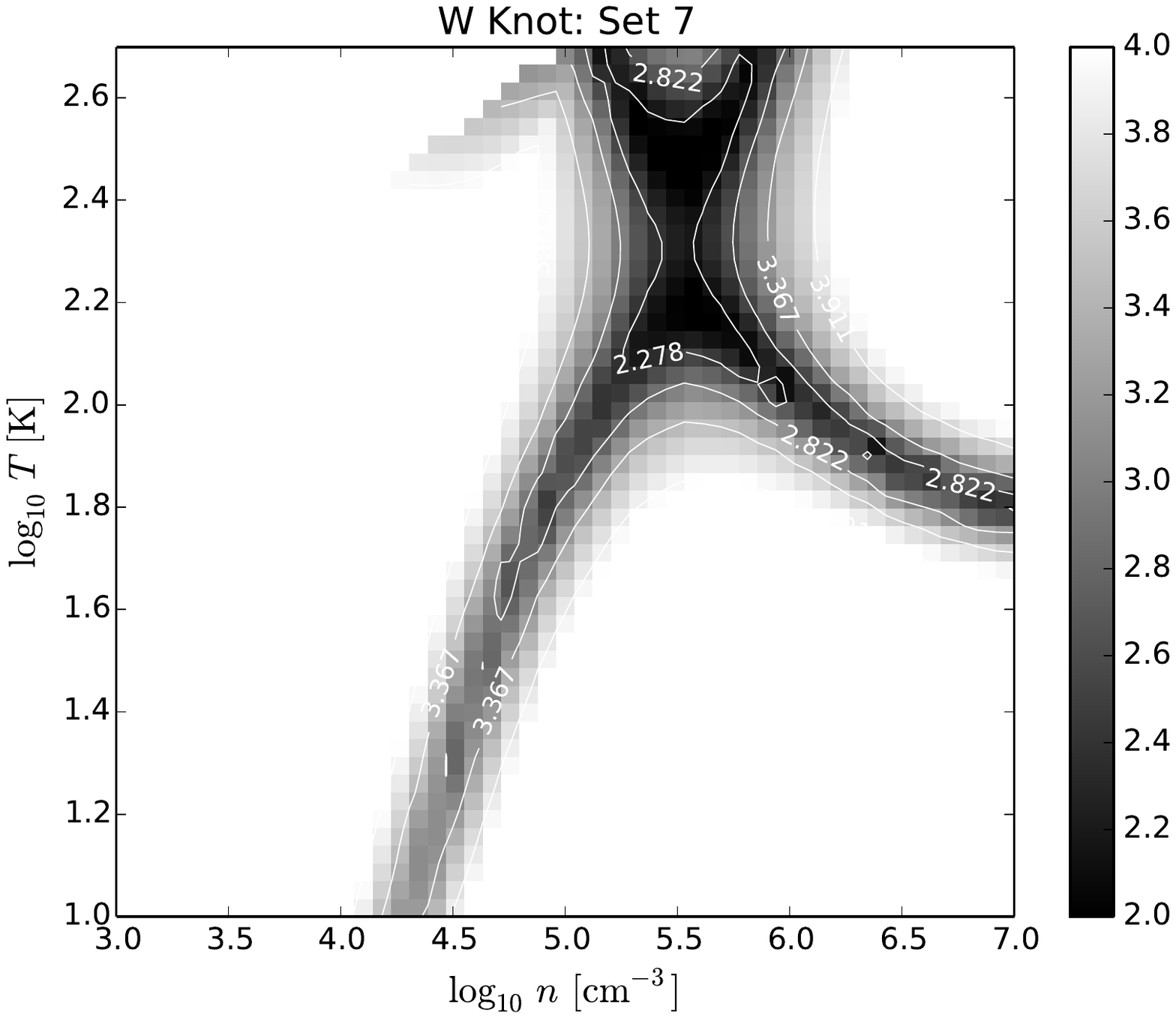}
}%
\centerline{%
\includegraphics[width=0.4\textwidth]{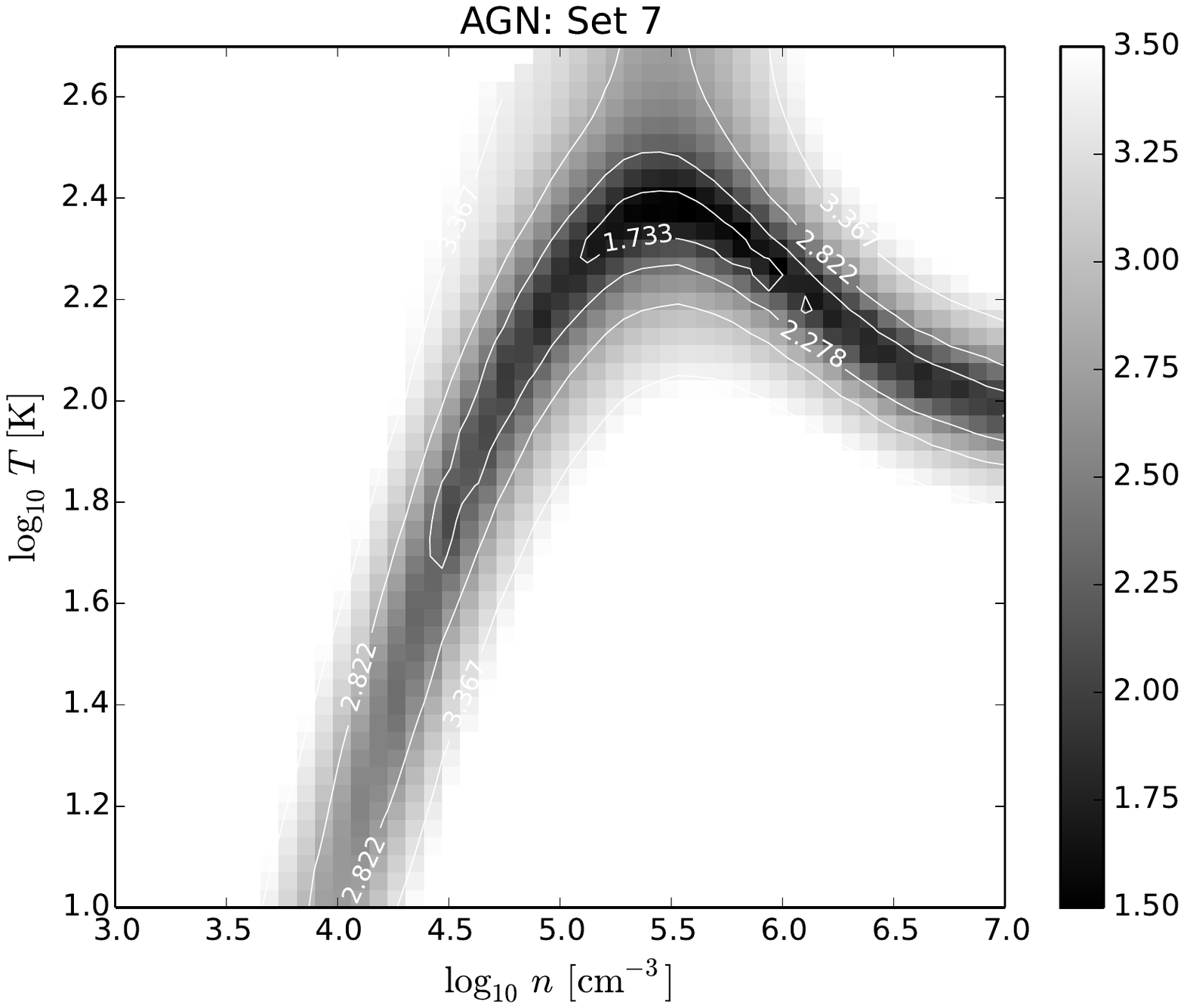}
\includegraphics[width=0.4\textwidth]{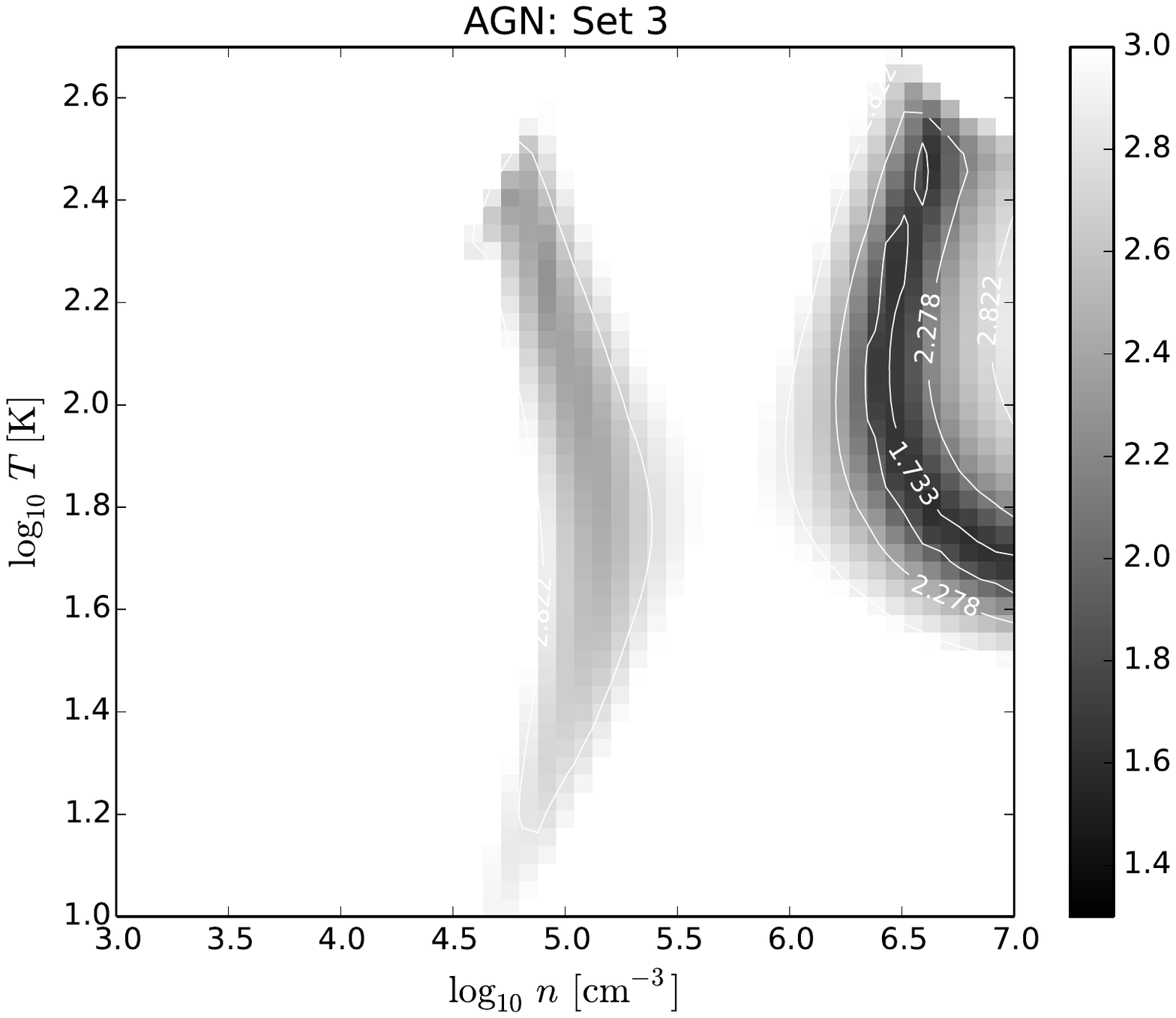}
\includegraphics[width=0.4\textwidth]{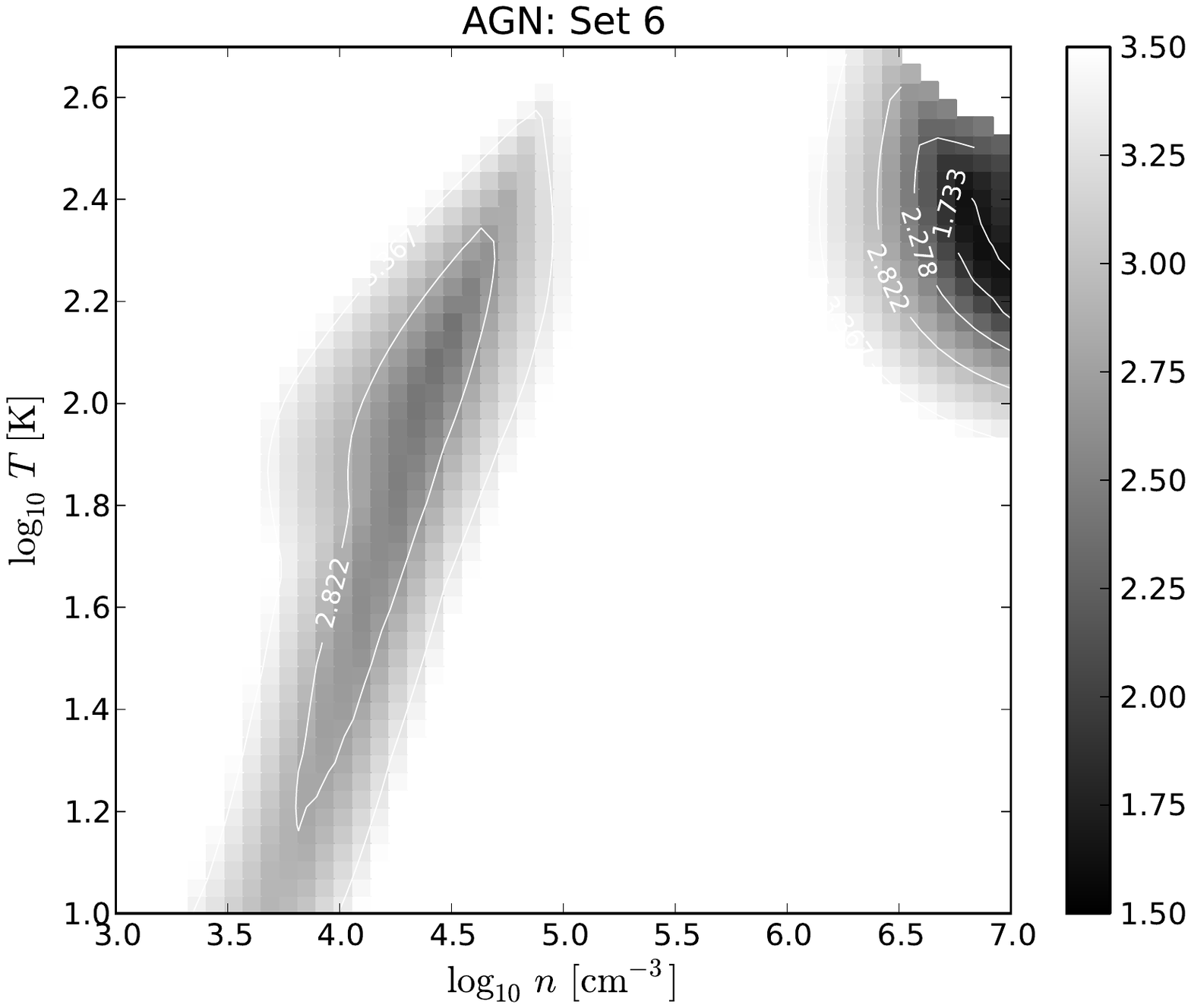}
}%
\centerline{%
\includegraphics[width=0.4\textwidth]{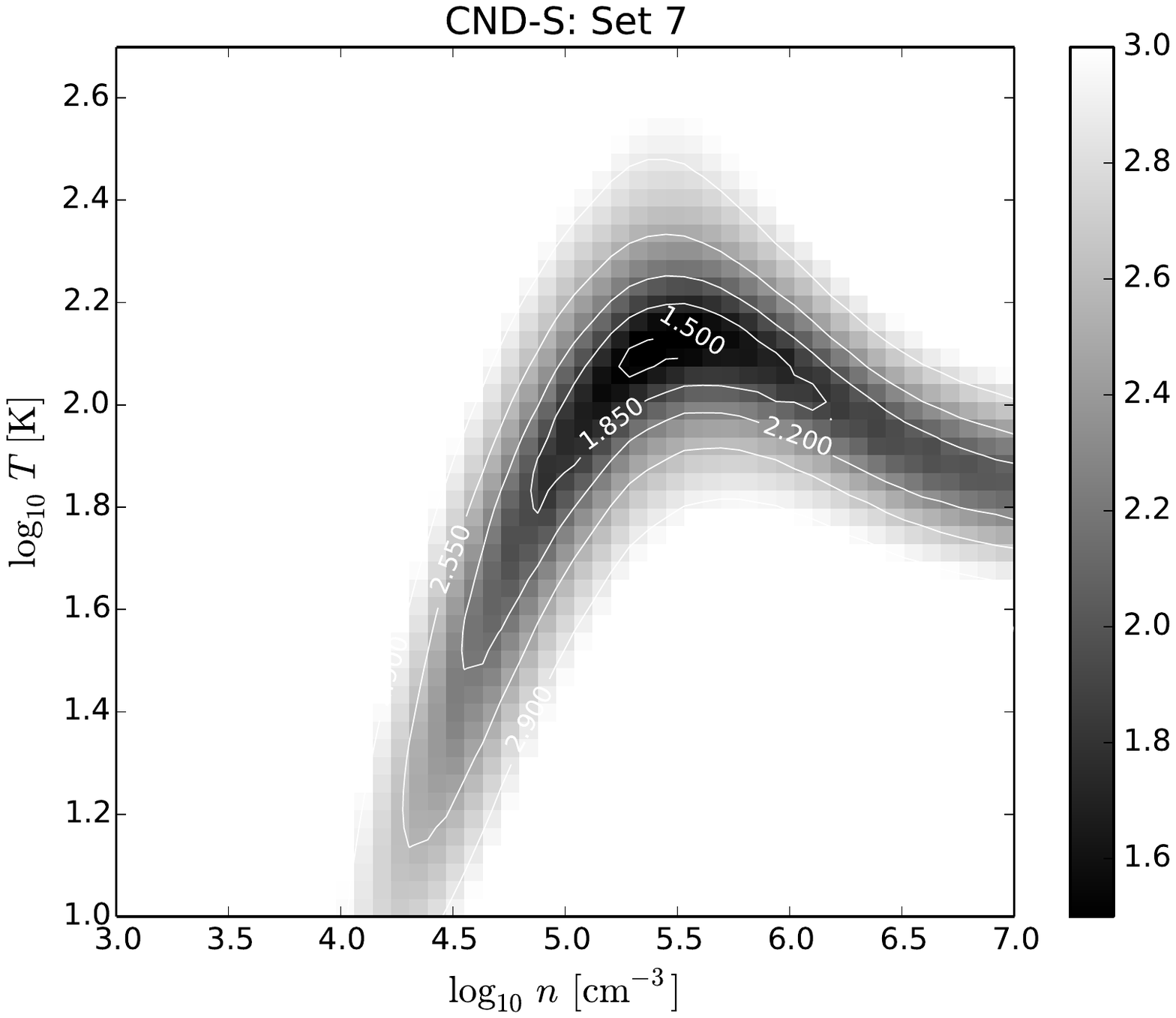}
\includegraphics[width=0.4\textwidth]{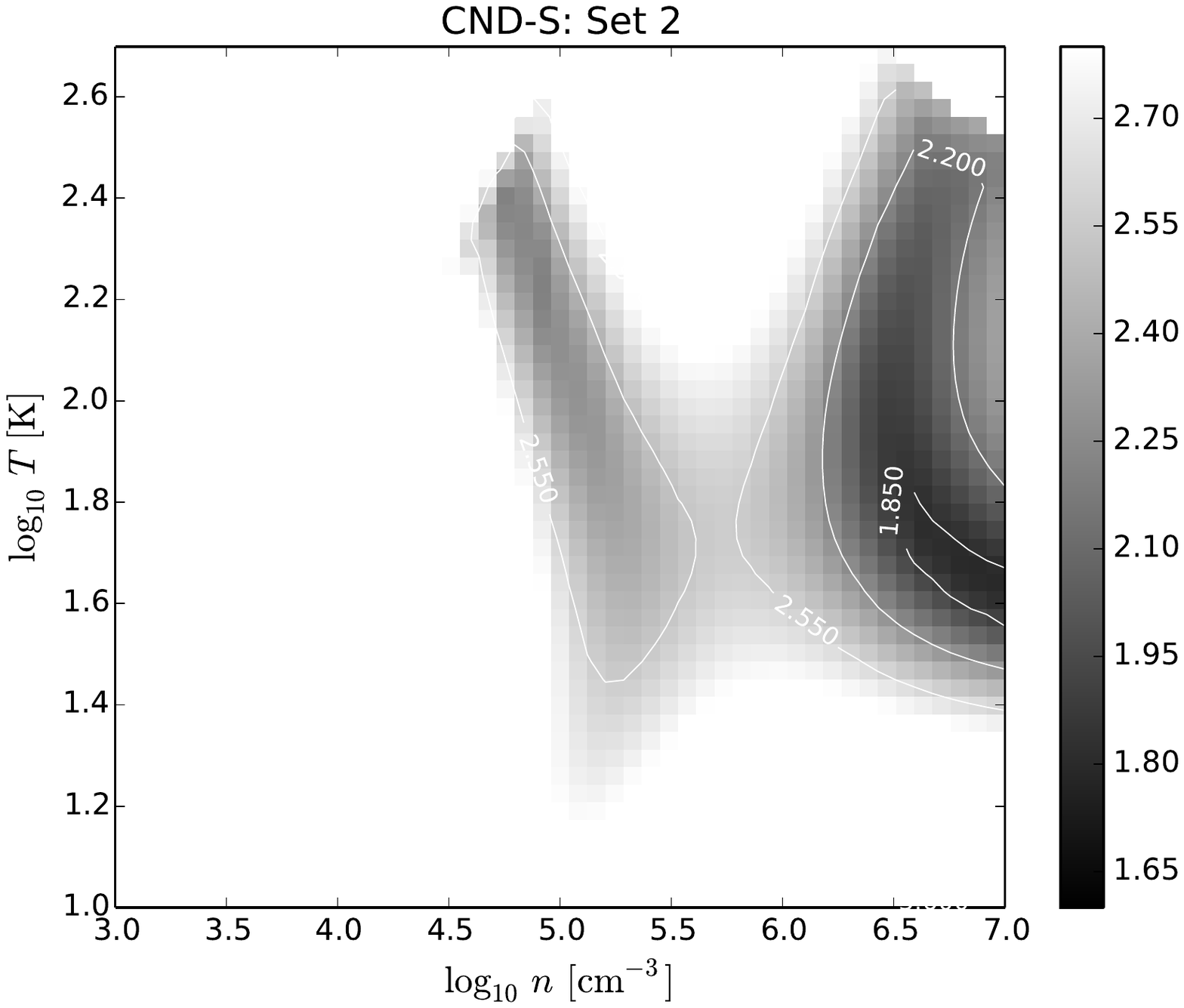}
\includegraphics[width=0.4\textwidth]{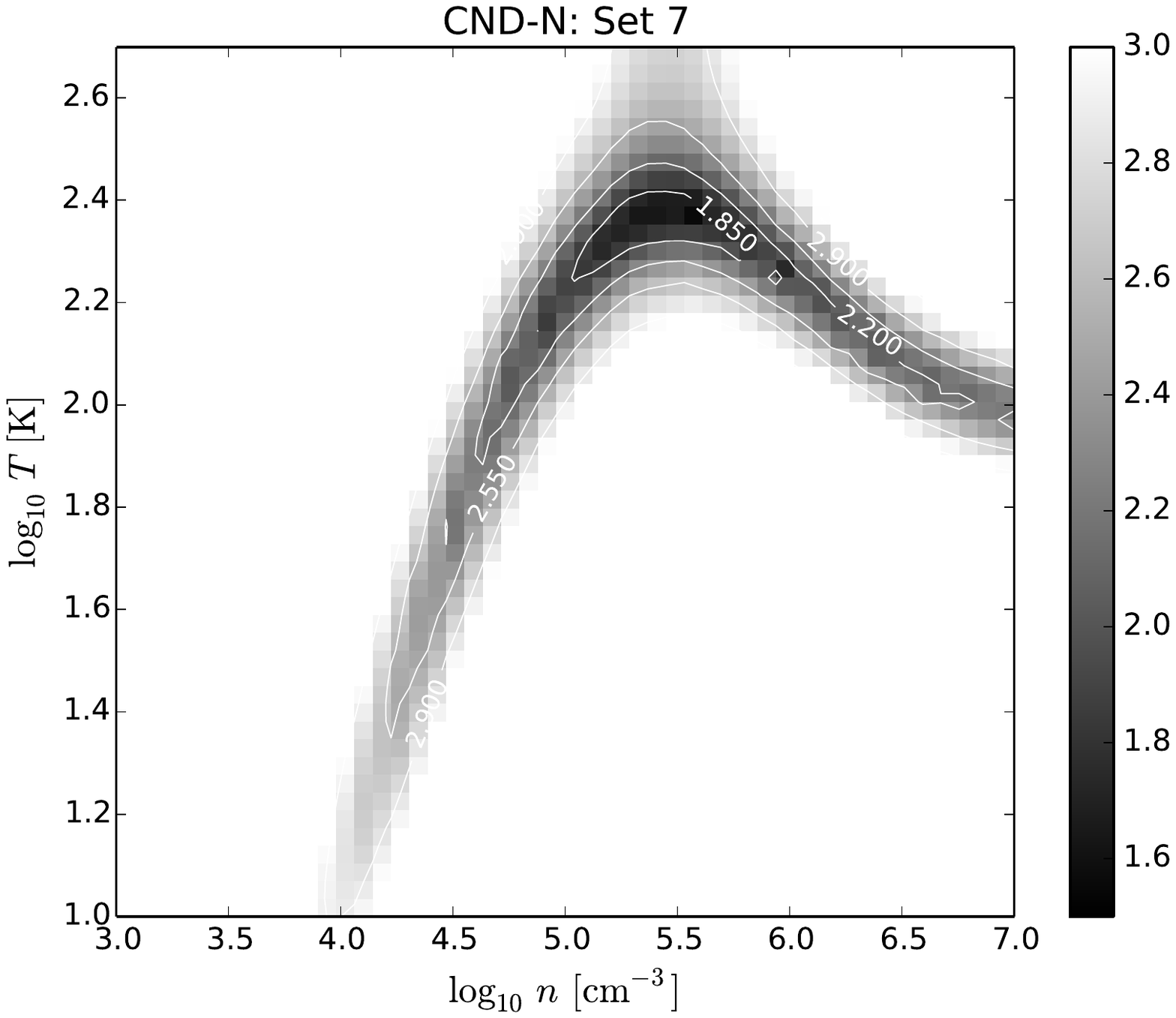}
}%
\caption{Log of $\chi^2$ fits results from the RADEX simulations of the excitation conditions within the E Knot (Top Left and Middle), the W Knot (Top Right), the AGN (Middle Row), the CND-S (Bottom Left and Middle), the CND-N (Bottom Right) using all the available transitions. }
\label{indep_radex}
\end{figure*}

Table~\ref{radex_all} reports a summary of all our results and Figure~\ref{indep_radex} shows our best fits. 
We discuss region by region below.

\begin{table}
\caption{The seven sets of RADEX models using all available transitions. Columns 2--5 are the adopted column densities, in units of cm$^{-2}$ 
where a(b) stands for a$\times$10$^b$.}
\label{sets}
\begin{tabular}{ccccc}
\hline
Set & CO & HCO$^+$ & HCN & CS \\
\hline
1 & 5(17) & 1(12) & 1(13) & 1(13) \\
2 & 5(17) & 1(13) & 1(14) & 1(14) \\
3 & 1(18) & 1(13) & 1(14) & 1(14) \\
4 & 1(18) & 1(14) & 1(15) & 1(15) \\
5 & 1(18) & 1(14) & 1(14) & 1(15) \\
6 & 1(19) & 1(14) & 1(15) & 1(15) \\
7 & 1(19) & 1(13) & 1(15) & 1(14) \\
\end{tabular}
\end{table}

\begin{table*}
\caption{Physical and chemical characteristics of the 5 subregions within the CND as derived by a RADEX analysis of the all the available independent ratios, $and$ using the lower limit of the gas density as derived by the CO RADEX fitting as a constraint. For the Sets parameters (Column 2), see Table~\ref{sets}. a(b) stands for a $\times$ 10$^{b}$.}
\label{radex_all}
\begin{tabular}{c|c|ccccccc}
\hline
Region & Set  & N(CO)/N(HCO$^+$)  & N(CO)/N(HCN) & N(HCN)/N(HCO$^+$) & N(CO)/N(CS) & $n$(H$_2$) (cm$^{-3}$) & $T_k$ (K)  \\
\hline
E Knot &  1 &  5(5) & 5(4) & 10 & 5(4) & 10$^6$ & 80--160 \\
       &  7 &  1(6) &  1(4) & 1(2) & 1(5)  & 10$^5$--10$^6$ & 100 \\
W Knot &  7 &  1(6) &  1(4) & 1(2) & 1(5) & $\sim$10$^5$ & $>$ 100 \\
AGN &     7 & 1(6) &  1(4) & 1(2) & 1(5) &  10$^5$--5$\times$10$^5$ & 200 \\
    &     3 & 1(5) &  1(4) & 10 & 1(4) & $\sim$2$\times$10$^6$ & 60--250 \\
    &     6 & 1(5) & 1(4) & 10 & 1(4)  & $>$ 5$\times$10$^6$ & 160--250 \\     
CND-S &   7 & 1(6) &  1(4) & 1(2) & 1(5) & 10$^5$ & 100-120 \\
      &   2 & 5(5) & 5(4) & 10 & 5(3) & $>$ 2$\times$10$^6$ & 40--60 \\
CND-N&    7 & 1(6) &  1(4) & 1(2) & 1(5)  &10$^5$--10$^6$ & 160--240 \\
\hline
\end{tabular}
\end{table*}

For the E Knot Sets 1, 2, 6 and 7 (see Table~\ref{sets}) give the same reduced $\chi^2$ (see Figure~\ref{indep_radex}); one of the results from the RADEX CO analysis was that the best fit gas densities were always above 10$^4$ cm$^{-3}$: we therefore use this lower limit as a further constraint and  
we are left with Sets 1 and 7, because the others all give gas densities below 10$^4$ cm$^{-3}$.  Set 7 gives multiple solutions in the $n_{\rm H}-T_k$ plane: the high temperature ($\sim$ 400 K) solution, although 
in agreement with the RADEX results from \citet{Krip11} (which was however an average across a larger area), is the less
likely given the findings from the CO RADEX analysis. For the W Knot Set 7 has the lowest reduced $\chi^2$ fit, with gas density and temperature solutions consistent with the CO RADEX analysis.  For the AGN, Sets 3, 6, and Set 7 give the same reduced $\chi^2$ and more than one solution in the density-temperature plane.  
For the CND-S the best fits are given by Set 2 and Set 7.
Finally, for the CND-N the best fit is only given by Set 7.

All regions, but the CND-N and the W Knot, can be fitted by more than one Set, but equally, all regions can be fitted by the {\em same} Set, implying that we cannot exclude that all the regions are chemically similar, with common ratios of 
HCN/HCO$^+$ = 100, CO/HCN = 10$^4$, CO/HCO$^+$ = 10$^6$, and CO/CS = 10$^5$. In this scenario we note that the HCN/HCO$^+$ abundance ratio is very high compared to those found in extragalactic environments, although chemically an abundance ratio of 100 is in fact feasible in both dense (10$^4$ cm$^{-3}$) PDR gas and star forming (n$_H$ $\geq$ 10$^5$ cm$^{-3}$) gas \citep{Baye08,Baye09}, due to the high abundance of HCN. Hence it may be an indication of very dense gas present.  
We also find that the AGN and the CND-N are the hottest regions and that the lowest limit for the gas density is 10$^5$ cm$^{-3}$. We note again here that vibrational excitation by infrared photons may provide an alternative explanation for the observed ratios in such hot gas; in fact several authors (e.g. \cite{SaKa10}) found that HCN, and possibly other species including HCO$^+$ and CO, can be excited by infrared pumping in AGN- or starburst-dominated galaxies, and very recently, \cite{gonza14} reported far-infrared pumping of H$_2$O by dust in NGC1068. 
If one takes into consideration the CO column densities as derived by the RADEX CO only analysis, then Set 1 is the best fit for the E Knot, while both Set 6 and 7 fit the AGN well, with Set 6 giving an higher average density and temperature, which may be more consistent with the environment of the AGN. In this scenario then the chemical ratios for the E Knot are 10, 5$\times$10$^4$, 5$\times$10$^5$, and 5$\times$10$^4$ for respectively HCN/HCO$^+$, CO/HCN, CO/HCO$^+$, and CO/CS, while for the AGN they are 10, 10$^4$, 10$^5$, and 10$^4$.    

This excitation analysis does not allow us to distinguish whether the differences in intensity ratios across the CND are due to different energetics or simply differences in the density of the regions. We again underline that we have used a large number of ratios involving several transitions for which we only had low resolution data (100 pc and lower) and the ALMA data were all degraded to 100 pc resolution. In particular, the use of the CS(7-6)/(2-1) ratio is dubious considering that the CS(2-1) observations were obtained with a beam three times the average of that for the other lines (see Table~\ref{res}).  Ultimately, that this analysis does not conclusively reveal a chemical differentiation may just be a consequence of the fact that it uses data with a {\it degraded} resolution, while in fact the observed 
molecular line ratios show dramatic changes of up to an order of magnitude inside the CND on the {\em spatial scales} probed by ALMA (see Paper I). Hence, in the next section we reduce the number of molecular ratios to those obtained at the highest spatial resolution in our analysis to test the robustness of our results so far.

\subsubsection{RADEX analysis of ALMA data only}

\begin{figure*}[tbh!]
\centerline{%
\includegraphics[width=0.4\textwidth]{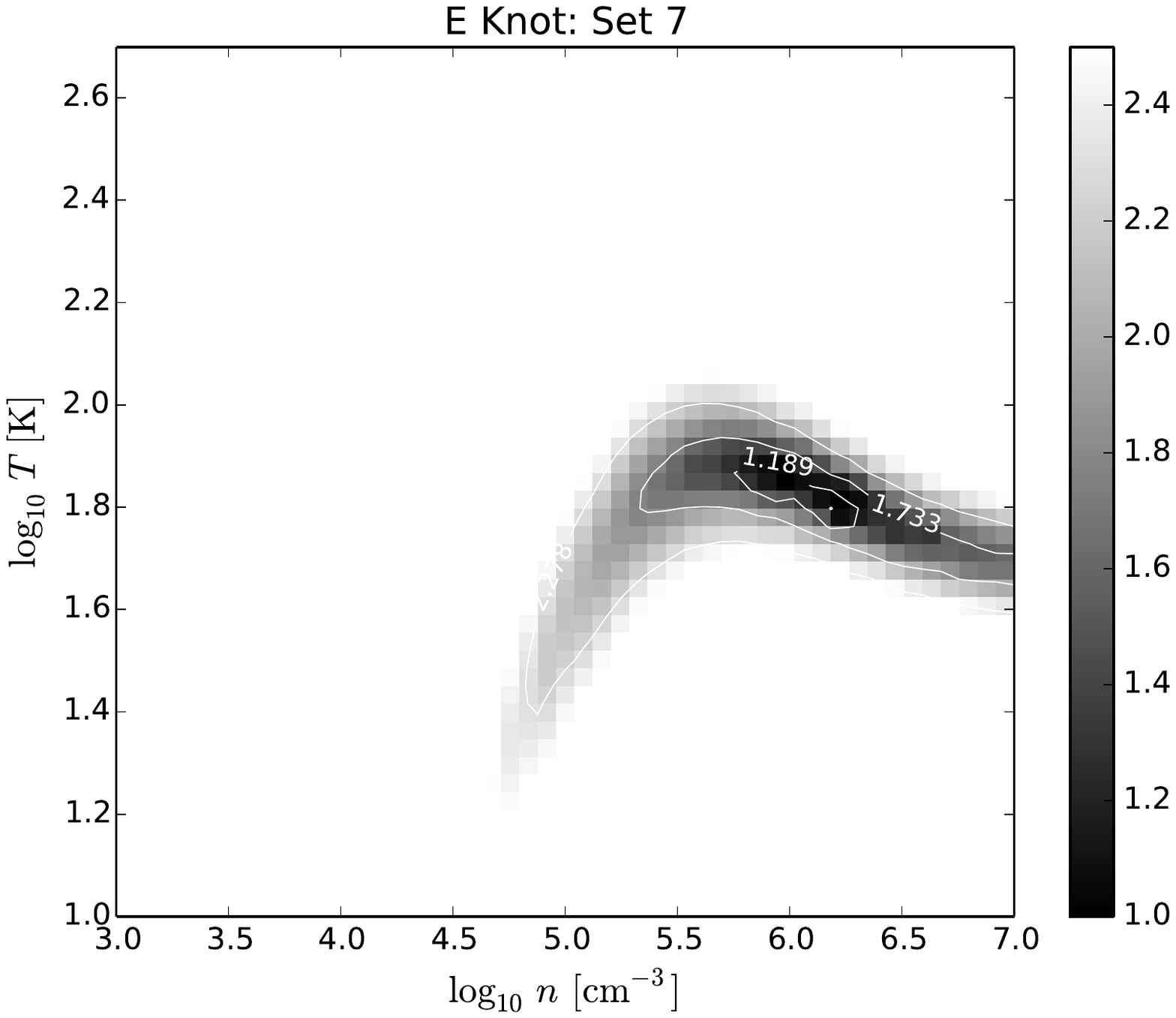}
\includegraphics[width=0.4\textwidth]{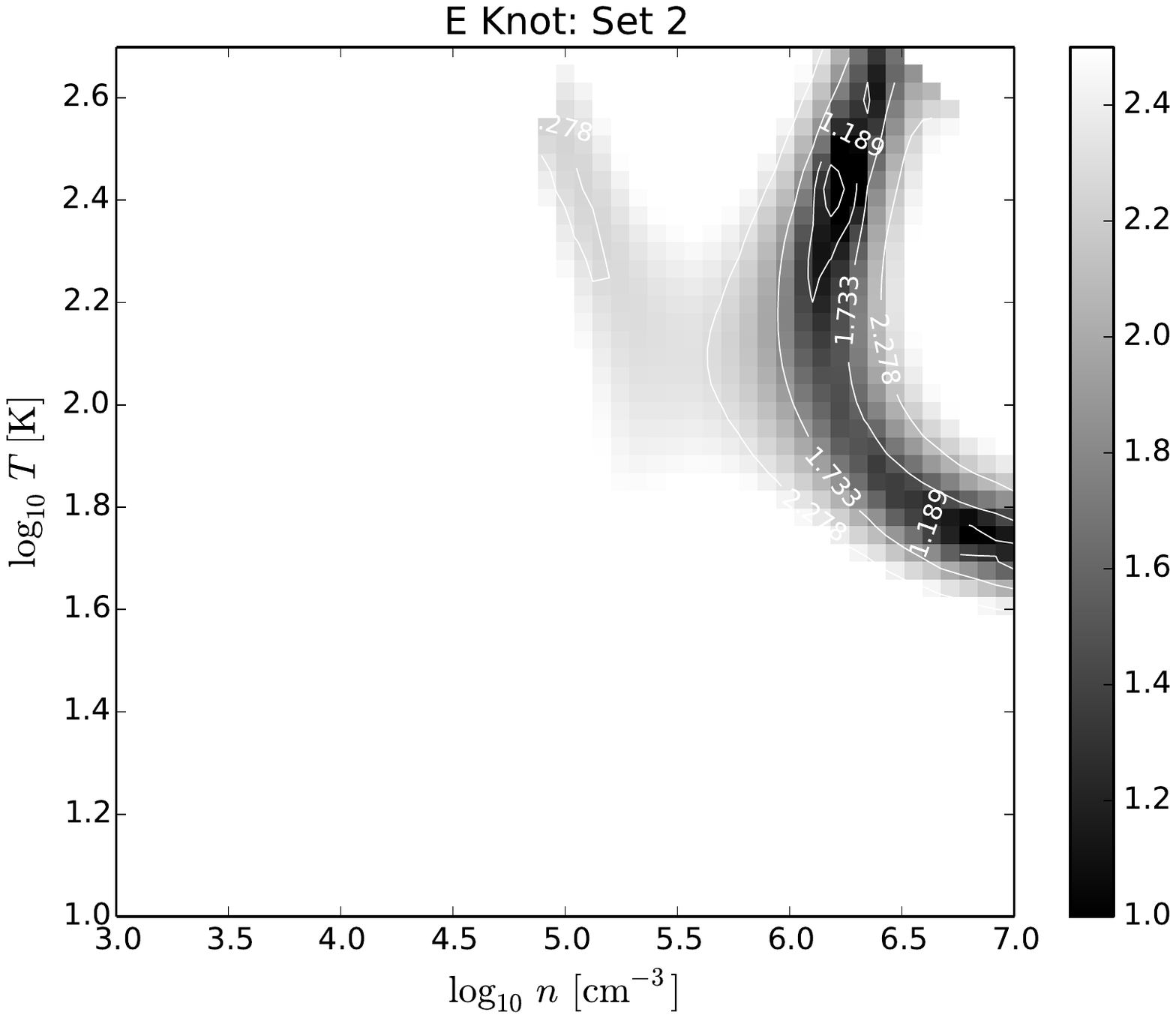}
\includegraphics[width=0.4\textwidth]{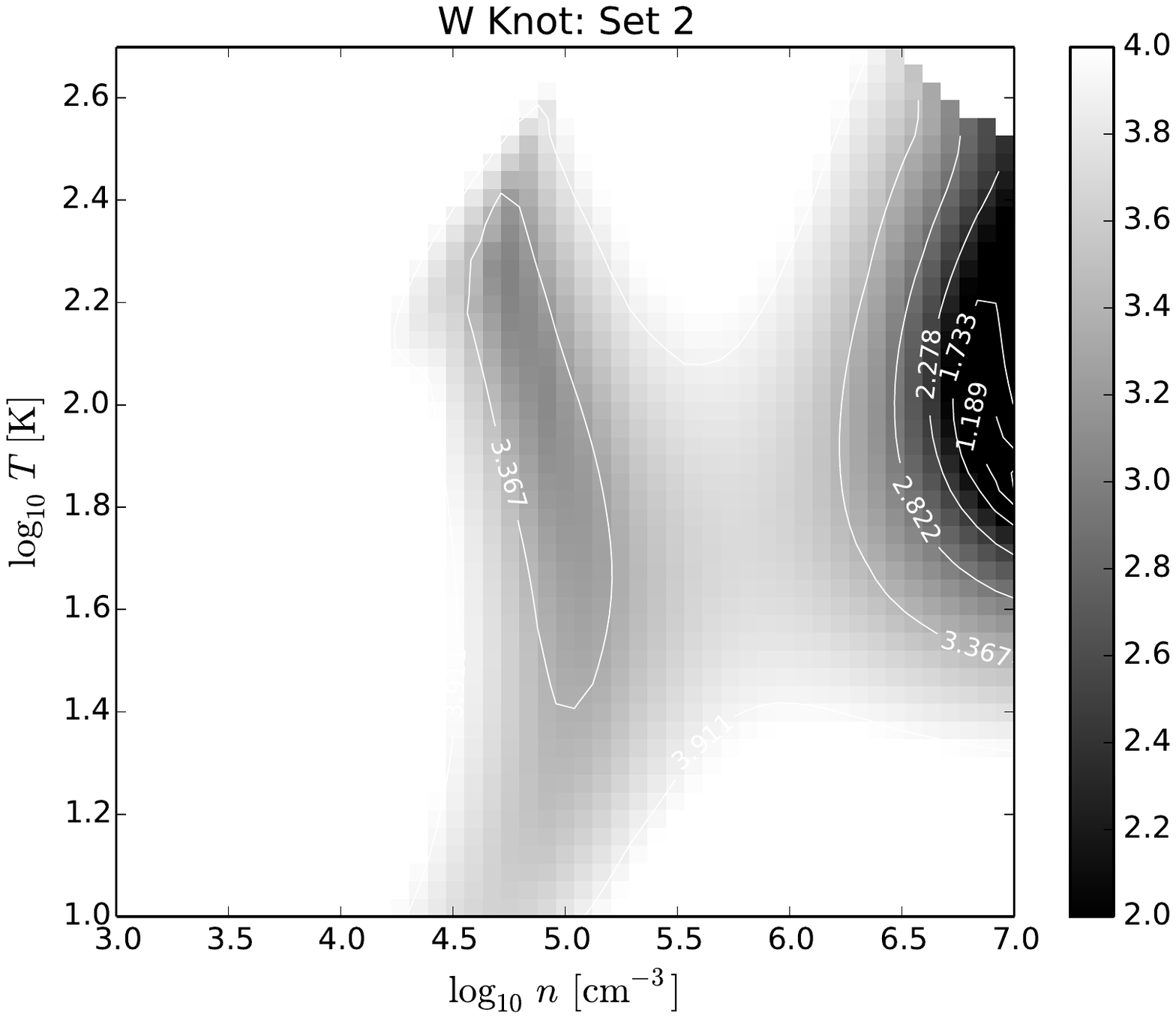}
}%
\centerline{%
\includegraphics[width=0.4\textwidth]{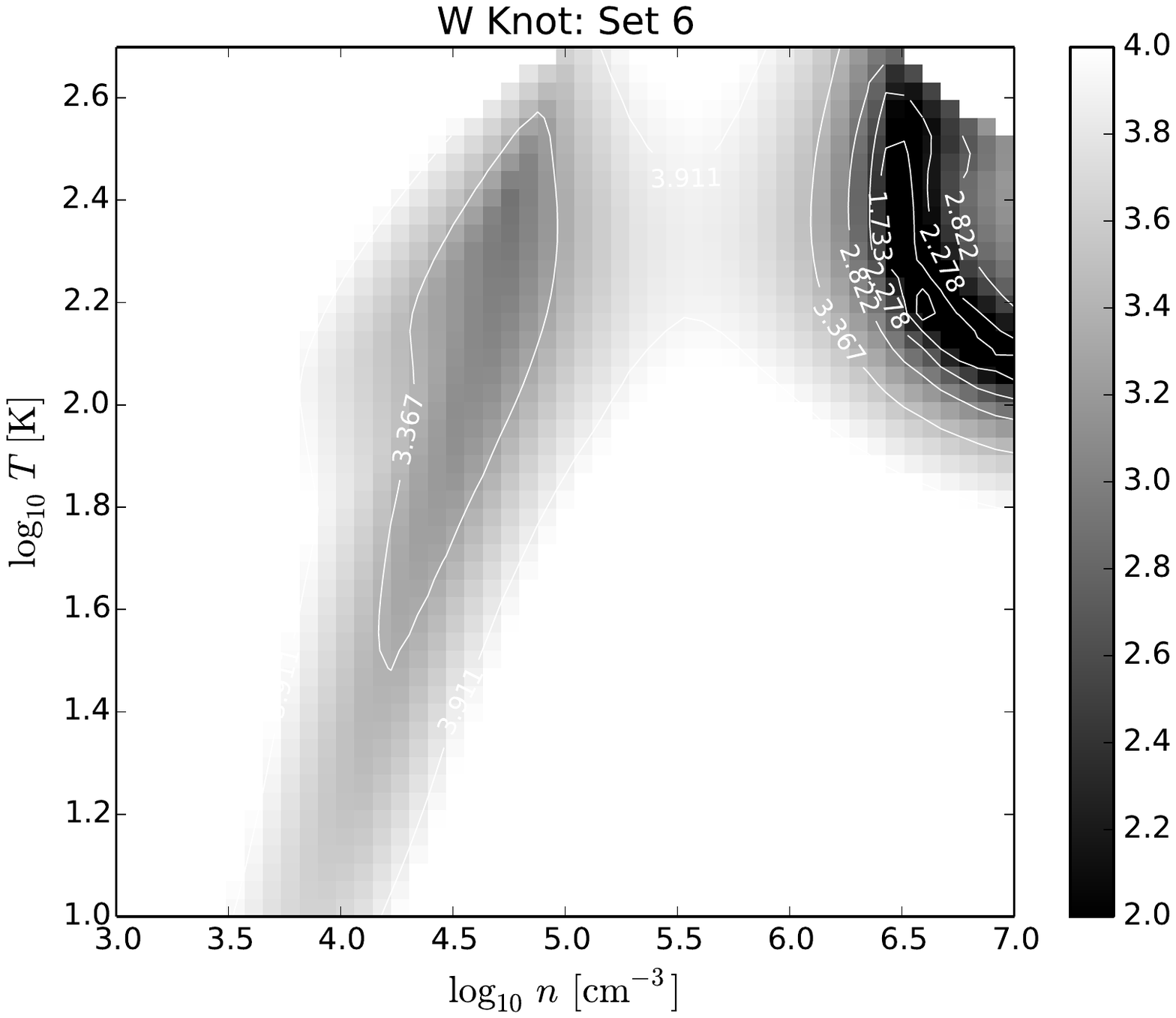}
\includegraphics[width=0.4\textwidth]{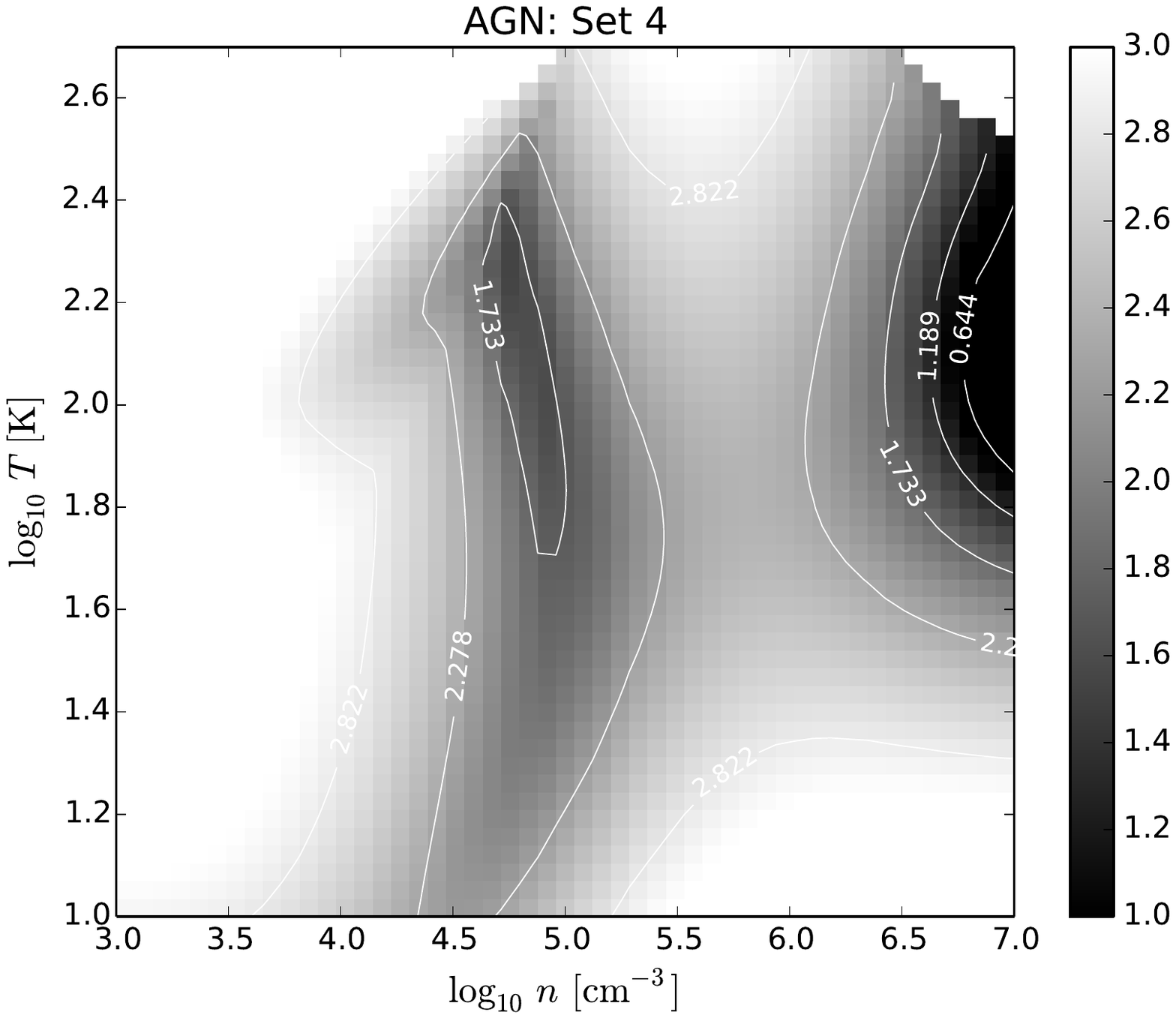}
\includegraphics[width=0.4\textwidth]{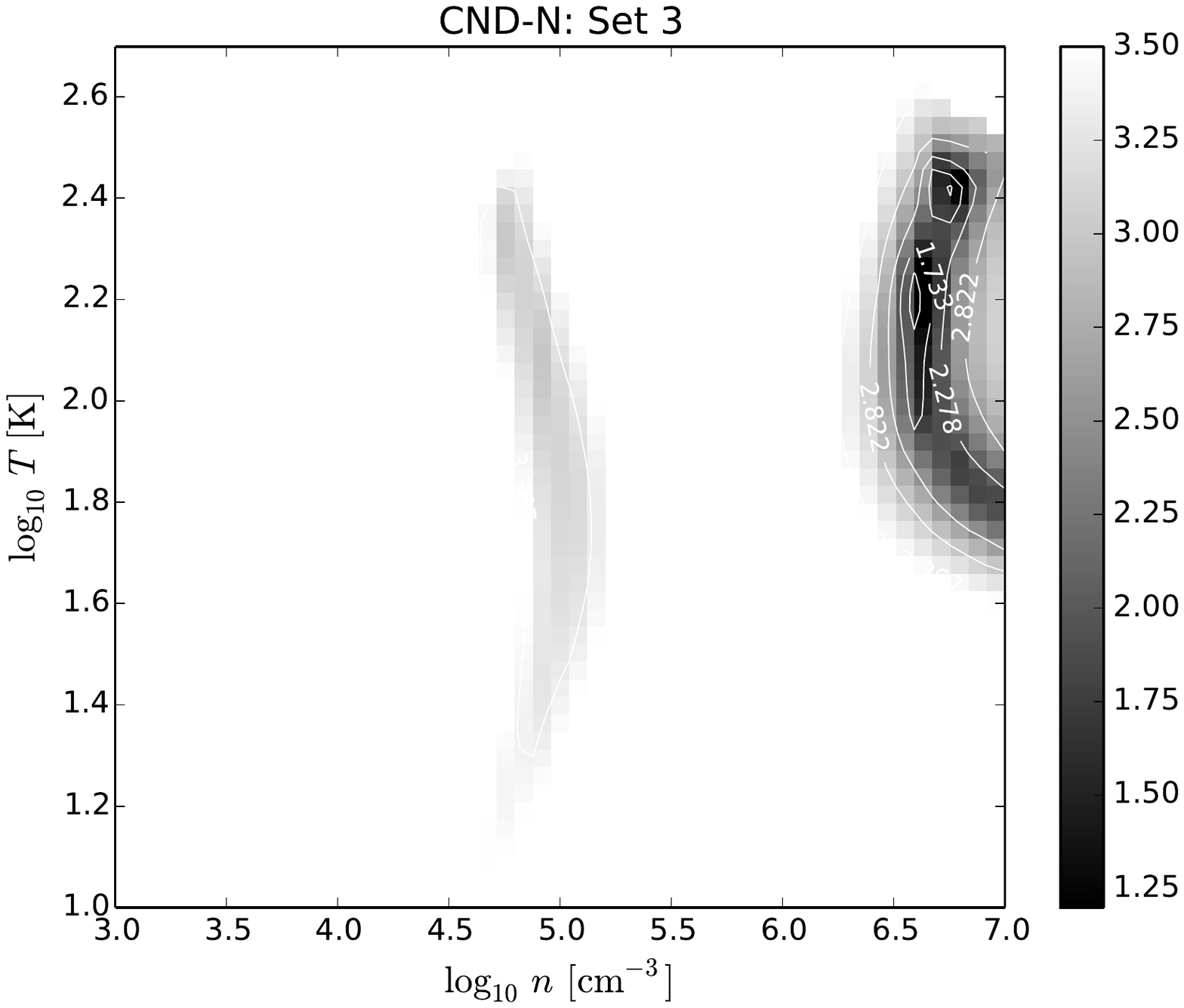}
}%
\caption{Log of reduced $\chi^2$ fits results from the RADEX simulations of the excitation conditions within the E Knot (Top Left and Middle), the W Knot (Top Right and Bottom Left), the AGN (Bottom Middle), the CND-S (Bottom Right) using ALMA data only (apart from one isotopic ratio - see text). }
\label{highres}
\end{figure*}

For this grid of models we used only ALMA data, apart from the intensities for the isotopic ratios. We used the following intensity ratios: CO(6-5/3-2); HCN(4-3)/HCO$^+$(4-3), HCN(4-3)/CO(3-2), CS(7-6)/CO(3-2),  $^{12}$CO(3-2)/$^{13}$CO(3-2), where the latter has been included in order to help constraining the N(X)/$\Delta(V)$.  
In general we find that this grid gives a better reduced log $\chi^2$. More importantly, by reducing the size of the emission region it is more likely that the emission may arise from gas with similar physical conditions. Figure~\ref{highres} and Table~\ref{tbhighres} present the best fits for four of our regions. For the CND-S we could not perform this analysis as the CS(7-6) line intensity was too weak at its original spatial resolution.
We note that (i) for this RADEX analysis the best fits are never given by a unique Set; (ii) the solutions for the gas densities are all consistent with the {\em higher} end of the density solutions found from the previous multi-species RADEX analysis, and especially the CO RADEX analysis; this is consistent both with the fact that HCN and CS are tracers of high density gas (unlike CO) and with the fact that a 35 pc beam is more likely to encompass an average higher density gas than a 100 pc one. 
(iii) Apart from the E Knot, none of the Sets producing the best fits coincide with those found in the previous RADEX analysis, again underlining the need to perform RADEX analysis with data
of the same (high) spatial resolution if one wants to characterize the chemical differentiation across the CND. 

For the E Knot both Set 2 and Set 7 give a best fit (see Table~\ref{sets}). Both sets constrain the kinetic temperature once again to around 60 K, confirming that this region is the coolest within the CND. 
For the W Knot, the best fits are given by either  Set 2 or Set 6,  with Set 6 giving best fits gas density and temperature closer to the previous RADEX analyses fits; the AGN is best fitted by Set 4 and the best fit density and temperature is in agreement with previous RADEX analyses; the CND-N is best fitted by Set 3. 

The column density ratios are listed in Table 9. Favouring the solutions from this analysis, we find that the AGN has the lowest CO/HCO$^+$, CO/HCN and CO/CS ratios; the E Knot and the W Knot have the highest CO/HCO$^+$ and CO/HCN ratios (if Set 2 is favoured); while HCN/HCO$^+$ is constant, at 10, across the CND with the possible exception of the E Knot where it is 100 (if Set 7 is favoured). In order to use these ratios as tracers of the dominating energetic process(es) within each CND component, they need to be compared with chemical models, as we shall do in Section 4. We emphasize here that, while this analysis benefitted from the use of all (but one) ratios of transitions observed at the highest spatial resolution, it however still suffers from the use of a single transition for 
three of the four molecules. This potentially leads to biases, especially with respect to gas density estimation. 

\begin{table*}
\caption{Physical and chemical characteristics of the 5 subregions within the CND as derived by a RADEX analysis of only the high resolution ALMA ratios, $and$ using the lower limit of the gas density as derived by the CO RADEX fitting as a constraint. For the Sets parameters (Column 2), see Table~\ref{sets}. a(b) stands for a $\times$ 10$^{b}$}
\label{tbhighres}
\begin{tabular}{c|c|ccccccc}
\hline
Region    & Set&  N(CO)/N(HCO$^+$)  & N(CO)/N(HCN) & N(HCN)/N(HCO$^+$) & N(CO)/N(CS) & $n$(H$_2$) (cm$^{-3}$) & $T_k$ (K)  \\
\hline
E Knot    & 7 & 1(6) &  1(4) & 1(2) & 1(5)  & 5$\times$10$^5$--10$^6$ & 60 \\
          & 2 & 5(5) &  5(4) & 10 &   5(3)  & $>$10$^6$  & 60 \\ 
W Knot    & 2 & 5(5) &  5(4) & 10 &   5(3)  & $>$5$\times$10$^6$ & 60--100\\
          & 6 & 1(5) &  1(4) & 10 &   1(4)  & 2$\times$10$^6$ & 150 \\ 
AGN       & 4 & 1(4) &  1(3) & 10   & 1(3)  & $>$3$\times$10$^6$      & 80--250 \\
CND-N     & 3 & 1(5) &  1(4) & 10   & 1(4)  & 3$\times$10$^6$         & 100-160 \\
\hline
\end{tabular}
\end{table*}

\subsection{RADEX analysis of the SB ring} 

\begin{figure*}[tbh!]
\centering
\includegraphics[width=15cm]{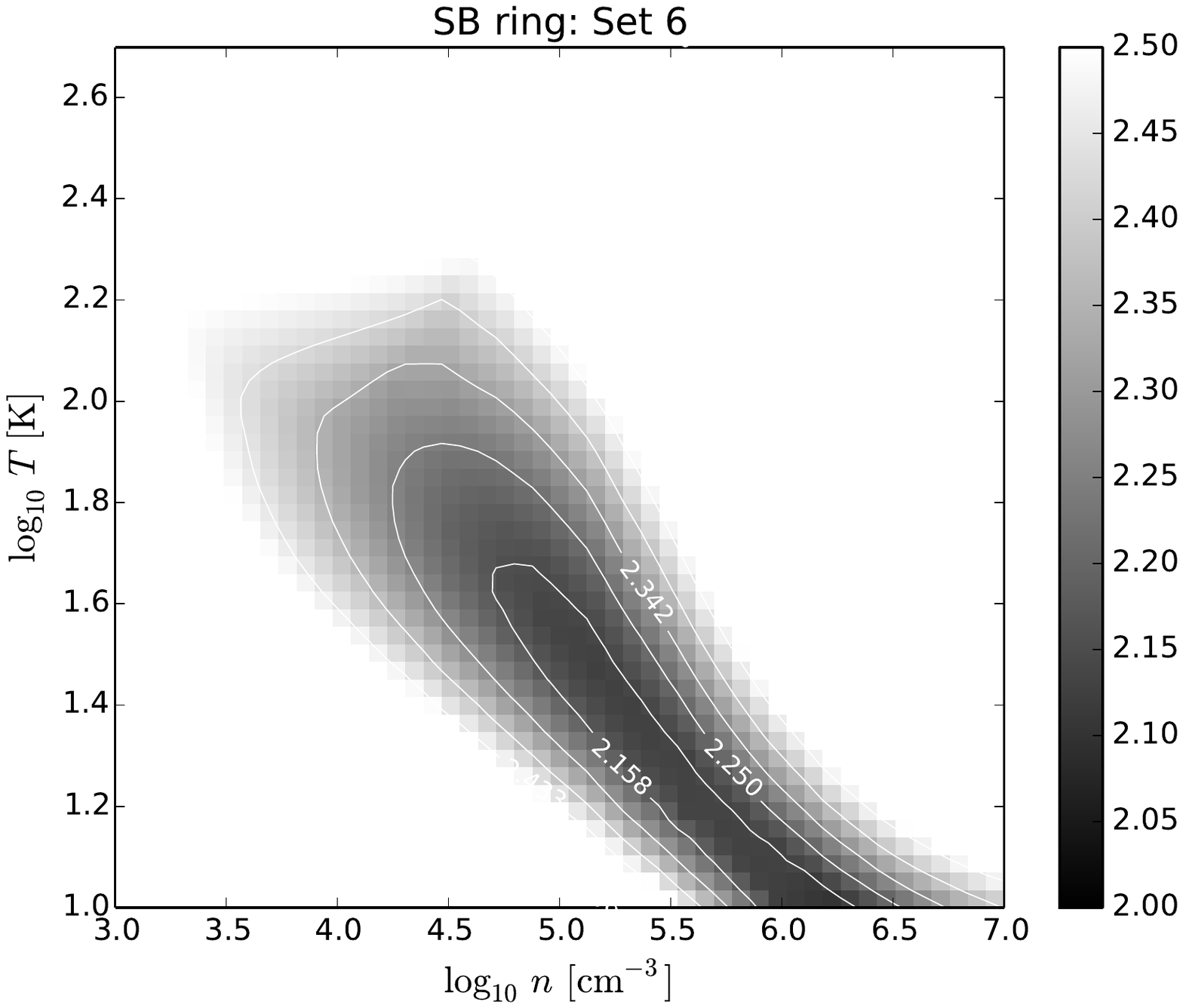}
\caption{Log of the reduced $\chi^2$ fit results from the RADEX simulations of the excitation conditions of the SB ring.}
\label{sbr_radex}
\end{figure*}

For the starburst ring only the following observations are available: CO (3-2), CO (1-0), HCO$^+$ (4-3), HCO$^+$ (1-0), HCN (4-3), and HCN (1-0). 
We therefore perform the RADEX analysis on the following ratios: CO(3-2)/(1-0), HCO$^+$(4-3)/(1-0), HCN(4-3)/(1-0), HCN(1-0)/CO(1-0), HCN(1-0)/HCO$^+$(1-0), the three isotopic line intensity ratios for the CO transitions, H$^{13}$CO$^+$(1-0)/H$^{12}$CO$^+$(1-0) and H$^{13}$CN(1-0)/H$^{12}$CN(1-0). The isotopic ratios are again from \citet[][Usero, priv. communication]{Papa99,User04}. 
We use the same sets as in Table ~\ref{sets} and find that the best fits are for Sets 6 (Figure~\ref{sbr_radex}) and 7, both giving a gas density ranging from 3$\times$10$^4$ to 3$\times$10$^6$ cm$^{-3}$; the temperature is less constrained with a maximum value around 60 K. 

In terms of molecular ratios, if we take the results from the RADEX fit in Section 3.2.3, then the SB ring 
has similar ratios as the W Knot or the E Knot, with however, quite different physical conditions. From Figure~\ref{sbr_radex}, it is clear that, once we exclude temperatures below $\sim$30 K (being the SB ring a site of high star formation activity),  then the gas density and temperature are quite well constrained to 5$\times$10$^4$--10$^5$ cm$^{-3}$ and $\sim$ 40 K.  

Without high spatial resolution data for the SB ring, we cannot quantitatively determine the chemical differentiation between the CND and the SB ring, but it seems that the SB ring has an 
average temperature much lower than that in the CND, implying 
that the differences in line intensities may be, at least partly, due to excitation effects i.e a lower gas temperature may lead to a lower HCO$^+$(4-3) and HCN(4-3) intensity, for example, which would justify the lower HCN(4-3)/HCN(1-0) and HCO$^+$(4-3)/HCO$^+$(1-0) ratios (see Table 4). A larger number of transitions at high spatial resolution will be required to determine, with RADEX, whether this scenario is valid. 

\section{Chemical modelling}

\citet{Alad13} observed the nucleus of NGC1068 with the IRAM 30-m single dish telescope in the $\lambda$ $\sim$ 3\,mm range 
in several molecules including,
CO, HCO$^+$ and HCN. Despite the lack of spatial resolution they attempted to disentangle the different gas components and energetics 
by the use of time dependent gas-grain chemical and PDR models.  Their main conclusions were that molecules that are usually assumed to be PDR tracers can in fact also 
arise from (shielded from the UV) dense gas. They also found that a high cosmic ray ionisation, possibly together with an enhanced (compared to a standard) radiation field, was needed in order to explain most of the molecular abundances, and they found that shocks were not necessary (but they could not rule them out). While their modelling constrained some of the energetics, it could not provide any physical information on 
the different gas components.  Here we use the UCL\_CHEM model \citep{Viti04,Viti11} 
to determine the best physical characteristics of each individual component within the CND  by comparing the ratio of theoretical column densities with the ratio of observed column densities derived from the RADEX analysis.

UCL\_CHEM is a 1D time dependent gas-grain chemical and shock model. It is an ab initio model in that it computes the evolution as a function of time and density  of chemical abundances of
the gas and on the ices starting from a complete diffuse and atomic gas. For the present calculations the code was used as a single point mode (i.e depth-independent) and was ran in two Phases in time (which we shall call Phase I and Phase II) so that: the initial density in Phase I was $\sim$ 100 cm$^{-3}$. During Phase I, the high density gas is achieved by means of a free-fall (or modified) collapse, in a time determined by the final density (which we varied). The temperature during this phase is kept constant at 10 K, and the cosmic ray ionization rates and radiation fields are at their standard Galactic values ($\zeta$ = 5$\times$10$^{-17}$ s$^{-1}$ and 1 Draine, or 2.74$\times$10$^{-3}$ erg/s/cm$^{2}$, respectively). Note that the collapse in Phase I is not meant to represent the formation of protostars, but it is simply a way to compute the chemistry of high density gas in a self-consistent way starting from a diffuse atomic gas, i.e without assuming the chemical composition at its final density. In reality we do not know how the gas reached its high density and we could have assumed any other number of density functions with time. For 
simplicity we adopt free-fall collapse.    Atoms and molecules are allowed to freeze onto the dust grains forming icy mantles. Non-thermal evaporation is also included \citep{Robe07}. Both gas and surface chemistry are self-consistently computed.  Once the gas is in dynamical equilibrium again, Phase I is either allowed to continue at constant density for a further million of years, or is stopped. In Phase II, 
UCL\_CHEM computes the chemical evolution of the gas and the dust after either an assumed burst of star formation or AGN activity has occurred. The temperature of the gas increases from 10~K  to a value set by the user, and sublimation from the icy mantles occurs. The chemical evolution of the gas is then followed for about $5\times10^6$ years. In both phases of the UCL\_CHEM model the chemical network is based on a modified UMIST data base \citep{Wood07}.     The surface reactions included in this model are assumed to be mainly hydrogenation reactions, allowing chemical saturation when possible. 

There are several free parameters involved in the setting up of chemical models. We have kept fixed the initial elemental abundances at the standard solar metallicity ISM abundances compared to 
the total number of hydrogen nuclei (here we list the most abundant ones, namely, C, O, N and S which are, respectively, 2.7$\times$10$^{-4}$, 4.9$\times$10$^{-4}$, 6.8$\times$10$^{-5}$ and 1.3$\times$10$^{-5}$; see \citet{Aspl09} for a complete list) but explored the parameter space for the quantities listed in Table~\ref{model_grid}. In particular, we varied: (i) the final gas density at the end of the collapse (note that for the models where a shock is simulated this density corresponds to the pre-shock density);(ii) the temperature (100 K and 200 K) during Phase II; the range of final densities and temperatures in Phase II explored reflect the best fits found in this and previous studies of NGC~1068.
(iii) the radiation field ($G_0$); although the range of radiation fields investigated is somewhat arbitrary, our aim was to explore the influence of 
relatively high 
values, typical of starburst galaxies, on the observed molecules, and hence we picked three representative values of a low (1 Draine), medium (10 Draine) and high (500 Draine) radiation field. (iv) the cosmic ray ionization rate ($\zeta$);  note that in our chemical models the cosmic ray ionization flux is also used to `mimic' an enhanced X-ray flux, which is why we choose such a large range. This approximation has its limitations because X-rays ionise and heat gas much
 more efficiently than cosmic rays; however the trends in some ions, including HCO$^+$, are the same for CRs and X-rays
for a fixed gas temperature (which is what is assumed in these models);   (v) the length of time at which the dense gas is quiescent in Phase I, where either Phase I was stopped once the final density was reached (we shall call these the S models), or we allowed Phase I to run for a further 1--2$\times10^6$  years (L models). Phase I in the S models lasts for different times depending on the final density (for reference, 5 $\times$ 10$^{6}$ yrs for a final density of 10$^4$ cm$^{-3}$ up to 5.4 $\times$ 10$^{6}$ yrs for a final density of 2 $\times$ 10$^6$ cm$^{-3}$). However, since models differing only in this parameter gave very similar solutions we shall not discuss this parameter any further.  (vi) Finally we also ran some models simulating the presence of shocks (in Phase II). For these models we adopted a representative shock velocity of 40 km~s$^{-1}$. The shocks in UCL\_CHEM are treated in a parametric form as in \citet{Jim08}, which considers a plane-parallel C-Shock propagating with a velocity v$_s$ through the ambient medium. 
Details of this version of UCL\_CHEM can be found in \citet{Viti11}. 
  
\begin{table*}
\caption{Grid of chemical models. Note that each model has been ran for two different visual extinctions, 2 and 10 mag, as well as for different Phase I times (see text). 
The cosmic ray ionization rate, the radiation field, the gas temperature, and the gas density of Phase II are listed in columns 2--5.}  
\label{model_grid}
\begin{tabular}{|c|ccccc|}
\hline
Model n$^o$ &  $\zeta$ (s$^{-1}$) & $G_0$ (Draine) & T (K) & n$_{final/preshock}$ (cm$^{-3})$ & Shock \\
\hline
1 & 1&  1 & 100 & 10$^4$  & N \\
2 & 1& 1& 100 & 10$^5$  & N \\
3 & 1& 1& 100 & 10$^6$  & N \\
4 & 10& 1& 100 & 10$^4$  & N \\
5 & 10& 1& 100 & 10$^5$  & N \\
6 & 10& 1& 100 & 10$^6$  & N \\
7 & 1& 10& 100 & 10$^4$  & N \\
8 & 1& 10& 100 & 10$^5$  & N \\
9 & 1& 10& 100 & 10$^6$  & N \\
10 & 10& 10 & 100 & 10$^4$  & N \\
11 & 1& 500 & 100 & 10$^4$  & N \\
12 & 500& 1& 100 & 10$^4$  & N \\
13 & 5000& 1& 100& 10$^4$  & N \\
14 & 10$^5$ & 1& 100 & 10$^5$  & N \\
15 & 10& 1& 100 & 2$\times$10$^6$  & N \\
16 & 1 & 1 & 100 & 5$\times$10$^6$  & N \\
17 & 1 & 1 & 100 & 2$\times$10$^6$  & N \\
18 & 1 & 1 & 200 &  10$^5$  & N \\
19 & 1 & 1 &  200 &  10$^4$  & N \\
20 & 1 & 1 & 200 & 10$^6$  & N \\
21 & 10 & 1 & 200 & 10$^6$  & N \\
22 & 1 & 10 & 200 & 10$^6$  & N \\
23 & 10 & 1 & 200 & 10$^5$  & N \\
24 & 1 & 10 & 200 & 10$^5$  & N \\
25 & 1& 1 & -- & 10$^4$  & Y \\
26 & 10& 1 & -- & 10$^5$  & Y \\
27 & 1& 1 & -- & 10$^5$  & Y \\
\hline
\end{tabular}
\end{table*}
Before performing a qualitative comparison we need to explicitly note that 
{\it even at the highest spatial resolution of the ALMA data, the gas is extended to
 $\sim$ 35 pc and hence the emission cannot be from 
a single density component}; hence we do not expect one chemical model to fit all the molecular ratios. Moreover, 
without any further information of the size/geometry we are constrained in 
using ratios of fractional abundances (which do not necessarily represent the ratio of column densities if the emission region varies from molecule to molecule). Hence, we ran each model for different sizes so that to achieve two representative visual extinctions of 2 and 10 mags, for the {\it same} gas density. The exact choice of A$_{\rm v}$ is {\it not} important for these calculations because we are then using column density ratios. However it is important to choose an A$_{\rm v}$ below and one above the typical extinction at which the gas is shielded by UV radiation. 
Note that all the analysis and comparisons of the abundances from the chemical models are performed at the last time step of Phase II. In fact, as discussed later, and as also discussed in \citet{Alad13} and more generally in \citet{Baye08} and \citet{Meij13}, the effects of time dependence cannot be ignored in energetic extragalactic environments such as AGNs. 

\begin{figure*}[tbh!]
\centerline{%
\includegraphics[width=0.5\textwidth]{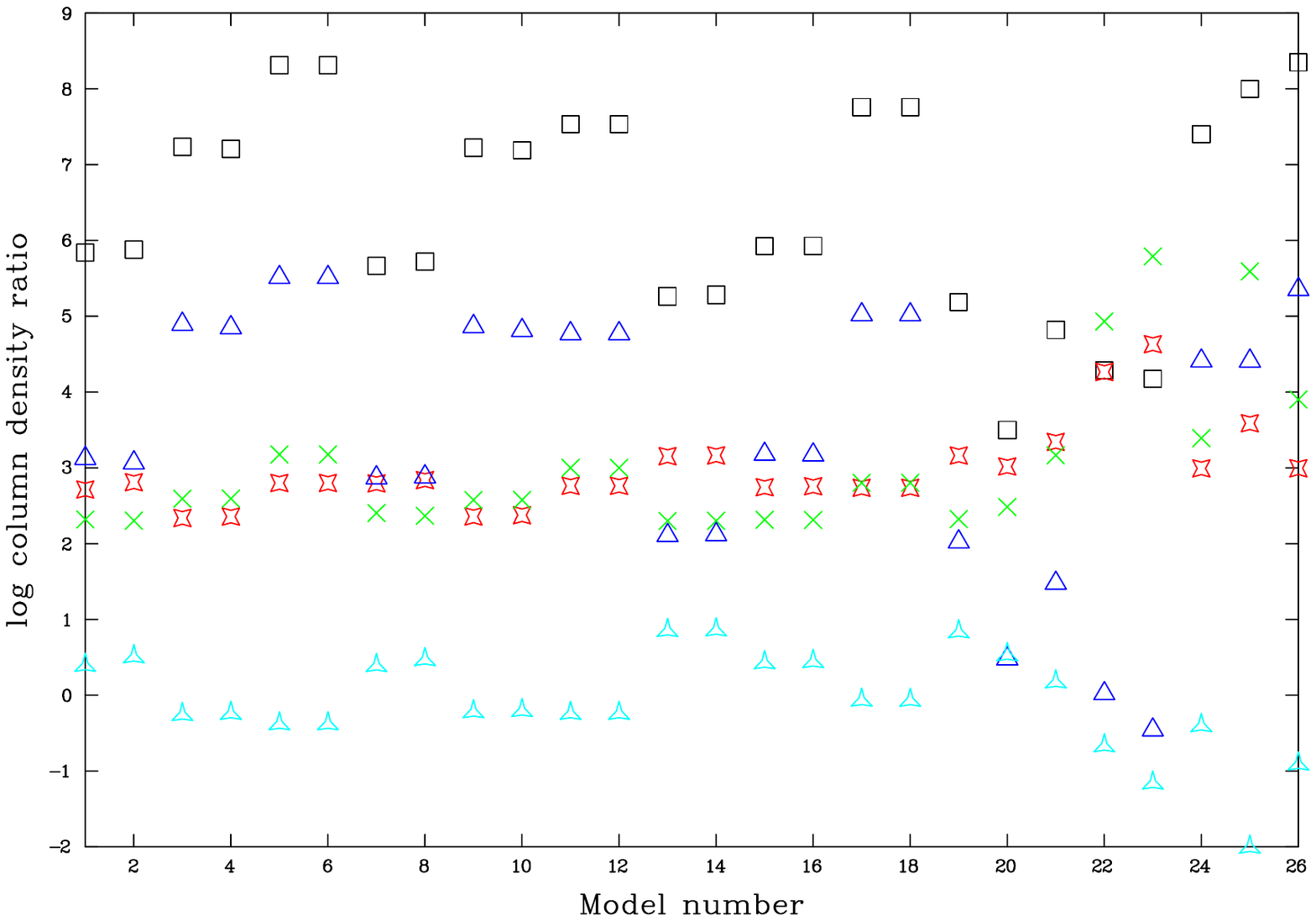}
\includegraphics[width=0.5\textwidth]{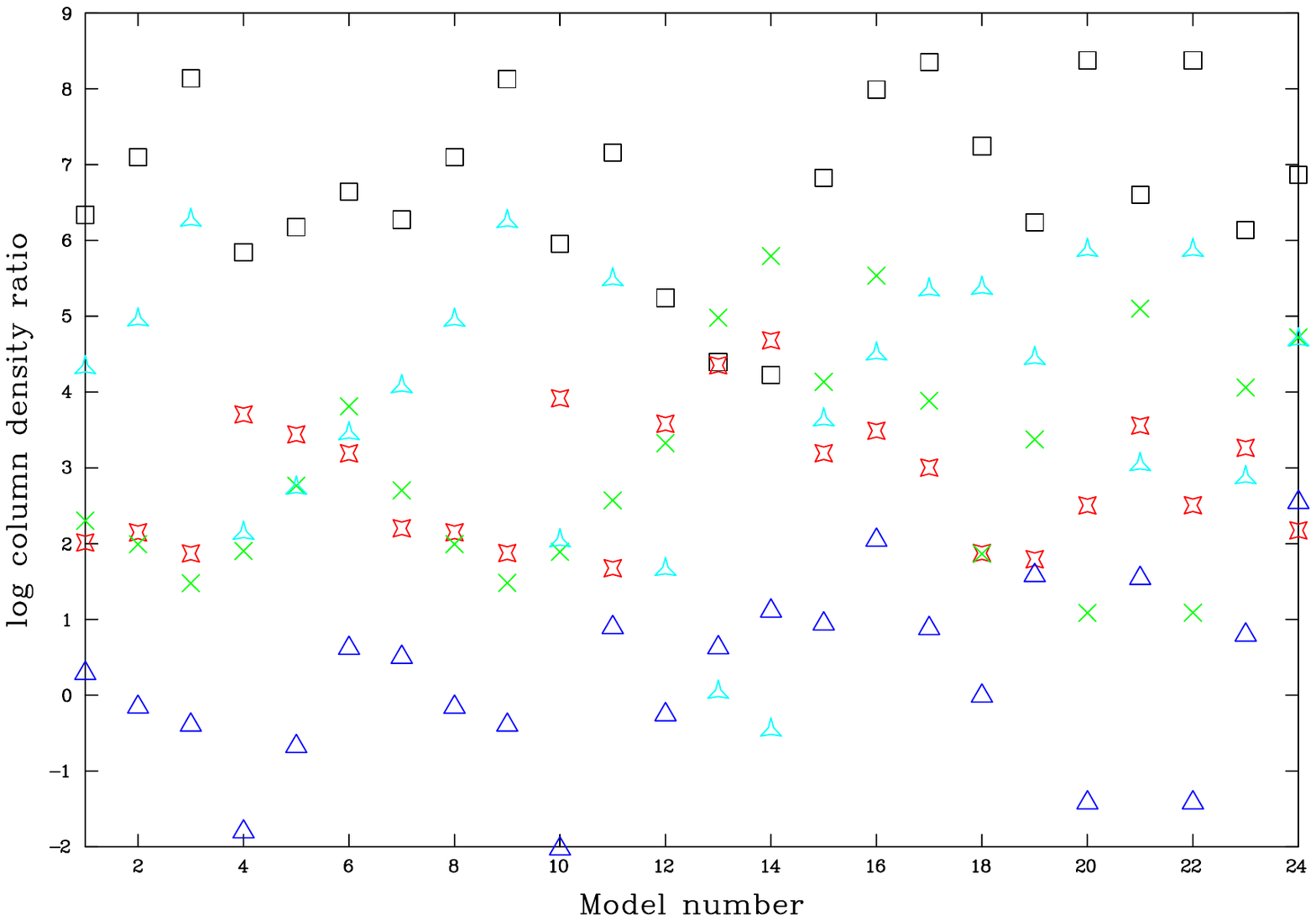}
}
\caption{Theoretical fractional abundances ratios for the grid of UCL\_CHEM models at low A$_V$ (2 mags; Left Panel) and at high A$_V$ (10 mags; Right Panel). The black squares are CO/HCO$^+$ ratios; red stars = CO/HCN; green crosses = CO/CS; cyan triangles = HCN/CS; blue triangles = HCN/HCO$^+$.}
\label{uclchem}
\end{figure*}

We shall now compare theoretical ratios (in log in Figure~\ref{uclchem}) with the RADEX solutions from Section 3.2.2 as well as Section 3.2.3.
As the chemical model is single point, the theoretical column densities have been estimated, a posteriori, from the fractional abundances using the `on the spot' approximation: 

\begin{equation}
N(X) = X \times {\rm A_v} \times N(H_{2}),
\end{equation}

\noindent
where A$_{\rm v}$ is the visual extinction and $N(H_{2})$ is equal to 1.6$\times$10$^{21}$ cm$^{-2}$, the hydrogen column density corresponding to A$_V$ = 1 mag \citep{dyswill97}. Hence the column density derived is $not$ cumulative but the column density at a particular visual extinction. Of course this column density would be equivalent to the cumulative one {\it if} the fractional abundance of species $X$ were the same at all the A$_{\rm v}$   (which is not the case). 
In Figure~\ref{uclchem} the left and right panels correspond, respectively, to models ran at low and high A$_{\rm v}$. It is worth noting that since we are using logarithmic values of column density ratios, we are `insensitive' to small differences. Note also that Models 25-27 are not shown in the Figure but will be discussed below.

Our multi-species RADEX analyses give several possible solutions for the molecular line ratios (e.g., Table 8) which we will now compare with the grid of chemical models 
from Figure~\ref{uclchem}. We report our results in Tables~\ref{summ_mod1} and ~\ref{summ_mod2}, where we compare the log of the molecular ratios from our chemical models with the solutions from Sections 3.2.2 and 3.2.3, respectively. 

\begin{table*}
\caption{Summary of the best fit chemical model(s) for the molecular ratios as derived from the RADEX analysis in Section 3.2.2. }
\label{summ_mod1}
\begin{tabular}{ccccccccc}
\hline
Region & Set  & log(Ratio)$_{radex}$  & Model & $A_V$ (mags) & $n(H_2)$ (cm$^{-3}$) & $T_k$ (K) & $G_0$ (Draine) & $\zeta$ (10$^{-17}$ s$^{-1}$) \\
\hline
All &7 & CO/HCO$^+$  =  6 & 9  & 2  & 10$^6$          & 100 & 10 & 1 \\
    &  &                  & 10 & 10 & 10$^4$          & 100 & 10 & 10 \\
    &  & CO/HCN      =  4 & 10 & 10 & 10$^4$          & 100 & 10 & 10 \\
    &  & HCN/HCO$^+$ =  2 & 17 & 2  & 2$\times$10$^6$ & 100 & 1  & 1 \\
    &  &                  & 16 & 10 & 5$\times$10$^6$ & 100 & 1  & 1 \\
    &  & CO/CS       =  5 & 14 & 2  & 10$^5$          & 100 & 1  & 10$^5$ \\
    &  &                  & 13 & 10 & 10$^4$          & 100 & 1  & 5$\times$10$^3$ \\
    &  & HCN/CS      =  1 & 25(early times) & -- & 10$^4$          & --- & 1  & 1  \\ 
\hline
E Knot & 1 & CO/HCO$^+$  = 5.7 & 5 & 2, 10 & 10$^5$ & 100 & 1 & 10 \\
      &    &                   & 20 & 2    & 10$^6$  & 200 & 1 & 1 \\
       &   & CO/HCN      = 4.7 & 15 & 2    & 2$\times$10$^6$ & 100 & 10& 1 \\
       &   &                  &  14 & 10   & 10$^5$          & 100 & 1  & 10$^5$ \\
       &   & HCN/HCO$^+$ = 1  &  3  & 2    & 10$^6$  & 100 & 1  & 1 \\
       &   &                  &  18 & 2    & 10$^5$  & 200 & 1  & 1 \\
       &   &                  &  15 & 10   & 2$\times$10$^6$ & 100 & 10& 1 \\
       &   & CO/CS       = 4.7&  24 & 10   & 10$^5$  & 200 & 10 & 1 \\
       &   & HCN/CS      = 0  &  14 & 2    & 10$^5$          & 100 & 1  & 10$^5$ \\
       &   &                  &  13 & 10  & 10$^4$          & 100 & 1  & 5$\times$10$^3$ \\
       &   &                  &  27(late times) & -- & 10$^5$          & --- & 1  & 1  \\ 
\hline
AGN & 3, 6 & CO/HCO$^+$  = 5 &   13 & 2    & 10$^4$          & 100 & 1  & 5$\times$10$^3$ \\
    &      & CO/HCN      = 4 &   10 & 10   & 10$^4$          & 100 & 10 & 10 \\
    &      & HCN/HCO$^+$ = 1 &   3  & 2    & 10$^6$  & 100 & 1  & 1 \\
    &      &                 &   18 & 2    & 10$^5$  & 200 & 1  & 1 \\
    &      &                 &   15 & 10   & 2$\times$10$^6$ & 100 & 10& 1 \\
    &      & CO/CS       = 4 &   18 & 2    & 10$^5$  & 200 & 1  & 1 \\
    &      &                 &   24 & 2    & 10$^5$  & 200 & 10 & 1 \\
    &      & HCN/CS      = 0 &   14 & 2    & 10$^5$          & 100 & 1  & 10$^5$ \\
       &   &                  &  13 & 10  & 10$^4$          & 100 & 1  & 5$\times$10$^3$ \\
       &   &                  &  27(late times) & -- & 10$^5$          & --- & 1  & 1  \\
\hline
CND-S & 2  & CO/HCO$^+$  = 5.7 & 5 & 2, 10 & 10$^5$ & 100 & 1 & 10 \\
      &    &                    & 20& 2    & 10$^6$  & 200 & 1 & 1 \\
    &   & CO/HCN      = 4.7     & 15& 2    & 2$\times$10$^6$ & 100 & 10& 1 \\
       &   &                    &  14 & 10   & 10$^5$          & 100 & 1  & 10$^5$ \\
    &   & HCN/HCO$^+$ = 1       &  3  & 2    & 10$^6$  & 100 & 1  & 1 \\
       &   &                    &  18 & 2    & 10$^5$  & 200 & 1  & 1 \\
       &   &                    &  15 & 10   & 2$\times$10$^6$ & 100 & 10& 1 \\
    &   & CO/CS       = 3.7     &  6  & 10   & 10$^6$  & 100 & 1  & 10 \\     
    &   & HCN/CS      = 0       & 14 & 2    & 10$^5$          & 100 & 1  & 10$^5$ \\
       &   &                  &  13 & 10  & 10$^4$          & 100 & 1  & 5$\times$10$^3$ \\
       &   &                  &  27(late times) & -- & 10$^5$          & --- & 1  & 1  \\

\hline
\end{tabular}
\end{table*}
Tables 11 and 12 are hard to interpret because not only each region can be fitted by more than one Set of column densities but also, as expected, there is no one single model that can explain the set of all ratios; this is of course consistent with the fact that we need more than one gas component to characterize each subregion. 
We now attempt to remove some of the chemical model solutions by taking into consideration the density constraints provided 
by the RADEX analyses: the latter exclude a density $\leq$ 10$^4$ cm$^{-3}$, hence we shall exclude {\it a priori} all models with a gas density of 10$^4$ cm$^{-3}$ to start with. For the matching of the
ratios involving HCO$^+$, we shall also exclude chemical models where neither the radiation field nor the cosmic ray ionization rate are enhanced: this implies that the CO/HCO$^+$ abundance ratio in all regions can be matched by Model 9 for Set 7. The ratio in the E Knot can also be matched by Model 5 for Sets 1 or 2, and in the W Knot and the CND-S 
by Model 5 for Set 2; and finally for the AGN by Model 15 for Set 4. The other ratio involving HCO$^+$ is the HCN/HCO$^+$ ratio: this ratio can only be matched by Model 15 for all our regions
with different Sets, implying that the gas emitting HCO$^+$ in the CND must be subjected to an enhanced cosmic ray ionization field but also that it must be quite dense.  We note that Set 7 cannot match the CO/HCN ratio which is well matched by Model 15 for all our regions but, again, for different sets. The CO/CS can only be matched by models with high temperatures (24 for the E Knot with Set 1, and 18 or 24 for the AGN for Sets 3 or 6) or very high densities (e.g., Model 6 for the CND-S for Set 2);  the HCN/CS ratio, for most best fit models, seems to require  the presence of shocks i.e {\it if} HCN and CS are co-spatial shocks must be invoked; the presence of shocks is consistent with the findings from Paper I. Considering the many possible combinations of chemical sets and chemical models one can have for each subregion we cannot draw any further conclusions. Nevertheless
the `take home' message that can be extracted from our chemical modelling is that it is that the chemistry is not the same in every subregion of the CND and 
that 
{\em provided} there is chemical differentiation across the CND then a simplified picture with three dominant gas phases emerges for all regions: a `low' density (10$^5$ cm$^{-3}$) region, plus a higher density ($\geq$ 10$^6$ cm$^{-3}$) region, both with enhanced (by a least a factor of 10) cosmic ray ionization rate (and/or X-ray activity), plus a component of shocked gas; conclusions on differences in gas densities, temperatures or energetics within the CND cannot be drawn from this preliminary modelling. While it may not be suprising that the chemistry differs across the CND, this is the first chemical study that, 
using high spatial resolution observations, coupled with a chemical model that includes gas-grain interactions,  
explicitly disentangle the different chemistries, albeit in a qualitative manner.

\begin{table*}[tbh!]
\caption{Summary of the best fit chemical model(s) for the molecular ratios in the different components in the CND using the best fit RADEX solutions from Section 3.2.3. }
\label{summ_mod2}
\begin{tabular}{ccccccccc}
\hline
Region & Set  & log(Ratio)$_{radex}$  & Model & $A_V$ (mags) & $n(H_2)$ (cm$^{-3}$) & $T_k$ (K) & $G_0$ (Draine) & $\zeta$ (10$^{-17}$ s$^{-1}$) \\
\hline
E Knot &7 & CO/HCO$^+$  =  6 & 9  & 2  & 10$^6$          & 100 & 10 & 1 \\
    &  &                  & 10 & 10 & 10$^4$          & 100 & 10 & 10 \\
    &  & CO/HCN      =  4 & 10 & 10 & 10$^4$          & 100 & 10 & 10 \\
    &  & HCN/HCO$^+$ =  2 & 17 & 2  & 2$\times$10$^6$ & 100 & 1  & 1 \\
    &  &                  & 16 & 10 & 5$\times$10$^6$ & 100 & 1  & 1 \\
    &  & CO/CS       =  5 & 14 & 2  & 10$^5$          & 100 & 1  & 10$^5$ \\
    &  &                  & 13 & 10 & 10$^4$          & 100 & 1  & 5$\times$10$^3$ \\
    &  & HCN/CS      =  1 & 25(early times) & -- & 10$^4$          & --- & 1  & 1  \\
\hline
E Knot, W Knot & 2  & CO/HCO$^+$  = 5.7 & 5 & 2, 10 & 10$^5$ & 100 & 1 & 10 \\
      &    &                    & 20& 2    & 10$^6$  & 200 & 1 & 1 \\
    &   & CO/HCN      = 4.7     & 15& 2    & 2$\times$10$^6$ & 100 & 10& 1 \\
       &   &                    &  14 & 10   & 10$^5$          & 100 & 1  & 10$^5$ \\
    &   & HCN/HCO$^+$ = 1       &  3  & 2    & 10$^6$  & 100 & 1  & 1 \\
       &   &                    &  18 & 2    & 10$^5$  & 200 & 1  & 1 \\
       &   &                    &  15 & 10   & 2$\times$10$^6$ & 100 & 10& 1 \\
    &   & CO/CS       = 3.7     &  6  & 10   & 10$^6$  & 100 & 1  & 10 \\
    &   & HCN/CS      = 0       & 14 & 2    & 10$^5$          & 100 & 1  & 10$^5$ \\
       &   &                  &  13 & 10  & 10$^4$          & 100 & 1  & 5$\times$10$^3$ \\
       &   &                  &  27(late times) & -- & 10$^5$          & --- & 1  & 1  \\
\hline
W Knot (CND-N) &  6 (3) & CO/HCO$^+$  = 5 &   13 & 2    & 10$^4$          & 100 & 1  & 5$\times$10$^3$ \\
    &      & CO/HCN      = 4 &   10 & 10   & 10$^4$          & 100 & 10 & 10 \\
    &      & HCN/HCO$^+$ = 1 &   3  & 2    & 10$^6$  & 100 & 1  & 1 \\
    &      &                 &   18 & 2    & 10$^5$  & 200 & 1  & 1 \\
    &      &                 &   15 & 10   & 2$\times$10$^6$ & 100 & 10& 1 \\
    &      & CO/CS       = 4 &   18 & 2    & 10$^5$  & 200 & 1  & 1 \\
    &      &                 &   24 & 2    & 10$^5$  & 200 & 10 & 1 \\
    &      & HCN/CS      = 0 &   14 & 2    & 10$^5$          & 100 & 1  & 10$^5$ \\
       &   &                  &  13 & 10  & 10$^4$          & 100 & 1  & 5$\times$10$^3$ \\
       &   &                  &  27(late times) & -- & 10$^5$          & --- & 1  & 1  \\
\hline
AGN    & 4 & CO/HCO$^+$  = 4 & 15 & 2 & 2$\times$10$^6$ & 100 & 10& 1 \\
       &      & CO/HCN   = 3 & 12 & 2 & 10$^4$          & 100 & 1 & 500 \\
       &      & HCN/HCO$^+$ = 1  &   3  & 2    & 10$^6$  & 100 & 1  & 1 \\
    &      &                 &   18 & 2    & 10$^5$  & 200 & 1  & 1 \\
    &      &                 &   15 & 10   & 2$\times$10$^6$ & 100 & 10& 1 \\
       &   & CO/CS       = 3 &   7  & 2  & 10$^4$          & 100 & 10 & 1 \\
       &   & HCN/CS      =  0 &   14 & 2    & 10$^5$          & 100 & 1  & 10$^5$ \\
       &   &                  &  13 & 10  & 10$^4$          & 100 & 1  & 5$\times$10$^3$ \\
       &   &                  &  27(late times) & -- & 10$^5$          & --- & 1  & 1  \\
\hline
\end{tabular}
\end{table*}

There are several caveats that should be considered when comparing chemical models with RADEX results: firstly that while RADEX provides a solution for the $average$ density and temperature, these chemical models calculate the chemistry at a fixed, homogeneous density and temperature, hence they do not calculate the $average$ molecular ratios across the regions, but the molecular ratio at a particular density, visual extinction and temperature. Hence it is not unreasonable to employ multiple chemical models to reproduce the chemical composition of our gas. 
Secondly, as RADEX is an excitation and radiative transfer model, the chemical effects of an enhanced UV, cosmic ray or X-ray irradiation 
on the chemistry are `embedded' in the input column densities.
While we have not, by any means, exhausted the parameter space of chemical modelling, we confirm that within each CND component more than one gas phase is present, and that more transitions 
of the same species need to be observed with ALMA resolution in order to determine the physical characteristics across the CND with confidence. 
Moreover we note that the models used here are time-dependent while we only compared ratios at chemical equilibrium. In fact some of our ratios, in particular HCN/HCO$^+$, are well known to be very time dependent in both dense gas (see Figure 1 from \citet{Baye08}) as well as in X-ray dominated environments (such as AGN) (see \citet{Meij13}) so multiple gas phases
 can co-exist spatially and temporally.  This attempt at chemical modelling was preliminary and a proper theoretical investigation including PDR, gas-grain, and XDR models will be presented in a forthcoming paper.

\section{Conclusions}

We performed a chemical analysis of the dense gas within the CND and the SB ring, with the aim of quantifying the chemical differentiation across the CND and attempt a determination of the chemical origin of such differentiation. From our analysis we can draw the following conclusions: 
\begin{itemize}
\item A simple LTE analysis shows that: the E Knot has a richer molecular content than the rest of the CND; the least chemically rich region is CND-S. The SB ring has a lower molecular content than the CND. From the CO/HCN ratios as derived from the transitions observed with ALMA we may deduce that
the E Knot is the most abundant in high density molecular tracers while the W Knot is the least abundant;  from the CO/HCO$^+$  ratio we see that the AGN and the E Knot are the regions where most energetic activity is happening;  from the CO/CS high $J$ column density ratio we also deduce that the E Knot may be the coolest region, while the AGN and the CND-S may be the hottest.    
\item  A rotation diagram analysis of CO leads to the following rotational temperatures: 58 K, 41 K, 50 K, 52 K, and 37 K for the E Knot, W Knot, AGN, CND-N and CND-S, respectively, with typical errors of 5-7 K. The rotational temperatures are lower limits, assuming that the emission is optically thin and in LTE.
\item A RADEX analysis of the CO ratios, including rare isotopologues, indicates that the lowest temperature component within the CND is the E Knot, which also has the lowest CO column density. The AGN, the W Knot and possibly the CND-N positions have the highest temperatures ($T_k$ $>$ 150 K). The average gas density is always above 10$^4$ cm$^{-3}$. 
\item A RADEX analysis of all the independent ratios with the species CO, HCN, HCO$^+$ and CS gives multiple solutions in terms of column densities as well as densities and temperatures. From this analysis alone we cannot discern whether the differences in molecular ratios across the CND are due to chemical diversity or differences in physical conditions or a combination of the two (which are not of course indepednent). 
\item A RADEX analysis performed with a reduced number of ratios using only ALMA data at the highest spatial resolution gives better contraints,
 in particular in terms of absolute values for the column densities, although the E Knot and the W Knot still have multiple solutions for the ratios of column densities. 
With this analysis we confirm our LTE findings that the AGN has the lowest CO/HCO$^+$, CO/HCN and CO/CS ratios, while the E Knot and W Knot have the highest CO/HCO$^+$ ratio and the W Knot (and possibly the E Knot) have the highest CO/HCN ratio.  This analysis gives a lower limit for the gas density of 5$\times$10$^5$ cm$^{-3}$ for the E Knot and higher densities for all the other components. The hottest region is the AGN while the coolest is the E Knot.
\item A RADEX analysis of the SB ring leads to the conclusion that it may be chemically similar to the W Knot, but with substantially lower gas density and temperature.
\item Finally, we attempt to determine the main driver of the chemistry in the CND by the use of a time dependent gas-grain chemical models, UCL\_CHEM. Our approach is to compare a relatively large grid of chemical models with solutions provided by the RADEX grids; this comparison proves to be rather difficult and does not lead to any strong quantitative conclusion. However, chemical modelling seems to indicate that there has to be a pronounced chemical differentiation across the CND and that each subregion could be characterized by a 3-phase component ISM, with two gas phases at different densities but both subjected to a high ionization rate, and one gas component comprising of shocked gas where most likely the CS (and possibly some of the HCN) arises from.  
\end{itemize}
In conclusion, the LTE, RADEX and chemical analysis all indicate that more than one gas-phase component is necessary to uniquely fit all the available molecular ratios. With the improved spatial resolution of ALMA we expect this result to be confirmed also by studies of other extragalactic systems. Chemically this is certainly not unexpected. For example, HCO$^+$ is usually the product of ion-molecule reactions involving C$^+$ and CO$^+$; these ions are only abundant in regions where CO can efficiently dissociate via reactions with He$^+$; while high fluxes of cosmic and/or X-rays can lead to a high ionization fraction even in dense environments, usually they also lead to low abundances of HCN and CS \citep[e.g.,][]{Baye11}. Even for the same molecular species, radiative transfer studies of multi-transition observations of nearby starburst galaxies have shown that different $J$ lines will peak at different gas densities and temperatures. Hence our somewhat degenerate results are consistent with multiple gas phases
spatially and temporally co-existing. Finally we emphasize that this analysis did not take into consideration infrared pumping which in fact may be responsible for the excitation of several of the observed transitions, as it has been recently shown to be the case for high-lying H$_2$O lines \citep{gonza14}.

It is clear from our preliminary analysis that a higher number of molecular transitions, including vibrationally excited ones,  at the ALMA resolution is necessary in order to determine the chemical and physical characteristics of each subregion within the CND. In a future paper we shall attempt a more thorough modelling which will include 2D maps as well as Principal Component Analysis to investigate the relationship between the different species across the structures in the maps, and a detailed analysis of the individual spectra by the use of a larger suite of chemical, PDR, XDR and non LVG line radiative transfer models. 
\begin{acknowledgements}
         We thank the staff of ALMA in Chile and the ARC-people at IRAM-Grenoble in France
for their invaluable help during the data reduction process. AU and PP acknowledge support from Spanish grants AYA2012-32295 and FIS2012-32096. This paper
makes use of the following ALMA data: ADS/JAO.ALMA$\#$2011.0.00083.S.
ALMA is a partnership of ESO (representing its member states), NSF (USA)
and NINS (Japan), together with NRC (Canada) and NSC and ASIAA (Taiwan),
in cooperation with the Republic of Chile.  The Joint ALMA Observatory is
operated by ESO, AUI/NRAO and NAOJ. The National Radio Astronomy
Observatory is a facility of the National Science Foundation operated under cooperative
agreement by Associated Universities, Inc. Some of the observations used were
carried out with the IRAM Plateau de Bure Interferometer. IRAM is supported by INSU/CNRS (France), MPG (Germany) and IGN (Spain).
SGB and IM acknowledge support from Spanish grants AYA2010-15169,  AYA2012-32295 and from the Junta de Andalucia through TIC-114 and the Excellence Project P08-TIC-03531. SGB and AF acknowledge support from MICIN within program CONSOLIDER INGENIO 2010, under grant `Molecular Astrophysics: The Herschel and ALMA Era--ASTROMOL' (ref CSD2009-00038).
\end{acknowledgements}

\bibliographystyle{aa}    
\bibliography{references}

\begin{thebibliography}{55}
\expandafter\ifx\csname natexlab\endcsname\relax\def\natexlab#1{#1}\fi

\bibitem[{{Aalto} {et~al.}(2011){Aalto}, {Costagliola}, {van der Tak}, \& {et
  al.}}]{Aalt11}
{Aalto}, S., {Costagliola}, F., {van der Tak}, F., \& {et al.} 2011, \aap, 527,
  69

\bibitem[{{Aalto} {et~al.}(2012){Aalto}, {Garc{\'{\i}}a-Burillo}, {Muller}, \&
  {et al.}}]{Aalt12}
{Aalto}, S., {Garc{\'{\i}}a-Burillo}, S., {Muller}, S., \& {et al.} 2012, \aap,
  537, 44

\bibitem[{{Aladro} {et~al.}(2013){Aladro}, {Viti}, {Bayet}, \& {et.
  al.}}]{Alad13}
{Aladro}, R., {Viti}, S., {Bayet}, E., \& {et. al.} 2013, \aap, 549, 39

\bibitem[{{Asplund} {et~al.}(2009){Asplund}, {Grevesse}, {Sauval}, \&
  {Scott}}]{Aspl09}
{Asplund}, M., {Grevesse}, N., {Sauval}, A.~J., \& {Scott}, P. 2009, \araa, 47,
  481

\bibitem[{{Bayet} {et~al.}(2008){Bayet}, {Viti}, {Williams}, \&
  {Rawlings}}]{Baye08}
{Bayet}, E., {Viti}, S., {Williams}, D.~A., \& {Rawlings}, J.~M.~C. 2008, \apj,
  676, 978

\bibitem[{{Bayet} {et~al.}(2009){Bayet}, {Viti}, {Williams}, {Rawlings}, \&
  {Bell}}]{Baye09}
{Bayet}, E., {Viti}, S., {Williams}, D.~A., {Rawlings}, J.~M.~C., \& {Bell}, T.
  2009, \apj, 696, 1466

\bibitem[{{Bayet} {et~al.}(2011){Bayet}, {Williams}, {Hartquist}, \&
  {Viti}}]{Baye11}
{Bayet}, E., {Williams}, D.~A., {Hartquist}, T.~W., \& {Viti}, S. 2011, \mnras,
  414, 627

\bibitem[{{Cicone} {et~al.}(2014){Cicone}, {Maiolino}, {Sturm}, \& {et.
  al.}}]{Cico14}
{Cicone}, C., {Maiolino}, R., {Sturm}, E., \& {et. al.} 2014, \aap, 562, 21

\bibitem[{{Combes}(2006)}]{Comb06}
{Combes}, F. 2006, Astrophysics Update, 2, 159

\bibitem[{{Combes} {et~al.}(2013){Combes}, {Garc{\'{\i}}a-Burillo}, {Casasola},
  \& {et al.}}]{Comb13}
{Combes}, F., {Garc{\'{\i}}a-Burillo}, S., {Casasola}, V., \& {et al.} 2013,
  \aap, 558, 124

\bibitem[{{Costagliola} {et~al.}(1011){Costagliola}, {Aalto}, {Rodriguez}, \&
  {et. al.}}]{Costa11}
{Costagliola}, F., {Aalto}, S., {Rodriguez}, M.~I., \& {et. al.} 1011, \aap,
  528, 30

\bibitem[{{Dyson} \& {Williams}(1997)}]{dyswill97}
{Dyson}, J.~E. \& {Williams}, D.~A. 1997, {The Physics of the Interstellar
  Medium} (Taylor \& Francis)

\bibitem[{{Feruglio} {et~al.}(2010){Feruglio}, {Maiolino}, {Piconcelli}, \&
  {et. al.}}]{Feru10}
{Feruglio}, C., {Maiolino}, R., {Piconcelli}, E., \& {et. al.} 2010, \aap, 518,
  155

\bibitem[{{Fuente} {et~al.}(2008){Fuente}, {Garc{\'{\i}}a-Burillo}, {Usero}, \&
  {et. al.}}]{Fuen08}
{Fuente}, A., {Garc{\'{\i}}a-Burillo}, S., {Usero}, M., \& {et. al.} 2008,
  \aap, 492, 675

\bibitem[{{Garc{\'{\i}}a-Burillo} {et~al.}(2008){Garc{\'{\i}}a-Burillo},
  {Combes}, {Grac{\'{\i}}a-Carpio}, \& {et. al.}}]{Garc08}
{Garc{\'{\i}}a-Burillo}, S., {Combes}, F., {Grac{\'{\i}}a-Carpio}, J., \& {et.
  al.} 2008, \aaps, 313, 261

\bibitem[{{Garc{\'{\i}}a-Burillo} {et~al.}(2014){Garc{\'{\i}}a-Burillo},
  {Combes}, {Usero}, \& {et. al.}}]{Garc14}
{Garc{\'{\i}}a-Burillo}, S., {Combes}, F., {Usero}, A., \& {et. al.} 2014,
  \aap, in press (Paper I)

\bibitem[{{Garc{\'{\i}}a-Burillo} {et~al.}(2002){Garc{\'{\i}}a-Burillo},
  {Mart{\'{\i}}n-Pintado}, {Fuente}, {Usero}, \& {Neri}}]{Garc02}
{Garc{\'{\i}}a-Burillo}, S., {Mart{\'{\i}}n-Pintado}, J., {Fuente}, A.,
  {Usero}, A., \& {Neri}, R. 2002, \apjl, 575, L55

\bibitem[{{Garc{\'{\i}}a-Burillo} {et~al.}(2010){Garc{\'{\i}}a-Burillo},
  {Usero}, {Fuente}, \& {et. al.}}]{Garc10}
{Garc{\'{\i}}a-Burillo}, S., {Usero}, A., {Fuente}, A., \& {et. al.} 2010,
  \aap, 519

\bibitem[{{Goldsmith} \& {Langer}(1999)}]{Gold99}
{Goldsmith}, P.~F. \& {Langer}, W.~D. 1999, \apj, 517, 209

\bibitem[{{Gonz{\' a}lez-Alfonso} {et~al.}(2014){Gonz{\' a}lez-Alfonso},
  {Fischer}, {Aalto}, \& {Falstad}}]{gonza14}
{Gonz{\' a}lez-Alfonso}, E., {Fischer}, J., {Aalto}, S., \& {Falstad}, N. 2014,
  \aap, in press

\bibitem[{{Hailey-Dunsheath} {et~al.}(2012){Hailey-Dunsheath}, {Sturm},
  {Fisher}, \& {et. al.}}]{Hai12}
{Hailey-Dunsheath}, S., {Sturm}, E., {Fisher}, J., \& {et. al.} 2012, \apj,
  755, 57

\bibitem[{{Henkel} {et~al.}(2014){Henkel}, {Asiri}, {Ao}, \& {et.
  al.}}]{Henk14}
{Henkel}, C., {Asiri}, H., {Ao}, Y., \& {et. al.} 2014, \aap, 565, 3

\bibitem[{{Israel}(2009)}]{Isra09}
{Israel}, F.~P. 2009, \aap, 493, 525

\bibitem[{{Jimenez-Serra} {et~al.}(2008){Jimenez-Serra}, {Caselli},
  {Mart{\'{\i}}n-Pintado}, \& {Hartquist}}]{Jim08}
{Jimenez-Serra}, I., {Caselli}, P., {Mart{\'{\i}}n-Pintado}, J., \&
  {Hartquist}, T.~W. 2008, \aap, 482, 549

\bibitem[{{Kamenetzky} {et~al.}(2011){Kamenetzky}, {Glenn}, {Maloney}, \& {et
  al.}}]{Kame11}
{Kamenetzky}, J., {Glenn}, J., {Maloney}, P.~R., \& {et al.} 2011, \apj, 731,
  83

\bibitem[{{Krips} {et~al.}(2011){Krips}, {Mart{\'{\i}}n}, {Eckart}, \& {et.
  al.}}]{Krip11}
{Krips}, M., {Mart{\'{\i}}n}, S., {Eckart}, A., \& {et. al.} 2011, \apj, 736,
  37

\bibitem[{{Krips} {et~al.}(2008){Krips}, {Neri}, {Garc{\'{\i}}a-Burillo},
  {Mart{\'{\i}}n}, {Combes}, {Graci{\'a}-Carpio}, \& {Eckart}}]{Krip08}
{Krips}, M., {Neri}, R., {Garc{\'{\i}}a-Burillo}, S., {et~al.} 2008, \apj, 677,
  262

\bibitem[{{Loenen} {et~al.}(2008){Loenen}, {Spaans}, {Baan}, \&
  {Meijerink}}]{Loen08}
{Loenen}, A.~F., {Spaans}, M., {Baan}, W.~A., \& {Meijerink}, R. 2008, \aap,
  488

\bibitem[{{Mart{\'{\i}}n} {et~al.}(2011){Mart{\'{\i}}n}, {Krips},
  {Mart{\'{\i}}n-Pintado}, \& {et. al.}}]{Mart11}
{Mart{\'{\i}}n}, S., {Krips}, M., {Mart{\'{\i}}n-Pintado}, J., \& {et. al.}
  2011, \aap, 527, 36

\bibitem[{{Meier} \& {Turner}(2005)}]{Meie05}
{Meier}, D.~S. \& {Turner}, J.~L. 2005, \apj, 618, 259

\bibitem[{{Meier} \& {Turner}(2012)}]{Meie12}
{Meier}, D.~S. \& {Turner}, J.~L. 2012, \apj, 755, 104

\bibitem[{{Meijerink} \& {Spaans}(2005)}]{Meij05}
{Meijerink}, R. \& {Spaans}, M. 2005, \aap, 436, 397

\bibitem[{{Meijerink} {et~al.}(2007){Meijerink}, {Spaans}, \&
  {Israel}}]{Meij07}
{Meijerink}, R., {Spaans}, M., \& {Israel}, F.~P. 2007, \aap, 461, 793

\bibitem[{{Meijerink} {et~al.}(2013){Meijerink}, {Spaans}, {Kamp}, \& {et.
  al.}}]{Meij13}
{Meijerink}, R., {Spaans}, M., {Kamp}, I., \& {et. al.} 2013, The Journal of
  Physical Chemistry A, 117, 9593

\bibitem[{{Meijerink} {et~al.}(2011){Meijerink}, {Spaans}, {Loenen}, \& {van
  der Werf}}]{Meij11}
{Meijerink}, R., {Spaans}, M., {Loenen}, A.~F., \& {van der Werf}, P.~P. 2011,
  \aap, 525, 119

\bibitem[{{Milam} {et~al.}(2005){Milam}, {Savage}, {Brewster}, \& {et.
  al.}}]{Mila05}
{Milam}, S.~N., {Savage}, C., {Brewster}, M.~A., \& {et. al.} 2005, \apj, 634,
  1126

\bibitem[{{Omont}(2007)}]{Omont07}
{Omont}, A. 2007, Reports on Progress in Physics, 70, 1099

\bibitem[{{Papadopoulos} \& {Seaquist}(1999)}]{Papa99}
{Papadopoulos}, P.~P. \& {Seaquist}, E.~R. 1999, \apjl, 516, 114

\bibitem[{{Roberts} {et~al.}(2007){Roberts}, {Rawlings}, {Viti}, \&
  {Williams}}]{Robe07}
{Roberts}, J.~F., {Rawlings}, J.~M.~C., {Viti}, S., \& {Williams}, D.~A. 2007,
  \mnras, 382, 733

\bibitem[{{Sakamoto} {et~al.}(2010){Sakamoto}, {Aalto}, {Evans}, \& {et
  al.}}]{SaKa10}
{Sakamoto}, S., {Aalto}, S., {Evans}, A.~S., \& {et al.} 2010, \apjl, 725

\bibitem[{{Schinnerer} {et~al.}(2000){Schinnerer}, {Eckart}, {Tacconi}, \& {et.
  al.}}]{Schi00}
{Schinnerer}, E., {Eckart}, A., {Tacconi}, L., \& {et. al.} 2000, \apj, 533,
  850

\bibitem[{{Sch{\"o}ier} {et~al.}(2005){Sch{\"o}ier}, {van der Tak}, {van
  Dishoeck}, \& {Black}}]{Scho05}
{Sch{\"o}ier}, F.~L., {van der Tak}, F.~F.~S., {van Dishoeck}, E.~F., \&
  {Black}, J.~H. 2005, \aap, 432, 369

\bibitem[{{Tacconi} {et~al.}(1997){Tacconi}, {Gallimore}, {Genzel}, \& {et.
  al.}}]{Tacc97}
{Tacconi}, L.~J., {Gallimore}, J.~F., {Genzel}, R.~C., \& {et. al.} 1997,
  \apss, 248, 59

\bibitem[{{Tacconi} {et~al.}(1994){Tacconi}, {Genzel}, {Blietz}, \& {et.
  al.}}]{Tacc94}
{Tacconi}, L.~J., {Genzel}, R., {Blietz}, M., \& {et. al.} 1994, \apjl, 426

\bibitem[{{Tsai} {et~al.}(2012){Tsai}, {Hwang}, {Matsushita}, \& {et.
  al.}}]{Tsai12}
{Tsai}, M., {Hwang}, C.-Y., {Matsushita}, S., \& {et. al.} 2012, \apj, 746, 129

\bibitem[{{Usero} {et~al.}(2004){Usero}, {Garc{\'{\i}}a-Burillo}, {Fuente},
  {Mart{\'{\i}}n-Pintado}, \& {Rodr{\'{\i}}guez-Fern{\' a}ndez}}]{User04}
{Usero}, A., {Garc{\'{\i}}a-Burillo}, S., {Fuente}, A.,
  {Mart{\'{\i}}n-Pintado}, J., \& {Rodr{\'{\i}}guez-Fern{\' a}ndez}, N.~J.
  2004, \aap, 419, 897

\bibitem[{{Usero} {et~al.}(2006){Usero}, {Garc{\'{\i}}a-Burillo},
  {Mart{\'{\i}}n-Pintado}, {Fuente}, \& {Neri}}]{User06}
{Usero}, A., {Garc{\'{\i}}a-Burillo}, S., {Mart{\'{\i}}n-Pintado}, J.,
  {Fuente}, A., \& {Neri}, R. 2006, \aap, 448, 457

\bibitem[{{van der Tak} {et~al.}(2007){van der Tak}, {Black}, {Sch{\"o}ier},
  {Jansen}, \& {van Dishoeck}}]{VanderTak07}
{van der Tak}, F.~F.~S., {Black}, J.~H., {Sch{\"o}ier}, F.~L., {Jansen}, D.~J.,
  \& {van Dishoeck}, E.~F. 2007, \aap, 468, 627

\bibitem[{{Veilleux} {et~al.}(2013){Veilleux}, {Mel{\'e}ndez}, {Sturm}, \& {et.
  al.}}]{Veil13}
{Veilleux}, S., {Mel{\'e}ndez}, M., {Sturm}, E., \& {et. al.} 2013, \apj, 776

\bibitem[{{Viti} {et~al.}(2004){Viti}, {Collings}, {Dever}, {McCoustra}, \&
  {Williams}}]{Viti04}
{Viti}, S., {Collings}, M.~P., {Dever}, J.~W., {McCoustra}, M.~R.~S., \&
  {Williams}, D.~A. 2004, \mnras, 354, 1141

\bibitem[{{Viti} {et~al.}(2011){Viti}, {Jimenez-Serra}, {Yates}, \& {et.
  al.}}]{Viti11}
{Viti}, S., {Jimenez-Serra}, I., {Yates}, J.~A., \& {et. al.} 2011, \apjl, 740

\bibitem[{{Wang}(2004)}]{Wang04}
{Wang}, J. 2004, \apjl, 614, L21

\bibitem[{{Watanabe} {et~al.}(2014){Watanabe}, {Sakai}, {Sorai}, \&
  {Yamamoto}}]{Watanabe14}
{Watanabe}, Y., {Sakai}, N., {Sorai}, K., \& {Yamamoto}, S. 2014, \apj, 788, 4

\bibitem[{{Woodall} {et~al.}(2007){Woodall}, {Ag{\'u}ndez}, {Markwick-Kemper},
  \& {Millar}}]{Wood07}
{Woodall}, J., {Ag{\'u}ndez}, M., {Markwick-Kemper}, A.~J., \& {Millar}, T.~J.
  2007, \aap, 466, 1197

\bibitem[{{Zinnecker} \& {Yorke}(2007)}]{Zinn07}
{Zinnecker}, H. \& {Yorke}, H.~W. 2007, \araa, 45, 481

\end{thebibliography}

\end{document}